\documentclass[twocolumn]{aastex7}

\defcitealias{Nguyen23}{N23}

\newcommand{\Msun}{\ensuremath{M_{\odot}}}
\newcommand{\Mbh}{\ensuremath{M_{\rm BH}}}
\newcommand{\Mnsc}{\ensuremath{M_{\rm NSC}}}

\newcommand{\Lsun}{\ensuremath{L_\odot}}

\renewcommand{\deg}{\ensuremath{^{\circ}}}
\newcommand{\ml}{\emph{M/L}}

\newcommand{\kms}{\ensuremath{\text{km~s}^{-1}}}
\newcommand{\se}{\ensuremath{\arcsec}}
\newcommand{\mm}{\ensuremath{\arcmin}}

\newcommand{\go}{\ensuremath{\text{G}_{\rm outer}}}

\newcommand{\hst}{\emph{HST}}
\newcommand{\jwst}{\emph{JWST}}

\hypersetup{colorlinks, linkcolor=blue, citecolor=cyan, urlcolor=magenta}

\usepackage{newtxtext,newtxmath}

\graphicspath{{./}{figures/}}

\received{May 5 2025}
\revised{June 18 2025}
\accepted{June 19 2025}

\shorttitle{IMBH Mass Measurements with HARMONI}
\shortauthors{D.\ D.\ Nguyen et al.} 

\begin{document} 

\title{Simulating intermediate black hole mass measurements for a sample of galaxies with nuclear star clusters using ELT/HARMONI high spatial resolution integral-field stellar kinematics}

\correspondingauthor{Dieu D.\ Nguyen}\email{dieun@umich.edu}
\author[0000-0002-5678-1008]{Dieu D.\ Nguyen}
\email{nddieuphys@gmail.com, dieun@umich.edu}
\affil{Department of Astronomy, University of Michigan, 1085 S University, Ann Arbor, MI 48109, USA}

\author[0000-0002-1283-8420]{Michele Cappellari}
\email{michele.cappellari@physics.ox.ac.uk}
\affiliation{Sub-Department of Astrophysics, Department of Physics, University of Oxford, Denys Wilkinson Building, Keble Road, Oxford, OX1 3RH, UK}

\author[0009-0006-5852-4538]{Hai N.\ Ngo}
\email{hai10hoalk@gmail.com}
\affiliation{Faculty of Physics -- Engineering Physics, University of Science, Vietnam National University in Ho Chi Minh City, Vietnam}

\author[0009-0004-3689-8577]{Tinh Q.\  T.\ Le}
\email{lethongquoctinh01@gmail.com}
\affiliation{Department of Physics, International University,Vietnam National University in Ho Chi Minh City, Vietnam}
\affiliation{International Centre for Interdisciplinary Science and Education, 07 Science Avenue, Ghenh Rang, 55121 Quy Nhon, Vietnam}
\affiliation{Vietnam National Space Center, Vietnam Academy of Science and Technology, 18 Hoang Quoc Viet, Cau Giay, Hanoi, Vietnam}

\author[0009-0009-0015-1208]{Tuan N.\ Le}
\email{tuan.le.nutshell@gmail.com}
\affiliation{University of Technology, Vietnam National University in Ho Chi Minh City, Vietnam}
\affiliation{International Centre for Interdisciplinary Science and Education, 07 Science Avenue, Ghenh Rang, 55121 Quy Nhon, Vietnam}

\author[0009-0007-3200-8751]{Khue N.\ H.\ Ho}
\email{khuehohuyngoc@gmail.com }
\affiliation{Department of Physics, International University,Vietnam National University in Ho Chi Minh City, Vietnam}

\author[0009-0002-0063-0857]{An K.\ Nguyen}
\email{tiffanylarcher559@gmail.com }
\affiliation{Department of Physics, International University,Vietnam National University in Ho Chi Minh City, Vietnam}

\author[0009-0007-6980-9480]{Phong T.\ On}
\email{onphong9502@gmail.com}
\affiliation{Department of Physics, International University,Vietnam National University in Ho Chi Minh City, Vietnam}

\author[0009-0008-6050-5736]{Huy G.\ Tong}
\email{tonggiahuy191203@gmail.com}
\affiliation{Faculty of Physics -- Engineering Physics, University of Science, Vietnam National University in Ho Chi Minh City, Vietnam}

\author[0000-0002-6694-5184]{Niranjan Thatte}
\email{niranjan.thatte@physics.ox.ac.uk}
\affiliation{Sub-Department of Astrophysics, Department of Physics, University of Oxford, Denys Wilkinson Building, Keble Road, Oxford, OX1 3RH, UK}

\author[0000-0002-4005-9619]{Miguel Pereira-Santaella}
\email{miguel.pereira@iff.csic.es}
\affiliation{Instituto de F\'isica Fundamental, CSIC, Calle Serrano 123, 28006 Madrid, Spain}

\begin{abstract}
    Understanding the demographics of intermediate-mass black holes (IMBHs, $M_{\rm BH} \approx 10^2-10^5$ M$_\odot$) in low-mass galaxies is key to constraining black hole seed formation models, but detecting them is challenging due to their small gravitational sphere of influence (SOI). The upcoming ELT/HARMONI instrument, with its high angular resolution, offers a promising solution. We present simulations assessing HARMONI's ability to measure IMBH masses in nuclear star clusters (NSCs) of nearby dwarf galaxies. We selected a sample of 44 candidates within 10 Mpc. For two representative targets, NGC 300 and NGC 3115 dw01, we generated mock HARMONI integral-field data cubes using realistic inputs derived from \hst\ imaging, stellar population models, and Jeans Anisotropic Models (JAM), assuming IMBH masses up to 1\% of the NSC mass. We simulated observations across six NIR gratings at 10 mas resolution. Analyzing the mock data with standard kinematic extraction (pPXF) and JAM models in a Bayesian framework, we demonstrate that HARMONI can resolve the IMBH SOI and accurately recover masses down to $\approx 0.5\%$ of the NSC mass within feasible exposure times. These results highlight HARMONI's potential to revolutionize IMBH studies.
\end{abstract}


\keywords{\uat{Astrophysical black holes}{98} --- \uat{Galaxy kinematics}{602} --- \uat{Galaxy dynamics}{591} --- \uat{Galaxy nuclei}{609} --- \uat{Galaxy spectroscopy}{2171} --- \uat{Astronomy data modeling}{1859}}


\section{Introduction}\label{intro}

Supermassive black holes (SMBHs, $M_{\rm BH}\gtrsim10^6$ \Msun) residing in the centers of massive galaxies ($M_{\star}\gtrsim10^{10}$ \Msun) exhibit well-established correlations with host galaxy properties, such as bulge stellar mass ($M_{\star}$) and velocity dispersion ($\sigma_\star$) \citep{Kormendy95, Magorrian98, Ferrarese00, Gebhardt00, Kormendy13}. These scaling relations suggest a fundamental co-evolution between SMBHs and their hosts \citep{Silk98, DiMatteo08, Fabian12, Krajnovic18a}, offering insights into galaxy assembly and the growth of cosmic structures \citep{Gultekin09, McConnell11, Netzer15, Nightingale23}.

However, a significant gap exists between the observed populations of stellar-mass black holes (s-BHs, $M_{\rm BH}\lesssim10^2$ \Msun), the remnants of massive stars \citep{Graham20}, and the dynamically confirmed SMBHs ($M_{\rm BH}\gtrsim10^5$ \Msun) \citep{Pechetti22}. The nature of objects within this intermediate-mass range (IMBHs, $M_{\rm BH}\approx10^{2-5}$ \Msun) remains unconstrained \citep{Nguyen17, Nguyen18, Nguyen19}.

Understanding the demographics of IMBHs is crucial for deciphering the origins of SMBHs. Current theories propose that SMBHs grow from ``seeds'' formed in the early Universe, which could be either light seeds ($M_{\bullet}\approx10^{2-3}$ \Msun) derived from the first stars \citep{vanWassenhove10, Volonteri12, Volonteri2021} or heavy seeds ($M_{\bullet}\approx10^{4-5}$ \Msun) formed via direct gas collapse \citep{Greene12, Bonoli14}. The abundance and mass distribution of IMBHs in present-day low-mass galaxies provide a key observational test for these seeding scenarios \citep{Mezcua17, Greene20, Inayoshi20}.

Detecting and characterizing IMBHs is observationally challenging due to their relatively small gravitational influence compared to SMBHs, requiring high spatial resolution kinematics. This difficulty explains the current uncertainty surrounding their existence and properties, motivating the simulations presented in this work.

\subsection{Previous IMBH Determinations and Observational Challenges}\label{previousIMBH}

Dynamical mass measurements, which infer mass from the motion of tracers (stars or gas) under gravity, exist for roughly 120 black holes \citep[BHs, see review by][]{Kormendy13}. However, only about 23 of these fall within the intermediate-mass black hole (IMBH) range ($M_{\rm BH}\sim10^{2-5}$ \Msun) \citep[see review by][]{Greene20}. These measurements utilize various kinematic tracers and techniques, including stellar dynamics \citep[e.g.,][]{Verolme02, Gebhardt2003, Nguyen17conf, Nguyen2025a}, ionized gas dynamics \citep[e.g.,][]{Barth2001, Walsh13}, molecular gas dynamics \citep[e.g.,][]{Davis13Nature, Nguyen19conf, Nguyen20}, atomic gas dynamics \citep[e.g.,][]{Nguyen21}, and maser disk kinematics \citep[e.g.,][]{Miyoshi95, Gonzalez-Alfonso}.

While SMBHs are common in massive galaxies  \citep{Kormendy13, McConnell2013, Saglia16}, the existence and characteristics of IMBHs in lower-mass galaxies are less certain. Some studies report detections \citep{Nguyen19, Nguyen22, Pechetti22}, while others provide only upper limits or suggest their absence \citep{Gebhardt01, Merritt2001, Valluri05, Nguyen14}. Indirect evidence, such as the off-nuclear X-ray source HLX-1 \citep{Straub14}, also points towards the existence of IMBHs. Promising locations for IMBH searches include nuclear star clusters (NSCs) in dwarf galaxies, globular clusters (GCs), and ultra-compact dwarf galaxies (UCDs) \citep{Gebhardt05, Noyola10, Seth14, Neumayer20}.

A major challenge in dynamically detecting IMBHs is resolving their gravitational sphere of influence (SOI), the region where the BH's gravity dominates. The SOI radius is approximately $R_{\rm SOI}\equiv GM_{\rm BH}/\sigma^2_\star$, where $\sigma_\star$ is the stellar velocity dispersion outside the SOI. For a typical IMBH ($M_{\rm BH}\approx10^5$ \Msun) in a dwarf galaxy ($\sigma_\star\approx10$ \kms) at a distance of $D\approx10$ Mpc, the SOI radius is very small, $R_{\rm SOI}\approx0\farcs09$ \citep{deZeeuw01}.

This small angular scale is difficult to resolve with current telescopes. Even facilities equipped with adaptive optics (AO), such as VLT/ERIS or Keck/OSIRIS (achieving a point spread function  (PSF) FWHM$_{\rm PSF}\approx0\farcs05$), struggle to resolve the SOI of IMBHs beyond the Local Group. The {\it James Webb Space Telescope} (\jwst), despite its excellent sensitivity, lacks the angular resolution needed for unambiguous kinematic detection of IMBHs in distant dwarf galaxies. As a result, the demographics of IMBHs remain poorly constrained.

Understanding the IMBH population is vital for several reasons. It helps distinguish between different BH seeding scenarios (e.g., light seeds from Population III stars versus heavy seeds from direct gas collapse), which predict varying IMBH occupation fractions in low-mass galaxies \citep{Gallo08, Nguyen18, Nguyen19}. It is also crucial for understanding BH--galaxy co-evolution and for predicting gravitational wave event rates for future observatories \citep{Kochanek16, Bailes21}. The limitations of current instruments highlight the need for next-generation facilities like the Extremely Large Telescope (ELT) with instruments such as the High Angular Resolution Monolithic Optical and Near-infrared Integral field spectrograph (HARMONI).

\subsection{The role of ELT/HARMONI}\label{harmoni_role}

Detecting IMBHs ($M_{\rm BH}\sim10^{2-5}$ \Msun) is challenging because their gravitational SOI is extremely small (\autoref{previousIMBH}), requiring observations with very high angular resolution. The HARMONI \citep{Thatte16, Thatte20, Thatte2024} on the ELT is well-suited for this task. Equipped with AO, HARMONI will provide excellent sensitivity along with high spectral ($\lambda/\Delta\lambda\sim3300-17,400$) and angular (FWHM$_{\rm PSF}\approx12$ milli-arcseconds, mas) resolution.

With these capabilities, HARMONI can resolve the SOI and measure detailed stellar kinematics within NSCs -- the likely hosts of IMBHs -- in galaxies out to $\approx$10 Mpc (this work). HARMONI will greatly expand the volume accessible for IMBH searches compared to current instruments and should provide the first robust constraints on IMBH demographics. This is crucial for distinguishing between BH seeding scenarios in the early Universe \citep{Greene20, Inayoshi20} and understanding the co-evolution of BHs and galaxies at low masses. This work complements our previous study on HARMONI's capability for measuring 
SMBHs in more massive galaxies \citep[][hereafter \citetalias{Nguyen23}]{Nguyen23}.

We outline the selection criteria for our HARMONI IMBH survey in \autoref{sample} and present the sample properties in \autoref{sample_properties}. The dynamical and photometric models are described in \autoref{dynamical_model} and \autoref{photometry}, respectively. \autoref{simulations} details the mock HARMONI IFS simulations using the HARMONI SIMulator \citep[{\tt HSIM}\footnote{v3.11; available from \url{https://github.com/HARMONI-ELT/HSIM}};][]{Zieleniewski15} and the kinematic extraction process. We discuss the recovery of IMBH masses and potential limitations in \autoref{discussion}, and summarize our findings in \autoref{conclusions}.

Throughout this paper, we adopt a flat $\Lambda$CDM cosmology with $H_0=70$ km s$^{-1}$ Mpc$^{-1}$, $\Omega_{\rm m,0}=0.3$, and $\Omega_{\rm \Lambda,0}=0.7$. All magnitudes use the AB system \citep{Oke74} and are corrected for foreground extinction \citep{Schlafly11} using the \citet{Cardelli89} law.

\section{Sample selection}\label{sample}

Our sample selection process aims to identify nearby, low-mass galaxies ($M_{\star}\lesssim5\times10^{10}$ \Msun) likely hosting NSCs and potentially IMBHs, suitable for observation with ELT/HARMONI. Unlike the selection for the MMBH survey of ultra-massive galaxies \citepalias{Nguyen23}, which relied heavily on homogeneous 2MASS $K_s$-band photometry \citep{Skrutskie06, Jarrett03} from the 2MASS Redshift Survey \citep[2MRS;][]{Huchra12}, we adopted a multi-stage approach due to the limitations of 2MASS for dwarf galaxies. The shallow depth and fixed aperture integration method of 2MASS can lead to unreliable total luminosities, effective radii ($R_{\rm e}$), and S\'ersic indices for compact, low-surface-brightness dwarf galaxies \citep{Schombert12}.

We began with the 2MRS catalog, estimating stellar masses from $K_s$-band luminosities assuming a constant mass-to-light ratio \ml$_K=1$ (\Msun/\Lsun) \citep{Nguyen18} and using distances from the Updated Nearby Galaxy Catalog \citep{Karachentsev13}. We then estimated a virial velocity dispersion proxy using the well-calibrated virial relation $\sigma_{\rm vir}^2 = G M_\star/(5R_{\rm e})$ \citep{Cappellari2006}, where $R_{\rm e}$ was taken from 2MRS. We applied the following initial cuts:
\begin{itemize}
    \item Virial dispersion proxy $\sigma_{\rm vir} \lesssim 70$ \kms. This threshold aims to select galaxies likely hosting IMBHs, based on extrapolating the $M_{\rm BH}$--$\sigma_\star$ relation \citep{Greene20}. Lower dispersions also imply larger angular sizes for the IMBH $R_{\rm SOI}$, making them easier to resolve kinematically.
    \item Distance $D \lesssim 10$ Mpc. This ensures the targets are close enough for HARMONI to potentially resolve the NSC structure and IMBH SOI.
    \item ELT Observability: Declination $|\delta + 24^\circ|< 45^\circ$ for good adaptive optics (AO) performance from the Armazones site.
    \item Galactic Plane Exclusion: Galactic latitude $|b| > 8^\circ$ to avoid regions of high dust extinction.
\end{itemize}
These criteria yielded an initial list of $\approx$500 galaxies.

We then visually inspected images for these $\approx$500 candidates, primarily using the Sloan Digital Sky Survey \citep[SDSS;][]{STScI2020} or Digitized Sky Survey \citep[DSS;][]{STScI2020} and higher-resolution  {\it Hubble Space Telescope} (\hst) archives when available, to identify galaxies clearly exhibiting a central nucleus (NSC).

Finally, to ensure completeness, we cross-referenced our visually selected nucleated galaxies with extensive literature compilations and studies focused on NSCs. This included photometric \citep{Boker02, Walcher05, Walcher06, Seth08, Georgiev14, Georgiev16, Pechetti20} and spectroscopic \citep{Ho09, Kacharov18, Nguyen18, Fahrion22, Baldassare22} surveys, as well as detailed case studies \citep{Hagele07, Shields08, deSwardt10, Fahrion20, Muller21, Pinna21, Nguyen22}. Any galaxy meeting our initial criteria ($D$, $\sigma_{\rm vir}$, observability) that was confirmed to host an NSC in the literature was added to our final sample, even if missed during the visual inspection phase.

\begin{figure*}
    \centering\includegraphics[width=0.99\textwidth]{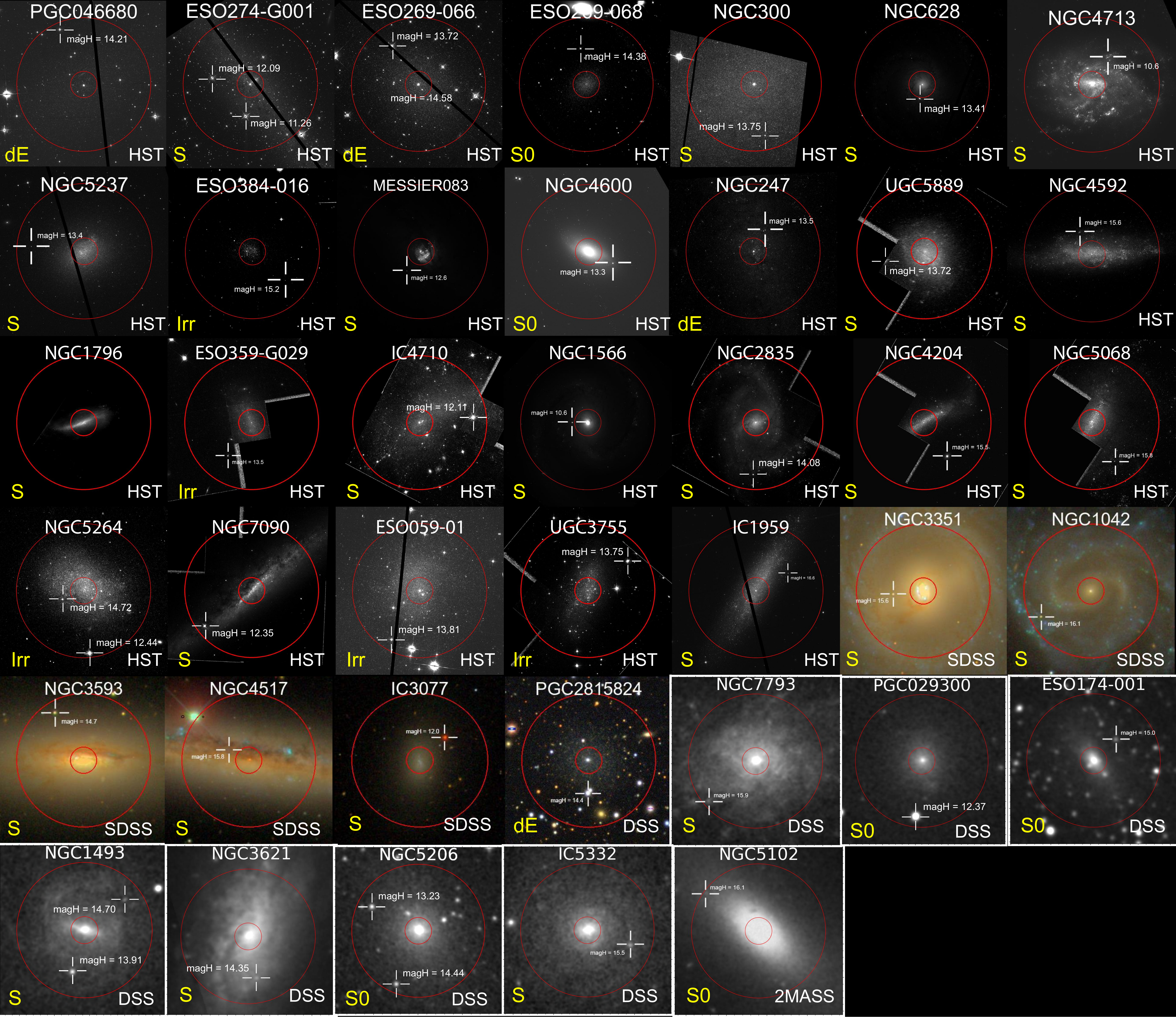}
    \caption{The 40 available observations out of the total number of 44 dwarf galaxies from our HARMONI IMBH sample, as captured in \hst, SDSS, DSS, and 2MASS images. To emphasize their NSCs at the core, the galaxy brightness has been adjusted. Each galaxy's name and galaxy types are also shown. Two red circles define the region around the galaxy's center, spanning from 12$\arcsec$ to 60$\arcsec$, where we search for a faint NGS essential for LTAO performance. The white crosses pinpoint the available NGS locations, specifying their apparent AB magnitudes measured in the $H$ band. Here, dE = dwarf Elliptical, S0 = lenticular, S = Sprial, Irr = Irregular.}
    \label{Tiptilt_star}
\end{figure*}

This process resulted in a final sample of 44 galaxies, which we refer to as the ``HARMONI IMBH survey''. This sample spans stellar masses $2\times10^7\lesssim M_{\star}\lesssim5\times10^{10}$ \Msun\ (corresponding to $K_s$ absolute magnitudes $-23.5\lesssim M_K\lesssim-15.4$ mag) and includes a mix of morphological types (6 dE, 7 S0, 27 S, 4 Irr; see \autoref{tab:parent_specs}). \autoref{Tiptilt_star} displays images of the sample galaxies, and their properties are summarized in \autoref{tab_imbhsample}. The selection criteria are summarized in \autoref{tab:parent_criteria}.

A crucial practical requirement for HARMONI observations using Laser Tomography Adaptive Optics (LTAO\footnote{LTAO mode on the ELT/HARMONI instrument needs at least one NGS to work simultaneously with six artificial off-axis Laser Guide Stars (LGS). This system allows the required NGS to be more than 10,000 times fainter than those needed for classical AO used on Gemini and VLT.}) is the availability of a nearby tip-tilt star. This natural guide star (NGS) is needed for correcting atmospheric turbulence and should ideally have an $H$-band AB magnitude $m_H<20.4$ mag, located within an annulus of 12$\arcsec$ to 60$\arcsec$ surrounding the science target \citep{Thatte20}. We searched Gaia and SDSS archives for suitable NGS candidates for our sample galaxies. Potential NGS are marked in \autoref{Tiptilt_star} with their estimated $H$-band AB magnitudes. SDSS magnitudes were converted following Section 2.1 of \citetalias{Nguyen23}, and Gaia Vegamags were converted using relations from the Gaia DR3 documentation \citep[Section 5.5.1;][]{Busso22}. Suitable NGS were found for the vast majority of targets, confirming the feasibility of LTAO observations.

\begin{table}
\caption{Targets selection criteria for low-mass galaxies hosting NSCs}
\centering
\begin{tabular}{rl}
\hline
Distant range: & $D\lesssim10$ Mpc \\
Galaxies $K$s-band absolute magnitude: & $-23.5\leq M_K\leq-15.4$ mag \\
Central stellar-velocity dispersion range: & $\sigma_{\star,\rm c}\lesssim70$ \kms \\
 (peak of NSC, when available) & \\
Virial dispersion proxy: & $\sigma_{\rm vir} \lesssim 70$ \kms \\
  & \\
ELT Observability (Armazones site): & $|\delta+24^\circ|<45^\circ$ \\
Galaxy zone of avoidance: & $|b|>8^\circ$ \\
\hline
\label{tab:parent_criteria}
\end{tabular}
\end{table}

\begin{table}
\caption{Main characteristics of the HARMONI IMBH sample}
\centering
\begin{tabular}{rl}
\hline
Galaxy stellar-mass range: & $2\times10^7\lesssim M_{\star}\lesssim5\times10^{10}$ \Msun \\
NSC's mass range: & $10^5\lesssim M_{\rm NSC} \lesssim6\times10^7$ \Msun\\
NSC's effective radius range: & $3 \lesssim R_e \lesssim 27$ pc \\
\#Galaxies in the sample: & $N_{\rm gal}=44$ \\
\#dwarf Ellipticals ($T\le-3.5$): & $N_{\rm dE}=6$ ($\approx$13\%)\\
\#Lenticulars ($-3.5<T\le-0.5$): & $N_{\rm S0}=7$ ($\approx$16\%)\\
\#Spirals ($-0.5<T\le8$): & $N_{\rm S}=27$ ($\approx$60\%)\\
\#Irregulars: & $N_{\rm Irr}=4$ ($\approx$11\%)\\
\hline
\vspace{-3mm}
\label{tab:parent_specs}\\
\end{tabular}
     {\raggedright \,\,\,\,\,\, {\it Notes:} The galaxy {\it Hubble} type ($T$) is defined from HyperLeda: \url{https://leda.univ-lyon1.fr/search.html}. \par}
\end{table}

\begin{figure*}
\centering\includegraphics[width=0.99\textwidth]{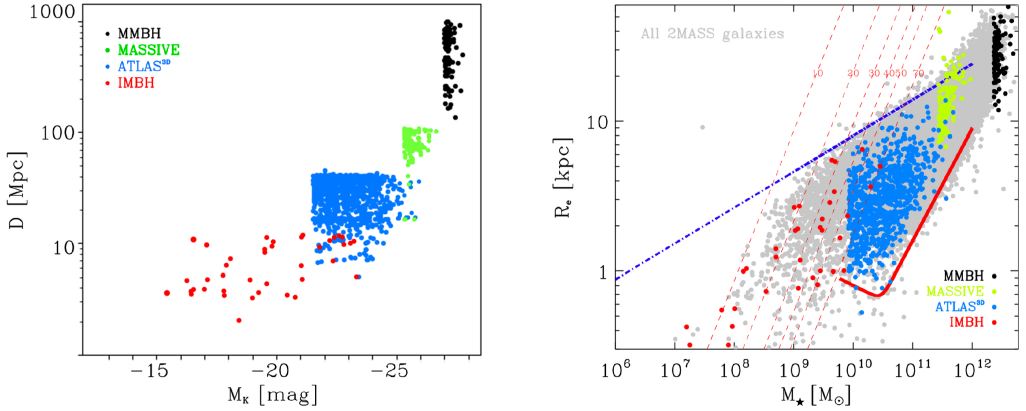}
    \caption{{\bf Left:} Distance vs. absolute $K$s-band magnitude of the HARMONI IMBH (red; this work), MASSIVE \citep[green;][]{Ma14}, ATLAS$^{\rm 3D}$ \citep[blue;][]{Cappellari11}, and MMBH (black; \citetalias{Nguyen23}) surveys. There is an overlapping region between the HARMONI IMBH sample and the ATLAS$^{\rm 3D}$ survey (i.e. $\approx$1.8 $M_K$ mag or $10^{10}<M_{\star}<5\times10^{10}$ \Msun). The plotted distance is the angular size distance $D=D_A$.        {\bf Right:} The mass--size diagram of the 2MRS sources (grey dots) shows a wide range of stellar mass: $10^7<M_{\star}<6\times10^{12}$ \Msun. The inclined-red-dashed lines indicate constant $\sigma_\star$ of 10, 20, 30, 40, 50, 70 \kms. The ATLAS$^{\rm 3D}$, MASSIVE, MMBH, and HARMONI IMBH samples occupy different regions of the diagram. The thick-solid red curve defines the zone of exclusion described by eq. (4) of \citet{Cappellari13b} in the previously explored stellar-mass range of $5\times10^9\leq M_{\star}\lesssim10^{12}$ \Msun. The thick dash-dotted blue line shows the relation $(R_e/{\rm kpc})=8\times[M_{\star}/(10^{10}\;{\rm M}_\odot)]^{0.24}$, which provides a convenient approximation for the lower 99\% contour for the distribution of early-type galaxies \citep{Cappellari13b}.}
    \label{all_sample}
\end{figure*}

\section{Properties of the HARMONI IMBH Sample}\label{sample_properties}

\begin{figure*}
\centering\includegraphics[width=0.99\textwidth]{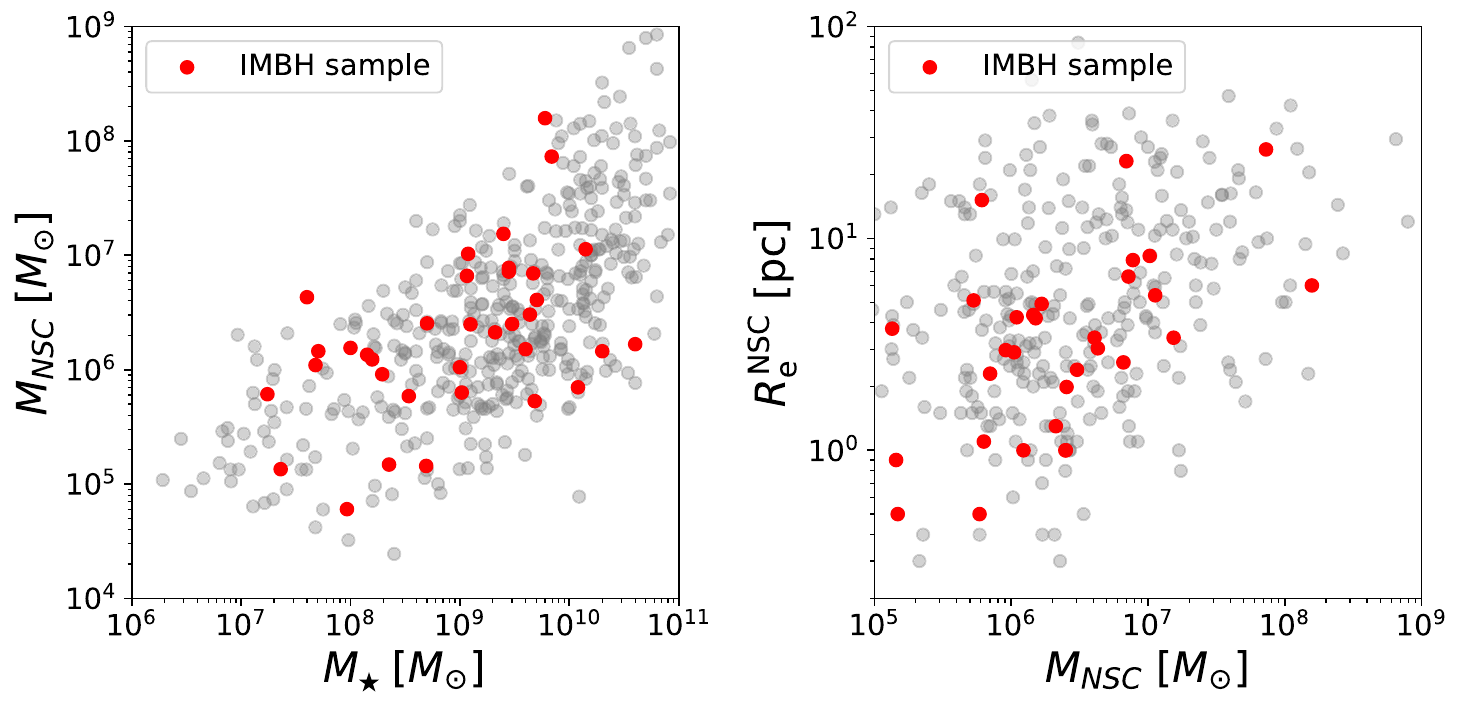}
    \caption{{\bf Left:} The $M_{\star}$--$M_{\rm NSC}$ relation for galaxies in which galaxy and NSC masses are available regardless of their Hubble types. The compilation of dynamical and spectroscopically modeled NSC masses from \citet{Erwin12}, \citet{Georgiev16}, \citet{Spengler17}, \citet{Ordenes-Briceno18b}, and \citet{Sanchez-Janssen19}.      {\bf Right:} The $(M_{\rm NSC},R_{\rm e})$ distribution for galaxies in which their NSC’s mass and effective radius are available. The data are taken from \citet{Walcher05}, \citet{Cote06}, \citet{Rossa06}, \citet{Erwin12}, \citet{Norris14}, \citet{Georgiev16}, \citet{Spengler17}, \citet{Baldassare22}, and \citet{Fahrion22}.}  
    \label{imbh_sample}
\end{figure*}

\subsection{HARMONI IMBH sample on the mass--size diagram}\label{mass_size_diagram}

The left panel of \autoref{all_sample} shows the distribution of the HARMONI IMBH survey targets (red dots) in the absolute magnitude ($M_K$) versus distance plane. This sample occupies the low-mass end of the galaxy population. For comparison, we also plot galaxies from larger surveys like ATLAS$^{\rm 3D}$ \citep{Cappellari11}, MASSIVE \citep{Ma14}, and MMBH \citepalias{Nguyen23}, highlighting the unique parameter space probed by the HARMONI IMBH sample.

The right panel of \autoref{all_sample} presents the mass--size diagram $(M_{\star},R_{\mathrm{e}})$ for the HARMONI IMBH sample, using available stellar masses ($M_{\star}$) and effective radii ($R_{\mathrm{e}}$). We overlay data from the 2MRS galaxy catalog \citepalias[compiled by][]{Nguyen23}, excluding sources with potentially unreliable $R_{\mathrm{e}}$ measurements ($R_{\mathrm{e}} < 5\arcsec$ or $R_{\mathrm{e}} < 0.3$ kpc, see \autoref{sample}). Data from the ATLAS$^{\rm 3D}$, MASSIVE, and MMBH surveys are also included. This comparison emphasizes that our sample targets galaxies with expected low stellar velocity dispersions ($\sigma_\star$) and, consequently, potentially low black hole masses (\Mbh). Lines of constant $\sigma_\star$ are shown for reference.

The HARMONI IMBH sample covers over four orders of magnitude in stellar mass. Its location on the mass-size diagram is consistent with the region occupied by dwarf and spiral galaxies, lying just below the approximate upper envelope defined by $(R_{\mathrm{e}}/{\rm kpc})=8\times[M_{\star}/(10^{10}\;{\rm M}_\odot)]^{0.24}$ (dashed blue line). The distribution resembles that of dwarf galaxies shown in figure 9 of \citet{Cappellari13b}.

For galaxies with $M_{\star} \gtrsim 10^9$ \Msun, various properties (stellar populations, gas fraction, dark matter fraction, density slope, IMF) tend to correlate with $\sigma_{\star}$ on the $(M_\star, R_\mathrm{e})$ diagram \citep[see review by][]{Cappellari2025}. Black hole masses also follow this trend, suggesting co-evolution \citep{Krajnovic18a}. However, below $M_{\star} \approx 10^9$ \Msun, the relationships between \Mbh, $\sigma_{\star}$, $M_{\star}$, and $R_{\mathrm{e}}$ are poorly understood due to the scarcity of IMBH measurements and limitations of surveys like 2MASS at low masses and small sizes (especially for $\sigma_{\star}<30$ \kms). Despite these limitations, the similarity between our sample's distribution and that of other dwarf galaxy samples with reliable $R_{\mathrm{e}}$ measurements suggests our selection is reasonably representative.

Recent studies \citep{Nguyen18, Nguyen19, Greene20} indicate that IMBHs in low-mass galaxies may deviate from the scaling relations established for more massive galaxies, hinting at different co-evolutionary pathways. While \citet{Mezcua17} discussed potential offsets in these relations at the low-mass end, observational challenges hinder definitive conclusions. Our HARMONI IMBH survey aims to overcome these limitations by providing high-resolution kinematic data for nearby IMBH candidates, shedding light on the connection between IMBHs and their host galaxies.

\subsection{Nuclear star clusters and central massive black holes}\label{nsc_bh}

Our HARMONI IMBH sample includes galaxies with NSCs. Among them, 31 NSCs have known masses and effective radii falling within the ranges of $10^5\lesssim M_{\rm NSC}\lesssim10^8$ \Msun\  and $3 \lesssim R_\mathrm{e}^\mathrm{NSC} \lesssim 27$ pc, respectively. For the remaining NSCs, this information is unavailable. We have compiled data on these NSCs and their host galaxies’ properties in \autoref{imbh_sample} alongside with those obtained from literature, including both photometric and spectroscopic surveys.

Out of the 44 selected targets, two IMBHs were estimated through the VLT/SINFONI stellar kinematics: NGC~5102 with $M_{\rm BH}\approx9.1\times10^5$ \Msun\ and NGC~5206 with $M_{\rm BH}\approx6.3\times10^5$ \Msun\ \citep{Nguyen18, Nguyen19}. \citet{Nguyen22} found a firm signature of an IMBH with $3\times10^5 \lesssim M_{\rm BH} \lesssim 4.3\times10^6$ \Msun\ in NGC~3593 based on the ALMA observation of circumnuclear CO(2-1)-molecular gas disc. Additional dynamical models presented in \citet{Neumayer12} suggested that the IMBH in NGC~300, with kinematics measured from VLT/UVES spectra \citep{Walcher05}, likely has a $M_{\rm BH}<10^5$ \Msun. Similarly, the IMBH in NGC~3621, with kinematics measured from the Keck Echellette Spectrograph \citep{Barth02}, suggest a $M_{\rm BH}<3\times10^6$ \Msun\ \citep{Barth09}.

In the regime of dwarf galaxies with $M_\star<7\times10^9$ \Msun, the behavior of the $M_{\rm BH}$--galaxy correlations remain largely unconstrained. It strongly depends on the mass distribution of the currently elusive IMBH population found in the nearby universe \citep[e.g.,][]{Mezcua17}. Recent work by \citet{Nguyen19} also provided evidence that the dynamical $M_{\rm BH}$ of five IMBHs in a sample of early-type galaxies within 3.5 Mpc, fall well below the predictions from the \citet{Kormendy13}, \citet{Sahu19a}, and \citet{Greene20} $M_{\rm BH}$--$\sigma_{\star}$ and $M_{\rm BH}$--$M_{\rm bulge}$ relations.

\subsection{Simulated targets}\label{simulated_targets}

We created mock integral field spectrograph (IFS) observations with {\tt HSIM} for two dwarf galaxies with NSCs: NGC~300 and NGC~3115, which are located at 2.2 Mpc \citep{Williams13} and 9.7 Mpc \citep{Jerjen00a, Jerjen00b, Karachentsev04}, respectively. These galaxies lie at the extreme ranges of our distance selection and are meant to represent our HARMONI IMBH survey’s overall characteristics.

{\bf NGC~300} is a SA(s)d galaxy, which inclination $i=(42.3\pm3.0)^{\circ}$ \citep{Kim04}, lacks of a bulge \citep{Bono10}, and has a total stellar mass of $M_{\star}\approx2.2\times10^9$ \Msun\ \citep{Kacharov18}. The galaxy exhibits a steadily increasing size of its NSC from ultraviolet (UV) to infrared (IR) wavelengths \citep{Carson15}, which has a central velocity dispersion of $\sigma_{\star}=13.3\pm2.0$ \kms\ and a dynamical mass of $M_{\rm NSC}\approx10^6$ \Msun, determined from the VLT/UVES spectroscopy \citep{Walcher05}, or $\sigma_{\star}=13.3\pm0.3$ \kms\ when measured from the VLT/X-Shooter spectroscopy \citep{Kacharov18}. 

The Jeans Anisotropic Model \citep[JAM,][]{Cappellari08} of the VLT/UVES stellar kinematics measurements \citep{Walcher05} found a $M/L_I\approx0.6$ (\Msun/\Lsun) and suggested an upper limit of $M_{\rm BH}<10^5$ \Msun\ (or $\approx$10\% of $M_{\rm NSC}$) for its IMBH, while its best-fit estimate is $M_{\rm BH}\approx10^2$ \Msun\ \citep{Neumayer12}. 

The star formation history (SFH) of NGC~300’s NSC had been independently analyzed by \citet{Walcher06} and \citet{Kacharov18}, utilizing different spectra, spectral ranges, and simple stellar population (SSP) models. Their findings reveal a relatively consistent stellar population. More than 50\% of the stars in the NSC formed over 5 Gyr ago with low metallicity ([Fe/H] $\sim-1$ dex). Subsequent star formation episodes involving young populations ($\approx$10 Myr) occurred until a few hundred million years ago, resulting in solar metallicity of 1 Gyr ago. {\bf Despite these later episodes, their contributions remain modest, accounting for only 10\% of the total luminosity and 1\% of the total stellar mass}. 
There is no evidence of emission lines in the spectra implies that the contribution of a very young population in NGC~300’s NSC is likely from an extended horizontal branch \citep{Conroy18} or blue straggler stars \citep{Schiavon07}.

{\bf NGC~3115 dw01 (PGC029300)} is classified as a dwarf elliptical \citep[dE;][]{Jerjen00b, Parodi02} with $M_{\star}\approx8.9\times10^8$ \Msun\, determined from its $B-V$ colour. The galaxy hosts an NSC with $R_\mathrm{e}^\mathrm{NSC}\approx6.61\pm0.09$ pc and $M_{\rm NSC}\approx7.2\times10^6$ \Msun, constrained from the \hst/ WFPC2 F814W images with a photometric $M/L_{\rm F814W}\approx1.4$ (\Msun/\Lsun) \citep{Pechetti20}.  Stellar kinematic measurements with the echelle spectrograph of the 4-meter telescope at Cerro Tololo Inter-American Observatory (CITO) yield a central velocity dispersion of $\sigma_{\star}=32\pm5$ \kms\ in the NSC. However, more global measurements within the galaxy’s effective radius ($R_\mathrm{e}\approx1.2$ kpc) result in a value of $\sigma_\mathrm{e}\approx45$ \kms\ \citep{Peterson93}.

Due to the absence of stellar populations and SFH in NGC~3115 dw01, we opted for simplicity in our {\tt HSIM} IFS simulations. We employed Stellar Population Synthesis (SPS) spectra with a fixed age of 5 Gyr, Solar metallicity ({\tt z002}), and inclination $i=42^{\circ}$, which are similar to the nucleus of NGC~300.

\section{Dynamical Model}\label{dynamical_model}

\subsection{Jeans Anisotropic Modelling}\label{jam}

\begin{figure*}
	\centering 
	\includegraphics[width=0.99\textwidth]{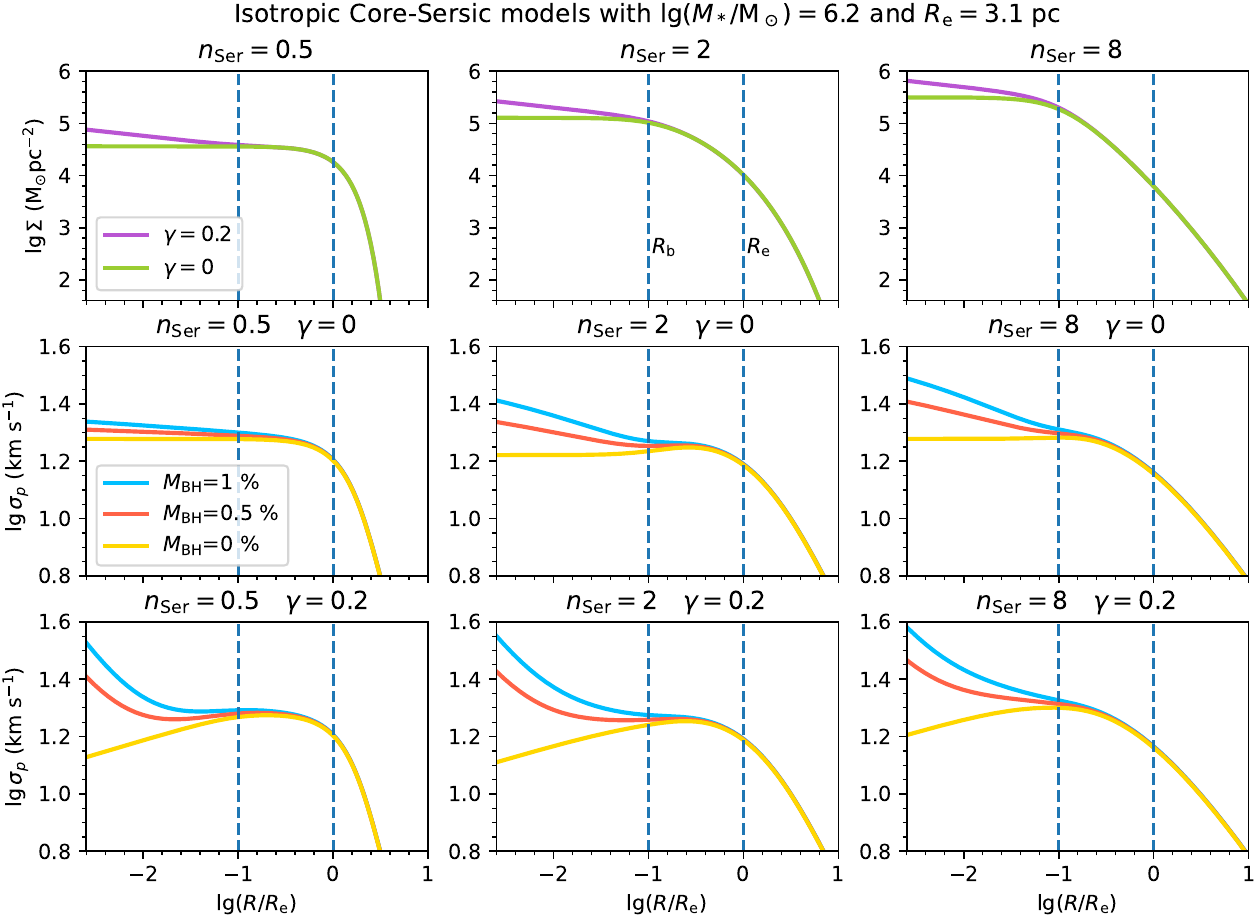}
	\caption{Atlas of the central surface-mass density ($\Sigma$, three top-row panels) and projected velocity dispersion ($\sigma_p$), the three middle-row and bottom-row panels) profiles resolved deeply into the resolvable region of HARMONI, which are predicted by various values of the Sérsic index ($n_{\rm Ser}$) and the core inner slope ($\gamma$) of the isotropic Core-Sérsic model. These theoretical profiles can be used to describe the qualitative behaviour of the $\sigma_p$ of stellar systems spanning from galaxies to NSC. For illustration, the values of $\Sigma$ and $\sigma_p$ were normalized to approximate the median values of the mass and radii of NSC (see text). We utilized three different Sérsic indices -- including the low $n_{\rm Ser}=0.5$ (corresponding to the Gaussian profile), intermediate $n_{\rm Ser}=2$, and high $n_{\rm Ser}=8$ of the Sérsic-index value -- and two different values of the core inner slope: $\gamma=0$ (green line) and $\gamma=0.2$ (purple line). To illustrate the effects of central IMBHs having at the central region of the projected velocity dispersion profiles, we modeled with three different BH masses: $M_{\rm BH} = 0$\% of $M_{\rm NSC}$ (no BH, yellow line), $M_{\rm BH} = 0.5$\% of $M_{\rm NSC}$ (or $M_{\rm BH} = 8\times10^3$ \Msun, red line), and $M_{\rm BH} = 1$\% of $M_{\rm NSC}$ (or $M_{\rm BH} = 1.6\times10^4$ \Msun, blue line). The two vertically dashed lines represent the break radius ($R_b$, inner line) and the effective radius ($R_{\rm e}$, outer line) of the NSC's isotropic Core-Sérsic model. The core inner slope $\gamma$ control the inner behaviour of the central surface-mass density and the projected velocity dispersion profiles, while the Sérsic index describes the outer profile.}
	\label{fig:jam_bh_nc_trends}
\end{figure*}

NSCs in dwarf galaxies, such as those targeted by our HARMONI IMBH survey, frequently display significant rotation \citep[e.g.,][]{Seth08, Seth10, Nguyen18, Thater23}. Observed ratios of rotational velocity to stellar velocity dispersion ($V/\sigma_{\star}$) typically range from 0.1 to 0.6 \citep{Pinna21}, indicating substantial ordered motion.

To model the kinematics of these rotating systems, we utilize the JAM  \citep{Cappellari08, Cappellari20}. Specifically, we adopt the JAM variant that assumes cylindrical alignment of the velocity ellipsoid (JAM$_{\rm cyl}$), an approach proven effective for modeling rotating NSCs \citep[e.g.,][figure~4]{Hartmann2011} and rapidly rotating galaxies \citep[e.g.,][figure~10]{Cappellari16}.

We implemented these models using the \textsc{JamPy} package\footnote{v7.2.0; available from \url{https://pypi.org/project/jampy/}}. In the \textsc{jam.axi.proj} procedure, we set {\tt align='cyl'} to enforce cylindrical alignment. We further assumed an oblate velocity ellipsoid, where the vertical dispersion differs from the equal radial and tangential dispersions ($\sigma_z \neq \sigma_R = \sigma_\phi$), by setting {\tt gamma=0} and {\tt kappa=1}.

A crucial aspect of this study is probing radii very close to the central BH. To accurately capture the BH's gravitational influence in this regime, we set the keyword \texttt{analytic\_los=False}. This treats the BH potential as a point mass, enabling the analytical calculation of the intrinsic second velocity moments ($\langle\sigma_z^2\rangle$ and $\langle\sigma_\phi^2\rangle$) in its vicinity, rather than using a Gaussian approximation. These analytically derived intrinsic moments are then numerically integrated along the line-of-sight to predict the observable kinematics.

\subsection{Predicted velocity dispersion profiles}\label{predicted_vrms}

Before presenting detailed simulations of the predicted kinematics due to IMBHs of different $M_{\rm NSC}$, here we try to understand the quantitative behaviour we should expect. This is  important because we have to decide how to extrapolate the inner surface brightness profile of NSCs, at radii much smaller than resolvable with \hst.

\begin{table*}
\caption{\hst/WFPC2 PC (PC1) observations (1 pixel = $0\farcs045$) of the two galaxies hosting NSCs chosen to perform {\tt HISM} simulations}  
\centering\begin{tabular}{cccccccccc}
 \hline\hline
Galaxy name      &$\alpha$(J2000)&$\delta$(J2000) &   UT Date    &   PID  &   PI     &  Filter &     Exptime     &Zeropoint&A$_{\rm \lambda}$\\
                           &      (h:m:s)         &($\deg:\mm:\se$)&                    &           &           &           &         (s)        &   (mag)   &      (mag)  \\
  (1)                    &           (2)           &           (3)          &      (4)           &  (5)     &   (6)   &    (7)   &          (8)         &   (9)        &      (10) \\	                
\hline
NGC~300           & 50:54:53.477    &$-$37:41:03.31 &2001 May 06 & 8599  &Boeker&F814W&$3\times213$ & 23.758 & 0.025\\
 \hline
NGC~3115 dw01&10:05:41.599&$-$07:58:53.40 &1995 Nov. 29 &  5999 &Phillips&F814W&$2\times160$ & 23.758 & 0.050 \\
            & 10:05:41.599&$-$07:58:53.40 &1995 Nov. 29 &  5999 &Phillips&F547M&$1\times160$ & 23.781 & 0.025 \\                                                
\hline
\end{tabular}\\
\label{hstdata}
\parbox[t]{0.99\textwidth}{\textit{Notes:} Column 1: The galaxy name. Columns 2 and 3: The galaxy’s position (R.A. and Decl.) from \hst/HLA. Column 4. The date when observations were performed. Columns 5 and 6: The project (PID) and principal investigator (PI) identification numbers. Column 7: The board-band filter used to take the data. Column 8: The exposure times of the observations show the number of exposures multiplied by the time spent on source for each exposure. Columns 9 and 10: The photometric zero-point and extinction value in each filter. }
\end{table*}

In their classic paper, \citet{Tremaine94} studied the qualitative behaviour of the projected velocity dispersion ($\sigma_p$) profiles in elliptical galaxies with central SMBHs. They described the predicted velocity-dispersion profile of isotropic galaxy models described by a double power-law density profile, with general inner density $\rho(r)\propto r^{-\gamma}$ and fixed outer density $\rho(r)\propto r^{-4}$ \citep{Dehnen93}, and a smooth transition between these two power-law regimes. Where, ${\gamma}$ is the power-law index.

Here, we illustrate the behaviour of the velocity dispersion for the more general Core-S\'ersic model \citep{Graham03, Trujillo04}, which parametrizes the projected surface brightness (not the intrinsic density) as follows
\begin{equation}\label{eq:core_sersic}
	I(R) = I' \left[1 + \left(\frac{R_b}{R}\right)^\alpha\right]^{\gamma/\alpha} 
	\exp\left[{-b_{n_{\rm Ser}} \left(\frac{R^\alpha + R_b^\alpha}{R_\mathrm{e}^\alpha}\right)^{1/(\alpha\, n_{\rm Ser})}}\right] 
\end{equation}  
where $I'$ is the normalization and $b_{n_{\rm Ser}}=Q^{-1}(2n_{\rm Ser},1/2)$ is a factor defined in such a way that the half-light radius $R_{\rm e }$ encloses half of the total light of the S\'ersic profile (without the core), and $Q^{-1}$ is the inverse of the regularized incomplete gamma function\footnote{Implemented in \href{https://docs.scipy.org/doc/scipy/reference/generated/scipy.special.gammainccinv.html}{scipy.special.gammainccinv}} \citep[eq.~14]{Zhu2025}. Here, $n_{\rm Ser}$ is the S\'ersic index, which controls the shape of the outer S\'ersic part. $R_b$ is the break radius, which is the point at which the surface brightness changes from the outer S\'ersic part to the inner power-law regime of the profile. The intensity $I_b$ at the break radius is given by \autoref{eq:core_sersic} with $R=R_b$ (we give it in magnitudes $\mu_b$ in \autoref{sersicfittab}). $\alpha$ is the sharpness parameter that controls the sharpness of the transition between the outer S\'ersic and inner power-law regimes. Outside the inner break at radius $R>R_b$ this is a \citet{Sersic68} profile of projected half-light radius $R_{\rm e}$, but gradually transitions to a power law surface brightness $\Sigma(R)\propto R^{-\gamma}$ at smaller radii $R\ll R_b$.

We adopt the Core-S\'ersic profile because it can describe the profile of real NSCs. However, we allow for a cusp in the inner slope to parametrize our ignorance of the profile of NSC at radii which are inaccessible with current instruments, but which will become observable with the ELT. In our analysis, we model the NSC in isolation, even though they are embedded in the large-scale gravitational potential of the host galaxy. However, the latter has an insignificant effect on the predicted $\sigma_p$ profiles at the centre of the NSC. In particular, in our case, $R_{\rm e}$ represents the size of the NSC, not the size of the host galaxy.

To compute the predictions for the $\sigma_p$ profiles we first used the \textsc{mge.fit\_1d} procedure in the \textsc{MgeFit} package\footnote{v5.0 of the Python package from \url{https://pypi.org/project/mgefit/}} \citep{Cappellari02} to fit the Core-S\'ersic profiles in the range $R_b/100<R<20R_e$ using 20 Gaussians. Then used the \textsc{jam\_sph\_proj} procedure in the \textsc{JamPy} package \citep[see footnote 7;][]{Cappellari08, Cappellari20} to compute the isotropic velocity dispersion for different values of the S\'ersic index $n_{\rm Ser}$ and the core inner slope $\gamma$. In all our tests we adopted a fixed $\alpha=2$ value, which produces a smooth transition between the outer S\'ersic and inner power-law profiles. We show the results in \autoref{fig:jam_bh_nc_trends}. The qualitative behaviour of the profiles is independent of the absolute normalization of their total masses $M_\star$ and $R_{\rm e}$. It can be applied unchanged to describe the $\sigma_p$ profile of NSC or giant elliptical galaxies. However, in this Fig., we adopt as mass and size reference the values $\lg(M_{\rm NSC}$/\Msun) = 6.2 and $R_e=3.1$ pc. These are the median values for NSC of late-type galaxies with $M_\star<10^9$ M$_\odot$ \citep[table~2]{Neumayer20}. 

In \autoref{fig:jam_bh_nc_trends}, we adopt either no BH, or BH masses of 0.5\% and 1\% of the mass of the NSC. These values are the same characteristic fractions observed for normal galaxies \citep[eq.~11]{Kormendy13}. However, nothing is currently known about the mass of IMBH in NSC. In fact, NSCs may not have IMBHs at all, or they may be dominated by the IMBH mass. But the qualitative behaviour remains unchanged.

For $\gamma>0$ and without BHs the $\sigma_p$ profile decreases towards the nucleus, regardless of $n_{\rm Ser}$. This is a well-known general feature of realistic galaxy models. It appeared in puzzling contrast with the early observations \citep{Binney80} until it became clear that all massive galaxies contain SMBHs. With an IMBH, the $\sigma_p$ profile starts increasing towards the centre, inside the IMBH's SOI, asymptotically approaching the Keplerian rise $\sigma_p\propto R^{-1/2}$ (for $0<\gamma<2$), regardless of the profile slope, as shown by \citet[eq.~52]{Tremaine94}.

\citet{Tremaine94} also pointed out that\footnote{\citet{Tremaine94} studied densities $\rho(r)\propto r^{-\gamma'}$, which we convert to the corresponding surface brightness $\Sigma(R)\propto R^{-\gamma}$ using $\gamma=\gamma'+1$.}, when $\gamma=0$, namely when the inner surface brightness approaches a constant value, the asymptotic behaviour becomes qualitatively different. In that case, without an IMBH, the $\sigma_p$ profile approaches a constant positive value, instead of dropping towards zero. We confirm the behaviour for the Core-S\'ersic profile with a flat inner core regardless of $n_{\rm Ser}$. With an IMBH, the $\sigma_p$ profile rises towards the centre, but it does so less steeply than the $\gamma>0$ case, as predicted for the double power-law models.

What is new in this study is the fact that we explore a range of S\'ersic indices, including the extreme case $n_{\rm Ser}=1/2$, which corresponds to a Gaussian profile, when $\gamma=0$. We find that in this extreme situation, the $\sigma_p$ profile becomes only weakly sensitive to the $M_{\rm BH}$. This is quite important for the present study, because we are approximating the surface brightness of the NSC using the multi-Gaussian expansion method \citep[MGE,][]{Emsellem94,Cappellari02}. And we are fitting photometric data that have a much lower resolution than the one we will be able to obtain with the ELT. This implies that our model may end up being described by a Gaussian in the central parts, well inside the width of the PSF of our photometry. This is precisely the key region, near the BH, where we want to produce an accurate $\sigma_p$ prediction.

In practice, we should not worry about NSC having flat surface brightness profiles and isotropic orbital distributions. This is because such models are unphysical and cannot exist in real stellar systems. Instead, if an IMBH is present, isotropic models produce a density cusp that rises as $\rho(r)\propto r^{-1/2}$ near the BH \citep{Tremaine94}, or $\Sigma(R)\propto R^{-3/2}$ in surface brightness \citep[see also][]{vanderMarel99}. One would require extreme tangential anisotropies for flat profiles $\gamma=0$ to be allowed, as indicated by the cusp-slope vs central anisotropy theorem \citep{An06}. We have no reason to think NSC satisfies this requirement. However, what is important is that, lacking real high-resolution photometry at the resolution of HARMONI, we extrapolate the observed surface brightness using a non-zero inner power slope, when constructing the mock kinematics of the NSCs. Failing to do so, we would generate predictions for $\sigma_p$ profiles that vastly underestimate the central rise that we expect from high-resolution HARMONI observations.

\subsection{MARCS synthetic library of stellar spectra}\label{MARCS}

To create mock {\tt HSIM} IFS simulations, we required stellar population synthesis (SPS) spectra\footnote{Available from \url{https://marcs.astro.uu.se/}} that contain information about the stellar populations \citep{Maraston11}, based on the Model Atmospheres with a Radiative and Convective Scheme (MARCS) synthetic library of theoretical spectra, originally developed by \citet{Gustafsson08}. Although MARCS synthetic library spectra are not as reliable as an empirical stellar library, when expecting the stellar kinematics of real galaxies, its minor imperfections are not an issue when generating mock spectra. The library offers broad wavelength coverage spanning from 1,300 \AA\ to 20 $\mu$m at high spectral resolution with $\sigma = 6.4$ \kms\ (equivalent to $R = \lambda/\Delta\lambda\sim$20,000). We assumed a Salpeter IMF, an age of 5 Gyr, and Solar metallicity ({\tt z002}) for both NGC~300’s NSC \citep{Kacharov18} and NGC~3115 dw01 (\autoref{simulated_targets}). The SPS spectra were truncated within the wavelength range of 1.0--2.5 $\mu$m to match the HARMONI/$J$, $H$, $K$, $H$-high, $K$-short, and $K$-long gratings.

\section{Galaxy photometry models}\label{photometry}

\begin{figure*}        
 \centering\includegraphics[width=0.99\textwidth]{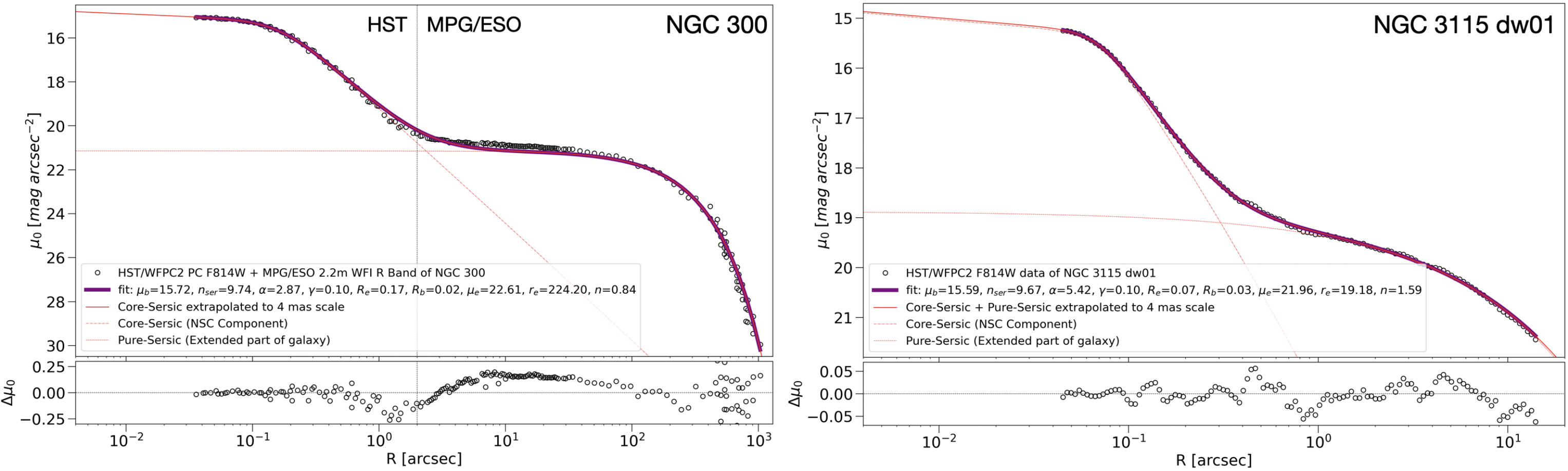}
    \caption{{\bf Upper panels:} The \hst/WFPC2 F814W surface brightness of NGC~300 (left) and NGC~3115 dw01 (right) constructed from {\tt IRAF ellipse} are plotted in open black dots, while their best-fit Core-Sérsic + pure-Sérsic surface brightness are plotted in purple solid-thick lines. Their best-fitting parameters are shown in the legends. Here, the Core-Sérsic profile is plotted in a red dashed line, while the pure-Sérsic profile is depicted in a red dotted line. For $r<0\farcs045$, these surface brightness are extrapolated to the 4 mas scale required for this IMBH survey’s imaging with MICADO (red solid-thin lines).  {\bf For NGC 300, the vertical black line indicates the radius at which we combined the narrow-field \hst/WFPC2 F814W data with the wide-field MPG/ESO 2.2-m WFI $R$-band data.}  {\bf Lower panels:}  The differences, {\tt data - model}, between the {\tt IRAF ellipse} surface brightness and their corresponding best-fit Core-Sérsic + pure-Sérsic models illustrate the fit’s goodness.}
    \label{iraf_profiles}
\end{figure*}

\begin{table*}
\caption{Best-fit Core-Sérsic ($\gamma=0.1$) + pure-Sérsic parameters of the \hst/WFPC2 surface brightness profiles.}    
\footnotesize
\centering\begin{tabular}{cccccccccc}
 \hline\hline
Galaxy name      & Filter  &        $\mu_b$       &$n_{\rm Ser}$&$\alpha$&     $R_{\rm e}$   &   $R_b$    &            $\mu_e$    &    $R_{\rm e}$    & $n$\\
                           &           &(mag/arcsec$^{-2}$)&                      &              &(arcsec)&(arcsec)&(mag/arcsec$^{-2}$)&(arcsec)&       \\
  (1)                    &  (2)     &            (3)             &      (4)            &  (5)        &   (6)          &      (7)        &           (8)              &      (9)        &(10) \\	                
\hline
NGC~300           &F814W& $15.72\pm0.22$  &$9.74\pm0.05$ & $2.87\pm0.05$ &$0.17\pm0.03$&$0.02\pm0.01$ &$22.61\pm0.45$&$224.20\pm1.52$&$0.84\pm0.11$\\
 \hline
NGC~3115 dw01&F814W& $15.59\pm0.21$  &$9.67\pm0.05$ & $5.42\pm0.12$ &$0.07\pm0.02$&$0.03\pm0.01$ &$21.96\pm0.20$&$19.18\pm0.25$&$1.59\pm0.07$\\
                          &F547M & $16.44\pm0.25$  &$9.51\pm0.06$ & $5.27\pm0.15$ &$0.06\pm0.02$&$0.03\pm0.01$ &$23.01\pm0.24$&$18.73\pm0.27$&$1.71\pm0.05$\\                
\hline
\end{tabular}\\
\label{sersicfittab}
\parbox[t]{0.99\textwidth}{\textit{Notes:} Column 1: The galaxy name. Column 2: The board-band filter used to take the data. Columns (3--7): The best-fit parameters of the Core-Sérsic profile for the NSC components, orderly the surface brightness density at the break radius, the Sérsic index of the outer Sérsic part, the sharpness parameter which indicates how fast/slow the profile changes from the outer Sérsic to inner power-law regime, the break radius where the profile changes from the outer Sérsic part to the inner power-law part, and the effective radius of the outer Sérsic part. Columns (8--10): The best-fit parameters of the pure-Sérsic profile for the galaxies' extended parts, orderly the surface brightness density at the effective radius, the effective radius, and the Sérsic index.}
\end{table*}

\subsection{HST images and their PSF models}\label{hst}

We summarized the \hst/Wide Field and Planetary Camera 2 (WFPC2) Planetary Camera (PC) images used in this study in \autoref{hstdata}. We accessed the reduced and drizzled images from the \hst/Hubble Legacy Archive (HLA), with a pixel scale of 0\farcs045.

For our analysis, we need the PSFs to deconvolve the \hst\ profiles into their intrinsic forms (\autoref{sb}).  We thus employed the simulated images of the \hst/WFPC2 PSFs in each filter using {\tt tiny1} and {\tt tiny2}  routines within the {\tt Tiny Tim} software package\footnote{\url{https://github.com/spacetelescope/tinytim}} \citep{Krist95, Krist11}, which creates a model \hst\ PSF based on the instrument, detector chip, detector chip position, and filter used in the observations. To ensure that the model PSFs were processed in the same way as the \hst\ images, we produced individual versions of each PSF corresponding to their realistic exposure in each filter and each galaxy on a subsampled grid with sub-pixel offsets, using the same four-point box dither pattern as the \hst/WFPC2 exposures. Next, in order to account for the effect of electrons leaking into neighbouring pixels on the CCD, each model PSF was convolved with the appropriate charge diffusion kernel. The same filters set of PSFs of each galaxy were then combined and resampled onto a final grid with a pixel size of 0\farcs045 using {\tt Drizzlepac}/{\tt AstroDrizzle} \citep{Avila12} to produce final the PSF image. 


\subsection{Surface brightness profiles and galaxy mass models}\label{sb}

We used the \hst\ data and their simulated  PSFs in \autoref{hst} to measure the galaxies’ surface brightness profiles. We fitted the model of \autoref{eq:core_sersic} and extrapolated these profiles using the best-fit parameters to create our mock surface brightness at higher resolution than \hst.

Our initial step involved extracting the one-dimensional (1D) surface brightness of NGC~300 and NGC~3115~dw01 by employing the {\tt Image Reduction and Analysis Facility (IRAF) ellipse}  task \citep{Jedrzejewski87} to conduct this task. The ellipse routine systematically integrated the flux of stars within concentric annuli, allowing for variations in position angles and ellipticities along the semi-major axis of the galaxy, although fixing these two parameters did not yield any significant changes in our results. Before extracting the intrinsically 1D surface brightness profiles in {\tt IRAF} for each image, we first performed spatial deconvolution using the corresponding PSF image generated in \autoref{hst}. Subsequently, we converted the average flux within each annulus, measured in counts/s, into surface brightness expressed in mag arcsec$^{-2}$. This conversion was carried out using the photometric information derived from the broad-band filters and camera specifications (\autoref{hstdata}).  The large-scale profile of NGC~300 ($r>2\arcsec$) was taken from \citet{Neumayer20}, which was extracted from the Wide Field Imager (WFI) observed in $R$ band with the MPG/ESO 2.2-metre telescope\footnote{Available at \url{https://www.eso.org/public/images/eso1037a/}}. Here, we calibrated our inner F814W profile ($r\leq2\arcsec$) with the same \hst\ profile from \citet{Carson15} using the AB magnitude system.

Next, we applied analytical functions to fit these 1D surface brightness of NGC~300 and NGC~3115 dw01. These functions involved combinations of a Core-S\'ersic model \citep{Graham03, Trujillo04} and a pure-S\'ersic model \citep{Sersic68}. Here, while the Core-S\'ersic model is used to fit the NSC component, the pure-S\'ersic model was applied to fit the extended component of the galaxy. This model fit also gives the S\'ersic index ($n_{\rm Ser}$) of the NSC, which is useful to predict the kinematics trend (see \autoref{predicted_vrms}). Furthermore, the pixel size of the \hst/WFPC2 is 45 mas. This means that a single pixel contains over 60 spaxels of the proposed 10 mas mode adopted for this HARMONI IMBH survey. In fact, it is more than 11 times larger than the pixel size for ELT/MICADO images (i.e., 4 mas). We thus do not know the stellar profile at this scale, and this region is indeed embedded deeply in the core of an NSC, which is controlled by the power-law behaviour of the Core-S\'ersic model. In this work, we adopted the power-law index $\gamma=0.1$ as in \citetalias{Nguyen23} for both galaxies, although our tests with varying $\gamma$ provide $\gamma=0.08$ for NGC~300 and $\gamma=0.13$ for NGC~3115 dw01 with their current \hst\ 1D surface brightness, implying our global choice of $\gamma=0.1$ for mock HARMONI IFS simulation in the subsequent Section is reasonable.

Additionally, the pure \citet{Sersic68} model has a form:  
\begin{equation} 
	I(R) = I_e\exp \Biggl\{-b_n\left[{\left(\frac{R}{r_e}\right)^{1/n}}-1\right]\Biggr\}	 
\end{equation} 
In which, we denoted $n$ is the S\'ersic index of the pure-S\'ersic profile to distinguish with that $n_{\rm Ser}$ of the Core-S\'ersic profile in equation (1), which controls the degree of curvature of the profile. $I_e$ is the intensity at $R_{\rm e}$, which is converted into surface brightness density $\mu_e$ shown in the legends of \autoref{iraf_profiles} and \autoref{sersicfittab}.  

We applied a non-linear least-squares algorithm using the {\tt MPFIT}\footnote{Available from \url{http://purl.com/net/mpfit}} function \citep{Markwardt09} to iteratively fit the combined function of a Core-S\'ersic and a pure-S\'ersic above for both NGC~300 and NGC~3115 dw01 to their corresponding spatially deconvolved profiles, thereby obtaining the best-fit parameters homogeneously.   After the fits, we extrapolated the best-fit surface brightness towards the galaxies' centers to a scale of 4 mas required for this IMBH survey’s imaging with MICADO (\autoref{iraf_profiles}). The best-fit parameters and their associated errors derived from these surface brightness fits are saved in \autoref{sersicfittab}.  

\begin{figure*}
    \centering\includegraphics[width=0.99\textwidth]{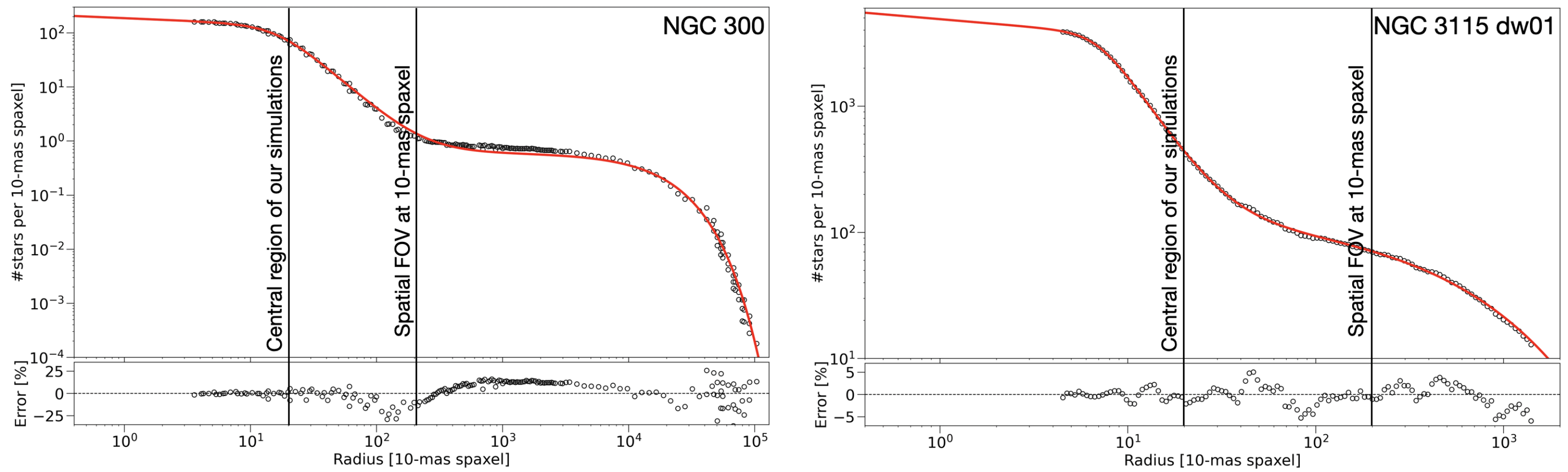}
    \caption{The upper panels present the estimated number of stars per 10 mas spaxel for the nuclei of NGC~300 (left) and NGC~3115 dw01 (right). The data points derived directly from the SB in \autoref{iraf_profiles} are shown in black circles at their best-fit models are shown in red lines. The lower panels of the figure display the residuals, calculated as ({\tt data - mode}), which provide a relative error in our estimation, helping to assess the accuracy of our calculations. The inner vertical solid line represents our spatially central regions of HARMONI simulations in \autoref{mocks} ($0\farcs4\times0\farcs4$), while that same outer line illustrates the full spatial FOV of HARMONI at our chosen 10-mas spaxel survey \citep[$2\farcs15\times1\farcs54$; see][]{Zieleniewski15}.}
    \label{discrete_starsx}
\end{figure*} 

Next, we employed these derived parameters to describe the two-dimensional (2D) luminosity density models extrapolated towards the central 4 mas for each galaxy. We assumed  a small flattening based on the fact that not all NSCs are perfectly spherical and non-rotating, characterized by the ratio between the semiminor and semimajor axes $q'=b/a\approx0.9$. This assumption also works for the extended bulge or disk of NGC~3115 dw01 as the galaxy morphology looks round through its \hst\ images. However, the extended scale of NGC~300 revealed through the MPG/ESO 2.2-m WFI $R$ band image suggests the galaxy's outer part ($r>2\arcsec$) can be approximately assumed as $q'=b/a\approx0.75$. Subsequently, we approximated these 2D luminosity density models using the MGE method \citep{Emsellem94, Cappellari02}. This involved using the \textsc{mge.fit\_sectors} routine from the \textsc{MgeFit} package \citep{Cappellari02} and included 15 Gaussians spanning radii from 4 mas to $\approx$1000$\arcsec$ for NGC~300 and $\approx$15$\arcsec$ for NGC~3115 dw01, as listed in \autoref{mgetab}.
 
Finally, we converted the luminosity surface density into mass surface density by assuming a constant mass-to-light ratio, $M/L_{\rm F814W, dyn}\approx M/L_I\approx0.6$ (\Msun/\Lsun) for NGC~300 \citep{Neumayer12} and $M/L_{\rm F814W, phot}\approx M/L_I\approx1.4$ (\Msun/\Lsun) for NGC~3115 dw01 derived in AB mag \citep{Pechetti20}.  To simplify our HARMONI IFS simulations, we disregarded (1) potential variations in $M/L_I$ due to differences in stellar populations \citep[e.g.,][]{Mitzkus17} and (2) the distribution of dark matter halos \citep{Navarro96}, given that the stellar desnity is expected to be orders of magnitude larger in the NSC. We focus on the stellar kinematics within the central part of $0\farcs4\times0\farcs4$ of HARMONI field-of-view (FOV), where the central BH’s and NSC's potential is the dominant factor.

\subsection{Effect of individual stars}\label{discrete_stars}

A critical consideration for our HARMONI IMBH survey, with its proposed $10\,\text{mas}$ spaxel resolution, is ensuring a sufficient number of stars within each spaxel. This is necessary to treat the stellar distribution as continuous rather than needing to model individual stars, a challenge that increases with closer targets. To evaluate this, we estimated the average number of stars per $10\,\text{mas}$ spaxel using the surface brightness profiles provided in \autoref{iraf_profiles}. Our method involved converting the surface brightness density from $\,$mag\,arcsec$^{-2}$ to surface luminosity density in $\,L_\odot\,\text{arcsec}^{-2}$, and then calculating the average number of stars within the $10\,\text{mas}$ spaxel area based on this luminosity density.

Our findings are illustrated in \autoref{discrete_starsx}, which shows the estimated number of stars per $10\,\text{mas}$ spaxel versus radius for NGC~300 and NGC~3115 dw01. In NGC~300, the stellar density per spaxel falls dramatically from over 100 in the nucleus to approximately one star at radii exceeding 200 spaxels (equivalent to the $\approx$$2\farcs15$ extent of our HARMONI FOV). NGC~3115 dw01, being significantly denser, shows a decrease from about 5000 central stars per spaxel to around 700 within the FOV. Notably, even in the difficult case of NGC~300 -- a close and well-resolved target -- the central $0\farcs4\times0\farcs4$ region maintains a density of roughly 100 stars per $10\,\text{mas}$ spaxel, supporting the validity of a continuous distribution assumption in such dense central areas. It is important to understand that this issue of discrete stars per single spaxel is primarily a concern only for the few innermost spaxels. For radii further out, we will employ Voronoi binning (see footnote 13, \autoref{extracted_kinematics}). This technique combines adjacent spaxels into larger spatial bins, thereby guaranteeing a sufficient number of stars within each bin for accurate integrated light measurements.

\section{HARMONI IFS Simulations}\label{simulations}

First, we describe the HARMONI instrument on ELT and HARMONI Simulator ({\tt HSIM}) in \autoref{HSIM}. Next, we combine the mass models of NGC~300 and NGC~3115 dw01 constructed in \autoref{photometry} with {\tt HSIM} to simulate their $J$ (1.046--1.324 $\mu$m), $H$ (1.435--1.815 $\mu$m), $K$ (1.951--2.469 $\mu$m), $H$-high (1.538--1.678 $\mu$m), $K$-short (2.017--2.201 $\mu$m), and $K$-long (2.199--2.400 $\mu$m) mock datacubes in \autoref{mocks}. Finally, we present the extracted kinematics from these mock {\tt HSIM} cubes in \autoref{extracted_kinematics}.

\subsection{HARMONI instrument on the ELT and {\tt HSIM} simulator}\label{HSIM}

HARMONI is the first optical and NIR IFS ranging from 0.458 to 2.469 $\mu$m and offering IFS at four distinct spatial scales (i.e., $4\times4$, $10\times10$, $20\times20$, and $30\times60$ mas$^2$) and three spectral resolving powers (i.e., $\lambda/\Delta\lambda\approx3,300$, $\approx$7,100 and $\approx$17,400) through 13 gratings. As integrated from 798 hexagonal segments with 1.4-meter diameter, HARMONI will gather spectral data from a field of $152\times214$ spaxels ($\approx$32,530 spectra) over a 39-meter primary mirror equipped with laser guide star AO. We note that HARMONI is currently undergoing a Rescoping Science Evaluation, and its detailed specifications may change. In this work, we use the last available stable specifications, as the instrument characteristics are not yet sufficiently settled to incorporate the latest potential changes into our simulations.

\citetalias{Nguyen23} demonstrated that HARMONI could robustly measure SMBH masses for massive galaxies up to $z\lesssim0.3$. Its scientific explorations can also extend from diffraction-limited imaging to ultrasensitive observations, including the study of morphology, spatially resolved populations and kinematics, abundances, and line ratios in distant sources, even in regions of faint SB \citep{Thatte16, Thatte20}. 

 \citet{Zieleniewski15} introduced a sophisticated {\tt HSIM} pipeline designed for simulating observations with the HARMONI instrument on the ELT. This software takes as input high spectral and spatial resolution IFS cubes without random noise (generated in \autoref{mocks}) and incorporates the physical properties of celestial targets to generate simulated data cubes. In the simulation process, {\tt HSIM} factors in the complex atmospheric conditions at the ELT’s observation site and realistic detector statistics/readout, thereby producing data that mimics actual observations. This paper is primarily focused on utilizing HARMONI’s LTAO IFS simulations to probe the nucleus stellar kinematics of nearby dwarf galaxies. We aim to search for the kinematic signatures of IMBHs and estimate their masses. These simulations also delineate the boundaries within which HARMONI can deliver feasible observables. Furthermore, they provide invaluable scientific insights, offering guidance for future research efforts that will use actual data obtained with HARMONI.

\begin{table}
\caption{Mock {\tt HSIM} IFS datacubes ({\tt DIT} = 15 min.)}   
\footnotesize
\centering\begin{tabular}{cccc}
 \hline\hline
Galaxy name      & {\tt HSIM} band &   Exptime    &   Sensitivity   \\
                           &                          & (minutes) &  (minutes)   \\
  (1)                    &  (2)     &            (3)             &      (4)       \\	                
\hline
NGC~300     &            $J$          &180={\tt DIT}$\times$12 &90={\tt DIT}$\times$6\\
            &      $H$, $H$-high      &210={\tt DIT}$\times$14 &120={\tt DIT}$\times$8\\
            &$K$, $K$-short, $K$-long &240={\tt DIT}$\times$16 &150={\tt DIT}$\times$10\\
 \hline
NGC~3115 dw01&                  $J$    &180={\tt DIT}$\times$12 &90={\tt DIT}$\times$6\\
             &            $H$, $H$-high&210={\tt DIT}$\times$14 &120={\tt DIT}$\times$8\\
             &$K$, $K$-short, $K$-long &240={\tt DIT}$\times$16 &150={\tt DIT}$\times$10\\                   
\hline
\end{tabular}\\
\label{hsimmock}
\parbox[t]{0.47\textwidth}{\textit{Notes:} Column 1: the galaxy name. Column 2: {\tt HSIM}  band chosen to perform IFS simulations. Column 3: real exposure time we put in {\tt HSIM} for our simulated kinematics maps presented in \autoref{ngc300_mock_kin} and \autoref{ngc3115dw01_mock_kin}, where the total integration time is determined as {\tt DIT}$\times$ {\tt NDIT}. Column 4: Sensitivity in terms of exposure time at which we test the lowest limit of S/N from the simulated IFS so that our {\tt pPXF} still extract accurate kinematics (will be discussed later in \autoref{sensitivity_limits}). We should note that the estimated time show in Columns 3 and 4 are the science time on targets without accounting for the target acquisition, overhead, and AO setup time.} 
\end{table}

\subsection{Simulations of Mock IFS Datacubes}\label{mocks}

We generated mock HARMONI IFS datacubes using the {\tt HSIM} pipeline. These simulations were conducted for three medium-resolution gratings ($J$, $H$, $K$, with $\lambda/\Delta\lambda \approx 7100$ and $\sigma_{\rm instr} \approx 18$ \kms) and three high-resolution gratings ($H$-high, $K$-short, $K$-long, with $\lambda/\Delta\lambda \approx 17400$ and $\sigma_{\rm instr} \approx 6$ \kms). The primary goal of these simulations was to assess the impact of kinematic errors and evaluate the feasibility of using different stellar spectral features for determining the black hole mass ($M_{\rm BH}$).

As discussed in \citetalias{Nguyen23} and \citet{CrespoGomez21}, the $K$-band spectrum (covering both $K$-short and $K$-long) contains several molecular absorption lines crucial for measuring stellar kinematics. Prominent among these are the CO absorption bandheads in the range of 2.29--2.47 $\mu$m, such as $^{12}$CO(2--0) at 2.293 $\mu$m and $^{12}$CO(3--1) at 2.312 $\mu$m. Additionally, atomic lines like \ion{Na}{1} $\lambda$2.207 $\mu$m, \ion{Ca}{1} $\lambda$2.263 $\mu$m, and \ion{Mg}{1} $\lambda$2.282 $\mu$m in the $K$-band are also important for this purpose.

In the $H$-band (including $H$-high), a rich set of atomic absorption lines is available (e.g., \ion{Mg}{1} $\lambda$1.487, 1.503, 1.575, 1.711 $\mu$m; \ion{Fe}{1} $\lambda$1.583 $\mu$m; \ion{Si}{1} $\lambda$1.589 $\mu$m). Strong CO absorption features are also present, including $^{12}$CO(3--0) $\lambda$1.540 $\mu$m, $^{12}$CO(4--1) $\lambda$1.561 $\mu$m, $^{12}$CO(5--2) $\lambda$1.577 $\mu$m, $^{12}$CO(6--3) $\lambda$1.602 $\mu$m, $^{12}$CO(7--4) $\lambda$1.622 $\mu$m, and $^{12}$CO(8--5) $\lambda$1.641 $\mu$m.

Furthermore, promising atomic and molecular absorption lines have been identified in the $J$-band IFS for potential use in future studies. These lines are primarily sensitive to metallicity and include species like \ion{Fe}{1}, \ion{Mg}{1}, \ion{Si}{1}, and \ion{Ti}{1}. Examples include \ion{Ti}{1} $\lambda$1.1896, 1.1953 $\mu$m; \ion{Si}{1} $\lambda$1.1988, 1.1995, 1.2035, 1.2107 $\mu$m; \ion{Mg}{1} $\lambda$1.1831, 1.2087 $\mu$m; and \ion{Fe}{1} $\lambda$1.1597, 1.1611, 1.1641, 1.1693, 1.1783, 1.1801, 1.1887, 1.1976 $\mu$m \citep{Rayner09, Lyubenova12}.

A representative subset of these potential atomic and molecular absorption lines is marked on the six {\tt HSIM} mock IFS spectra displayed in \autoref{spectrum_IMBH}. These lines were selected to guide the determination of the optimal wavelength range for extracting stellar kinematics, specifically aiming to avoid contamination from sky emission (such as OH lines in the $J$ band). For a more exhaustive list and visual representation of these stellar absorption tracers, interested readers are directed to \citet{Wallace96, Wallace97, Rayner09, Evans11, Lyubenova12, CrespoGomez21}.

It is important to note that the atomic absorption species and molecular absorption lines observed in the $J$, $H$, $H$-high, $K$-short, and the blue portion of the $K$-band IFS generally exhibit relatively narrow profiles and low intensity. This characteristic can lead to saturation at very high velocity dispersions, typical of the vicinity of massive galaxies hosting SMBHs \citepalias{Nguyen23}. However, this limitation is less critical for NSCs, which generally have significantly lower stellar velocity dispersions ($\sigma_\star$). While these lines may be saturated near SMBHs ($\sigma_\star \gtrsim 70$ \kms) due to line broadening and blending with noise, they are expected to serve as effective indicators for measuring stellar kinematics around IMBHs, where velocity dispersions are lower. In the context of this simulated work focusing on IMBHs, we will specifically evaluate the feasibility of employing these lines in the $J$, $K$-short, and the blue part of the $K$-band IFS for measuring stellar kinematics, as detailed in \autoref{extracted_kinematics}.

\subsubsection{Creation of the input noiseless {\tt HSIM} datacubes}\label{noiseless_inputcube}

To generate the input noiseless datacubes for {\tt HSIM}, we adopted the MARCS synthetic stellar spectra (\autoref{MARCS}) and assumed a Gaussian line-of-sight velocity distribution (LOSVD).

First, we used the \textsc{jam\_axi\_proj} routine from the \textsc{JamPy} package \citep{Cappellari08, Cappellari20} to compute the projected intrinsic first ($V$) and second ($V_{\rm rms}$) velocity moments. We assumed an oblate velocity ellipsoid (JAM keywords \texttt{gamma=0, kappa=1}) for predicting $V$. The velocity dispersion was then calculated as $\sigma_\star=\sqrt{V_{\rm rms}^2-V^2}$.

For each galaxy, we generated three kinematic models corresponding to different IMBH masses (\Mbh), expressed as fractions of the NSC mass (\Mnsc): 0\%, 0.5\%, and 1\%. These fractions correspond to $M_{\rm BH} = (0,\;5\times10^3,\;10^4)$ \Msun\ for NGC~300 and $M_{\rm BH} = (0,\; 3.5\times10^4,\;7\times10^4)$ \Msun\ for NGC~3115 dw01. The NSC masses, $M_{\rm NSC} = 10^6$ \Msun\ for NGC~300 and $M_{\rm NSC} = 7\times10^6$ \Msun\ for NGC~3115 dw01, were calculated using the MGE models representing the Core-Sérsic profile (\autoref{mgetab}) and the respective mass-to-light ratios: $M/L_I\approx0.6$ (\Msun/\Lsun) for NGC~300 \citep{Neumayer12} and $M/L_{\rm F814W}\approx1.4$ (\Msun/\Lsun) for NGC~3115 dw01 \citep{Pechetti20}. These values are consistent with the JAM model masses within 3\%. The JAM$_{\rm cyl}$ kinematic maps were computed on a $5\times5$ mas$^2$ grid, assuming isotropy ($\beta_z=0$) and an inclination $i=42^{\circ}$ (\autoref{simulated_targets}).

Given the proximity of the HARMONI IMBH sample ($D \lesssim 10$ Mpc), redshift effects are negligible. The input noiseless IFS datacube for {\tt HSIM} was then created for each spaxel $(x,y)$ using the following procedure:
\begin{enumerate}
    \item The chosen synthetic stellar spectrum (\autoref{MARCS}) was logarithmically rebinned to a velocity scale of {\tt velscale = }0.5 \kms\ per pixel.
    \item For each spatial position $(x,y)$, a kinematic Gaussian kernel was constructed using the mean velocity $V(x,y)$ and velocity dispersion $\sigma_\star(x,y)$ predicted by the JAM$_{\rm cyl}$ model. The kernel was sampled with a velocity step $\Delta V = 0.5$ \kms.
    \item The logarithmically rebinned spectrum from step (i) was convolved with the Gaussian kernel from step (ii) in log-wavelength space. The resulting spectrum represents the broadened stellar light at position $(x,y)$.
    \item The convolved spectrum was then resampled onto a linear wavelength grid. The wavelength step $\Delta\lambda$ was chosen to be at least two times smaller than the FWHM spectral resolution element of the highest-resolution HARMONI grating ($\Delta\lambda \approx 0.5$ \AA\ for the high-resolution gratings), ensuring proper sampling and conservation of spectral information during subsequent processing by {\tt HSIM}.
    \item Finally, the spectrum at $(x,y)$ was scaled to match the surface brightness predicted by the MGE photometric model (\autoref{mgetab}) at that location. The scaling factor was determined by comparing the integrated flux of the unscaled spectrum within the relevant photometric band (e.g., F814W) to the MGE surface brightness, using the \texttt{ppxf.ppxf\_util.mag\_spectrum} function from {\tt pPXF} \citep{Cappellari23}.
\end{enumerate}

\begin{figure*}
\centering\includegraphics[width=0.99\textwidth]{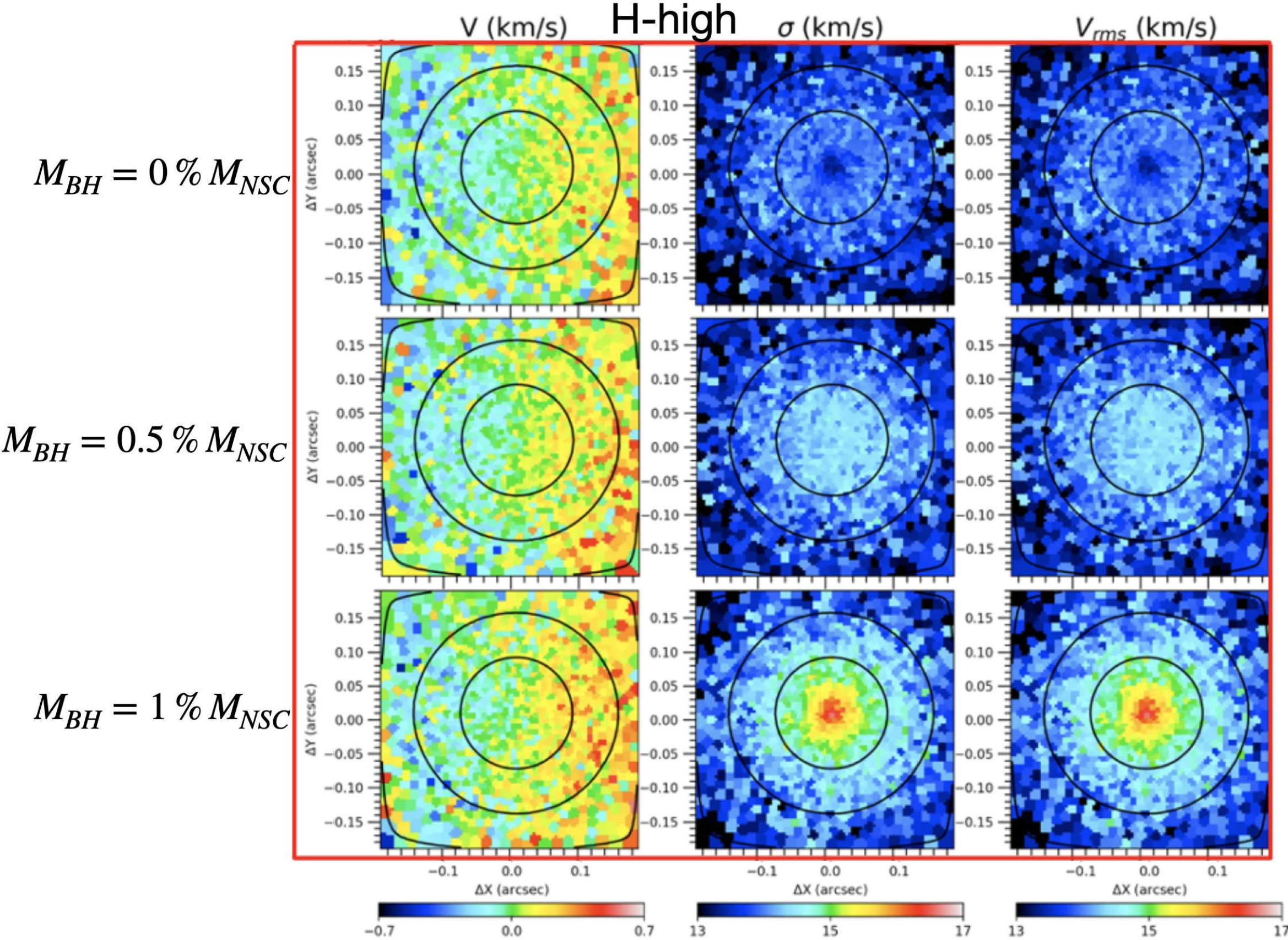}
    \caption{Stellar kinematic maps ($V$, $\sigma_{\star}$, and $V_{\rm rms}$) for the nucleus of NGC~300, derived from mock HARMONI $H$-high IFS observations. The simulations were performed using {\tt HSIM} \citep{Zieleniewski15}, based on input kinematics generated with JAM$_{\rm cyl}$ models \citep{Cappellari08} assuming three different central BH masses: $M_{\rm BH} = 0$ \Msun\ (top row), $5\times10^3$ \Msun\ (middle row), and $10^4$ \Msun\ (bottom row). Kinematics were extracted from the simulated noisy data cubes using {\tt pPXF} \citep{Cappellari23}, fitting the $H$-band spectral range $\lambda$1.45--1.74 $\mu$m. Contours represent isophotes from the collapsed data cube, spaced by 1 mag arcsec$^{-2}$. The color bars are fixed across the rows to facilitate comparison and illustrate the kinematic impact of the central IMBH. The results demonstrate HARMONI's capability to detect IMBH kinematic signatures in NSCs.} 
    \label{ngc300_mock_kin} 
\end{figure*}

\begin{figure*}
    \centering\includegraphics[width=0.99\textwidth]{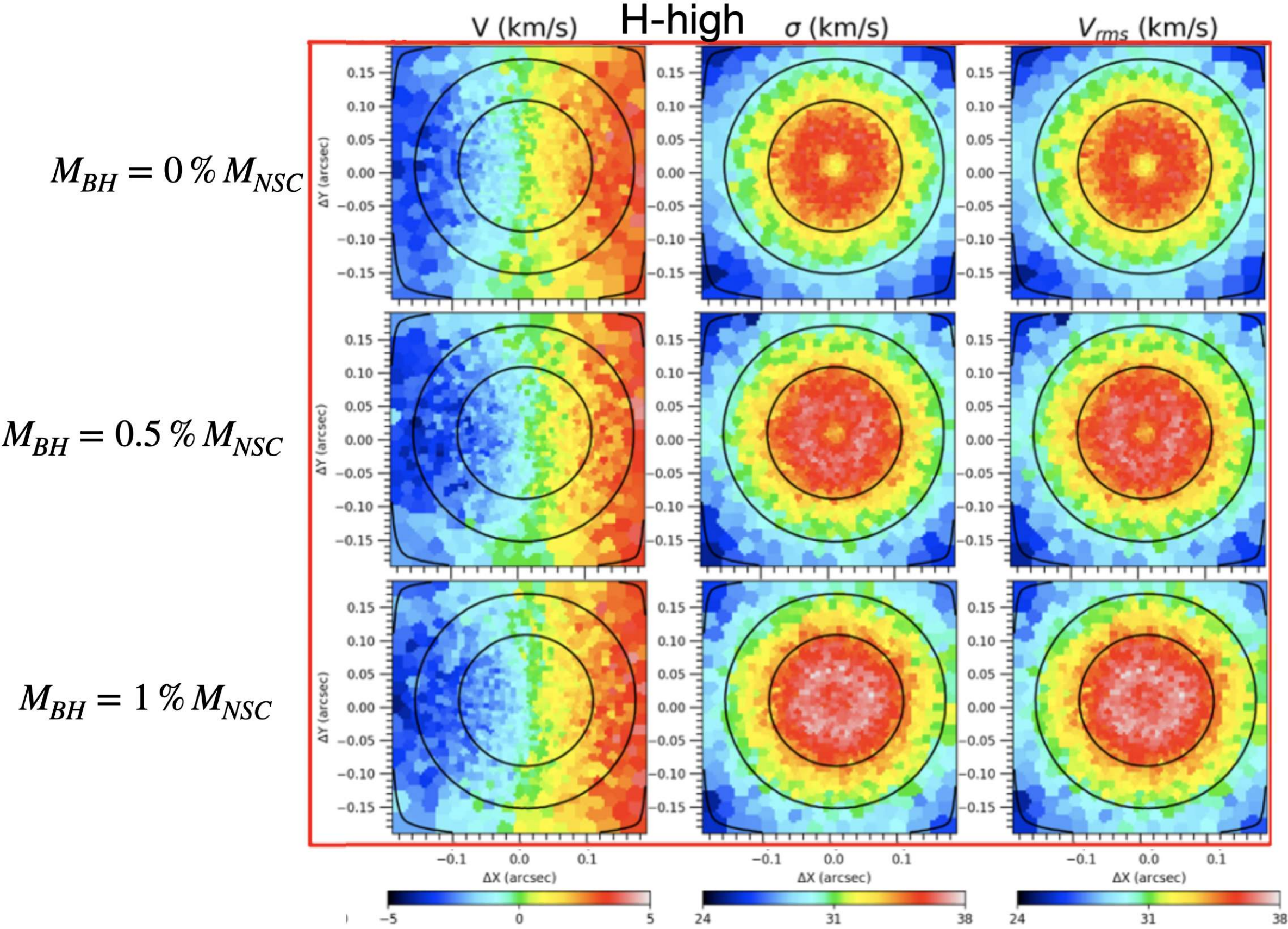}
    \caption{Same as \autoref{ngc300_mock_kin} but for NGC~3115 dw01 with three $M_{\rm BH}$: $ = 0$ \Msun\ (first row), $ = 3.5\times10^4$ \Msun\ (second row), and $ = 7\times10^4$ \Msun\ (third row).}
    \label{ngc3115dw01_mock_kin} 
 \end{figure*}

\subsubsection{{\tt HSIM} output datacubes (mock IFS HARMONI)}\label{mock_outputcube}

We used the input-noiseless cubes created in \autoref{noiseless_inputcube} as an input for {\tt HSIM} to produce the mock HARMONI IFS observations. For computational efficiency, we limited our IFS simulations within a central subset of the full HARMONI FoV measuring $0\farcs4\times0\farcs4$ region with a pixel size of $10\times10$ mas$^2$. This choice ensured that the stellar kinematics, primarily influenced by the central BHs in our HARMONI IMBH sample, could be robustly detected. The exposure time for each simulation was carefully adjusted based on the gratings to ensure a median S/N of the final HARMONI spectral sampling produced in output by HSIM at least 5 per spectral pixel in each spaxel at the measured stellar features. Subsequently, pixel binning was employed to increase the S/N further. To replicate the actual ELT observations, we incorporated multi-exposure frames and dithering by setting a Detector Integration Time ({\tt DIT}) of 900 s (15 minutes) for each exposure. The total exposure time was determined by the number of exposures ({\tt NDIT}), which is an integer specified in the {\tt HSIM} pipeline.

All essential properties of NGC~300 and NGC~3115 dw01 needed for the modelings are presented in \autoref{simulated_targets}, while the chosen grating IFS and {\tt HSIM} simulations are shown in \autoref{hsimmock}. Regarding the AO performance during {\tt HSIM} simulations, we applied the LTAO mode with an NGS of 17.5 mag in $H$-band within the radial distance of 30$\arcsec$, the optical (500 nm) zenith seeing of FWHM = $0\farcs64$, and airmass of 1.3. These parameters are defaulted in {\tt HSIM} to perform median observational conditions for the Armazones site.

\subsection{Stellar kinematics extraction from {\tt HSIM} mock datacubes} \label{extracted_kinematics}

To measure the stellar kinematics from the mock HARMONI data cubes generated in \autoref{mocks}, we utilized the spectral ranges defined by the HARMONI gratings themselves, as simulated by {\tt HSIM}. Within these ranges, we focused on specific wavelength intervals known to contain prominent stellar absorption features suitable for kinematic analysis, as discussed in \autoref{mocks} and illustrated in \autoref{spectrum_IMBH}. The key intervals used within each grating's range were:
\begin{itemize}
    \item Within the $K$ (1.951--2.469 $\mu$m) and $K$-long (2.199--2.400 $\mu$m) bands: $\lambda$2.28--2.40 $\mu$m (targeting CO bandheads).
    \item Within the $H$ (1.435--1.815 $\mu$m) and $H$-high (1.538--1.678 $\mu$m) bands: $\lambda$1.45--1.74 $\mu$m (targeting various atomic lines and CO features).
    \item Within the $J$-band (1.046--1.324 $\mu$m): $\lambda$1.16--1.218 $\mu$m (targeting atomic lines, avoiding strong OH sky emission).
    \item Within the $K$-short band (2.017--2.201 $\mu$m): $\lambda$2.05--2.20 $\mu$m (targeting atomic lines).
\end{itemize}
We verified the suitability of fitting only these restricted wavelength intervals by comparing the results with those obtained from fitting the full spectral range provided by each grating (excluding regions heavily contaminated by sky lines, particularly in the $J$-band). The differences in the extracted kinematics were minimal ($<$5\%), confirming that these selected features robustly capture the kinematic information.

Before extracting kinematics, we applied the adaptive Voronoi Binning method \citep[{\tt vorbin}\footnote{v3.1.5 available from \url{https://pypi.org/project/vorbin/}};][]{Cappellari03} to the spatial dimensions of the data cubes. This process groups adjacent spaxels into larger bins to achieve a targeted-S/N greater than 20 per bin per \AA\ within the selected spectral fitting ranges, thereby reducing uncertainties in the subsequent kinematic measurements.

Next, we extracted the stellar kinematics from the binned spectra using the penalized pixel-fitting method \citep[{\tt pPXF}\footnote{v8.2.1 available from \url{https://pypi.org/project/ppxf/}};][]{Cappellari23}. For each bin:
\begin{enumerate}
    \item The spectrum was logarithmically rebinned in wavelength.
    \item A set of stellar templates, specifically the MARCS \citep{Gustafsson08} version of the \citet{Maraston11} SPS models (using 13 templates with ages 3--15 Gyr and Solar metallicity, {\tt z002}), was fitted to the spectrum.
    \item The template continuum shape was corrected using Legendre polynomials ({\tt mdegree}=0, {\tt degree}=4).
    \item The LOSVD was parameterized by its first two moments (mean velocity $V$ and velocity dispersion $\sigma_\star$) by setting {\tt moments}=2.
    \item The instrumental broadening of HARMONI (using the constant dispersion adopted by {\tt HSIM}) was accounted for by convolving the templates during the fit.
\end{enumerate}
\autoref{spectrum_IMBH} shows examples of the best-fit SPS template overlaid on the simulated spectrum from the central bin for each of the six gratings, along with the fitting residuals ({\tt data-model}).

The resulting kinematic maps ($V$, $\sigma_\star$, and the root-mean-square velocity $V_{\rm rms}=\sqrt{V^2 +\sigma_\star^2}$) for the nuclei of NGC~300 and NGC~3115 dw01 are presented in \autoref{ngc300_mock_kin} and \autoref{ngc3115dw01_mock_kin} for the $H$-high grating, and in \autoref{ngc300_mock_kin_full} and \autoref{ngc3115dw01_mock_kin_full} for the other gratings. Since the rotation $V$ is small in these models ($V \approx 0.7$ \kms\ for NGC~300 and $V \approx 5.0$ \kms\ for NGC~3115 dw01), the $V_{\rm rms}$ maps are very similar to the $\sigma_\star$ maps. The kinematic results are consistent across all six gratings for simulations with the same input $M_{\rm BH}$, with differences in $V_{\rm rms}$ typically less than 3\%, validating the extraction procedure.

\begin{figure*}
    \centering\includegraphics[width=0.99\textwidth]{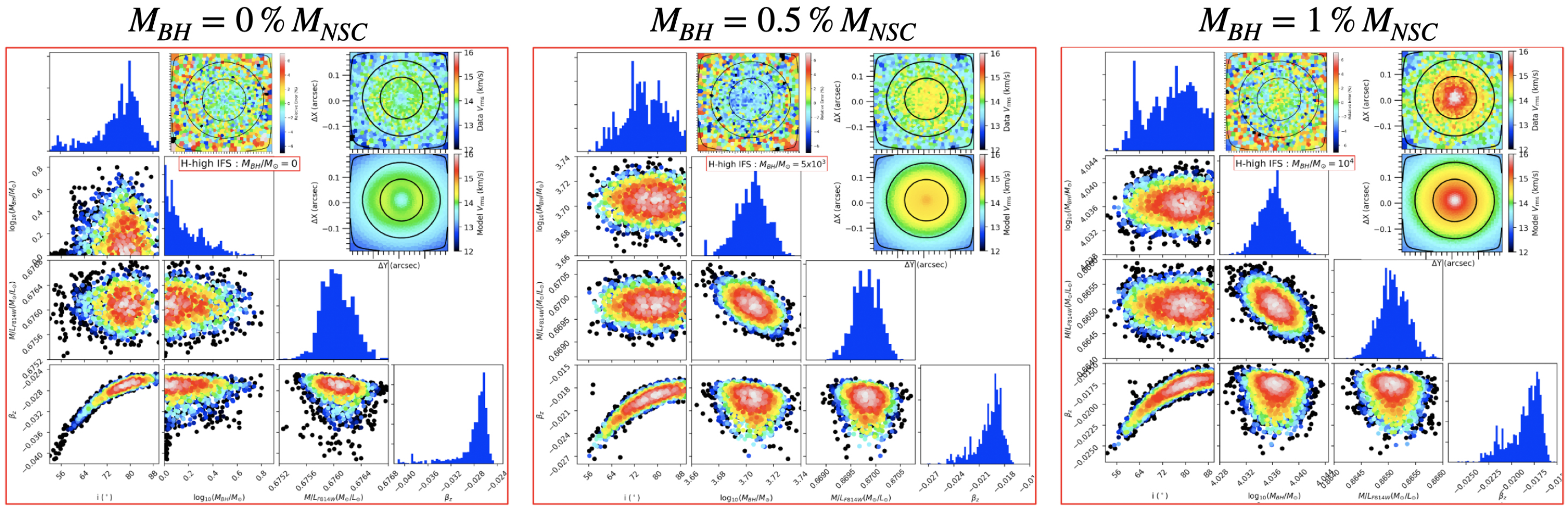}
    \caption{The posterior distributions obtained after the burn-in phase of the {\tt adamet} MCMC optimization process for the JAM$_{\rm cyl}$ models applied to the $H$-high {\tt HSIM} kinematics of NGC~300. These simulations were generated using the JAM$_{\rm cyl}$ models and feature three $M_{\rm BH}$: 0 \Msun\ (left), $5\times10^3$ \Msun\ (middle), and $ 10^4$ \Msun\ (right). Each red-square panel presents a set of four parameters ($i$, $M_{\rm BH}$, $M/L_{\rm F814W}$, and $\beta_z$), depicted as scatter plots illustrating their projected 2D distributions and histograms displaying their projected 1D distributions. In the top right corner, there are inset maps that depict the $V_{\rm rms}$ values. The top maps represent the simulated kinematic maps extracted from the simulated datacubes, while the bottom maps represent the kinematic maps recovered from the best-fit JAM$_{\rm cyl}$ model. These maps visually illustrate the level of agreement or disagreement at each spaxel between the simulated data and our best-fit model. The determination of the best-fit JAM$_{\rm cyl}$ model is based on the PDF with the highest likelihood.} 
    \label{ngc300_mock_kin_BHrecover_H-high}
\end{figure*}

\begin{figure*}
    \centering\includegraphics[width=0.99\textwidth]{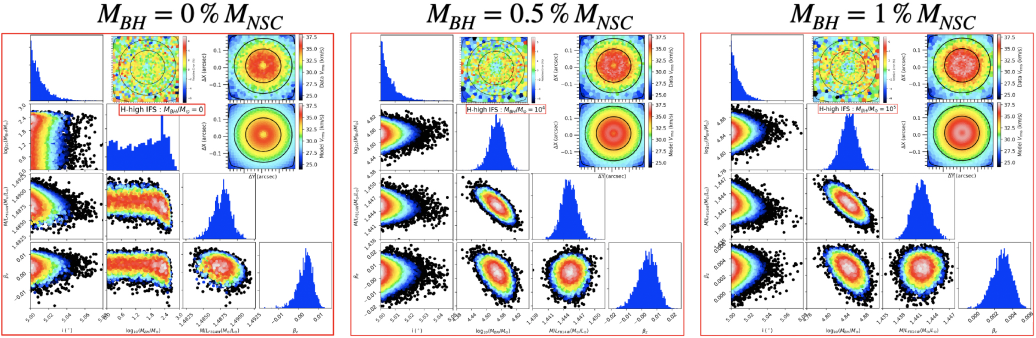}
    \caption{Same as \autoref{ngc300_mock_kin_BHrecover_H-high} but for the $H$-high {\tt HSIM} kinematics of NGC~3115 dw01 with three $M_{\rm BH}$: 0 \Msun\ (left), $3.5\times10^4$ \Msun\ (middle), $7\times10^4$ \Msun\ (right).} 
    \label{ngc3115dw01_mock_kin_BHrecover_H-high}
\end{figure*}

As theoretically predicted (\autoref{predicted_vrms} and \citealt{Tremaine94}), the kinematic maps clearly show the signature of the central IMBH. Models without a BH ($M_{\rm BH}=0$) exhibit a characteristic drop in $\sigma_\star$ and $V_{\rm rms}$ towards the center. In contrast, models with sufficiently massive IMBHs show a distinct central peak in $\sigma_\star$ and $V_{\rm rms}$ within the BH's SOI, with the peak becoming more pronounced as $M_{\rm BH}$ increases. This Keplerian-like rise in velocity dispersion is the key kinematic evidence for a central massive object.

{\bf In contrast, models that include a BH (i.e., $M_{\rm BH}=3.5\times10^4$ and $7\times10^4$\Msun) for NGC~3115 dw01 show a characteristic central drop in $\sigma_\star$ and $V_{\rm rms}$, although not as pronounced as in the $M_{\rm BH}=0$ case. This may be due to the combined effects of the galaxy's distance, a massive NSC, and the small assumed $M_{\rm BH}$}.  However, the clear difference between the kinematic maps with and without an IMBH, even for this most distant target NGC~3115 dw01 (at 9.7 Mpc), demonstrates HARMONI's capability to resolve the kinematic signatures of IMBHs in nearby NSCs. HARMONI can effectively distinguish between a scenario with no IMBH and one containing an IMBH with a mass of just 0.5-1\% of the NSC mass.

We note the presence of edge effects in the kinematic maps, particularly higher $\sigma_\star$ and $V_{\rm rms}$ values in the outermost bins. This is an artifact of the PSF convolution performed by {\tt HSIM} near the boundary of the simulated field of view ($0\farcs4\times0\farcs4$), resulting in the squared appearance of the outermost surface brightness contour. To avoid biasing the dynamical modeling, these affected edge bins were excluded during the $M_{\rm BH}$ recovery process (\autoref{bhrecovering}).

\section{Discussion}\label{discussion}

\subsection{Recovering Black Hole Masses}\label{bhrecovering}

Using the mock IFS data cubes from the six HARMONI gratings and their derived kinematic maps (\autoref{extracted_kinematics}), we fitted these maps ($V_{\rm rms}$) with JAM$_{\rm cyl}$ models. The goal was to determine the IMBH mass ($M_{\rm BH}$) and constrain other dynamical parameters: stellar anisotropy ($\beta_z$), the stellar mass-to-light ratio relative to the \hst/F814W photometry (\ml$_{\rm F814W}$), and the galaxy inclination ($i$). We sampled $M_{\rm BH}$ logarithmically (assuming a flat prior in $\lg M_{\rm BH}$), while the other three parameters were sampled linearly. The JAM$_{\rm cyl}$ models predict kinematics which are then compared to the mock $V_{\rm rms}$ maps, accounting for the HARMONI LTAO PSF (FWHM$_{\rm PSF}\approx12$ mas, \citetalias{Nguyen23}). We set the following search ranges for the parameters:
\begin{itemize}
    \item $\log_{10}(M_{\rm BH}/$\Msun): 0 to 6
    \item \ml$_{\rm F814W}$: 0.1 to 3 (\Msun/\Lsun)
    \item $\beta_z$: $-$1.0 to 0.99
    \item $i$: 5$^{\circ}$ to 90$^{\circ}$
\end{itemize}

We started the fitting process with initial guesses for the parameters. For NGC~300, we used (\ml$_{\rm F814W}$, $\beta_z$, $i$) = (0.6, 0, 42), and for NGC~3115 dw01, (\ml$_{\rm F814W}$, $\beta_z$, $i$) = (1.4, 0, 42). The initial guesses for $M_{\rm BH}$ were set to the values used in the simulations (\autoref{mocks}): $M_{\rm BH}=0,\;5\times10^{3},\;10^{4}$ \Msun\ for NGC~300 and $M_{\rm BH}=0,\;3.5\times10^{4},\;7\times10^{4}$ \Msun\ for NGC~3115 dw01, depending on the specific simulation being fitted.

To find the best-fit JAM$_{\rm cyl}$ model and explore the parameter space thoroughly, we employed a Markov Chain Monte Carlo (MCMC) simulation using the adaptive Metropolis algorithm \citep{Haario01} within the Bayesian framework provided by the {\tt adamet}\footnote{v2.0.9 available from \url{https://pypi.org/project/adamet/}} package \citep{Cappellari13a}. Each MCMC chain ran for $3\times10^4$ iterations. We discarded the first 20\% as the burn-in phase and constructed the full probability distribution function (PDF) from the remaining 80\%. The best-fit parameters correspond to the point of highest likelihood in the PDF. We determined the 1$\sigma$ (16--84 percentile range) and 3$\sigma$ (0.14--99.86 percentile range) confidence levels from the posterior PDF.

The best-fit parameters and their uncertainties for NGC~300 are shown in \autoref{ngc300_mock_kin_BHrecover_H-high}, \autoref{ngc300_mock_kin_BHrecover11}, and \autoref{ngc300_mock_kin_BHrecover22}, and for NGC~3115 dw01 in \autoref{ngc3115dw01_mock_kin_BHrecover_H-high}, \autoref{ngc3115dw01_mock_kin_BHrecover11}, and \autoref{ngc3115dw01_mock_kin_BHrecover22}. These figures display the 2D posterior distributions for pairs of parameters, where color indicates likelihood (white is highest, black is below 3$\sigma$). Histograms show the 1D marginalized distributions. The best-fit values and uncertainties derived from these distributions are listed in \autoref{tabA3} (NGC~300) and \autoref{tabA4} (NGC~3115 dw01).

Inset plots in the top-right corner of each panel in these figures compare the input mock $V_{\rm rms}$ map (top) with the best-fit JAM$_{\rm cyl}$ model prediction (bottom). Relative error maps, {\tt (data-model)/data}, are also shown (middle) to visualize the goodness of fit bin-by-bin. Overall, the models agree well with the mock kinematics within the central $0\farcs4\times0\farcs4$ region for all gratings.

The recovered $M_{\rm BH}$ and \ml$_{\rm F814W}$ values are consistent with the input values used in the simulations, typically within 5\%. This success is due to the high quality of the simulated kinematics and the fact that the IMBH SOI is well-resolved by HARMONI's 10 mas spaxels, even for the smallest simulated BH masses ($M_{\rm BH}=5\times10^{3}$ \Msun\ for NGC~300 and $M_{\rm BH}=3.5\times10^{4}$ \Msun\ for NGC~3115 dw01). The corresponding SOI radii are $R_{\rm SOI}=13$ mas for NGC~300 ($D=2.2$ Mpc, $\sigma_\star=13.3$ \kms) and $R_{\rm SOI}=35$ mas for NGC~3115 dw01 ($D=9.7$ Mpc, $\sigma_\star=32$ \kms).

The 2D posterior distributions show consistent features across different $M_{\rm BH}$ values and gratings. As expected, when $M_{\rm BH}>0$, there is a negative covariance between $M_{\rm BH}$ and \ml$_{\rm F814W}$, creating the characteristic ``banana shape" in the confidence contours. This degeneracy arises because a larger BH mass can be partially compensated by a smaller stellar \ml, and vice versa, while still fitting the mock kinematics.

The high spatial resolution of the simulations allows us to break this degeneracy effectively. For the simulations without a BH ($M_{\rm BH}=0$), the fits yield upper limits of $M_{\rm BH} \approx 10^2$ \Msun\ for NGC~300 and $M_{\rm BH} \approx 10^3$ \Msun\ for NGC~3115 dw01. This indicates HARMONI's ability to distinguish between the presence and absence of IMBHs above these mass thresholds at the respective distances (though see \autoref{mass-segregation} for caveats regarding mass segregation).

The anisotropy parameter $\beta_z$ is well-constrained around the input value of $\beta_z=0$ (isotropy), with best fits slightly favoring tangential orbits ($\beta_z \lesssim 0$) for NGC~300 and radial orbits ($\beta_z \gtrsim 0$) for NGC~3115 dw01, consistent with isotropy within errors. The inclination $i$, however, is less well-constrained. For NGC~300, it spans a wide range (40$^{\circ}$ to 90$^{\circ}$), while for NGC~3115 dw01, it is mostly constrained to low values ($\gtrsim 5^{\circ}$), except for the $J$-band fit. This poor constraint on inclination is expected because the models are nearly spherical and show little rotation; spherical systems look identical regardless of viewing angle.

 \begin{figure}
    \centering\includegraphics[width=0.47\textwidth]{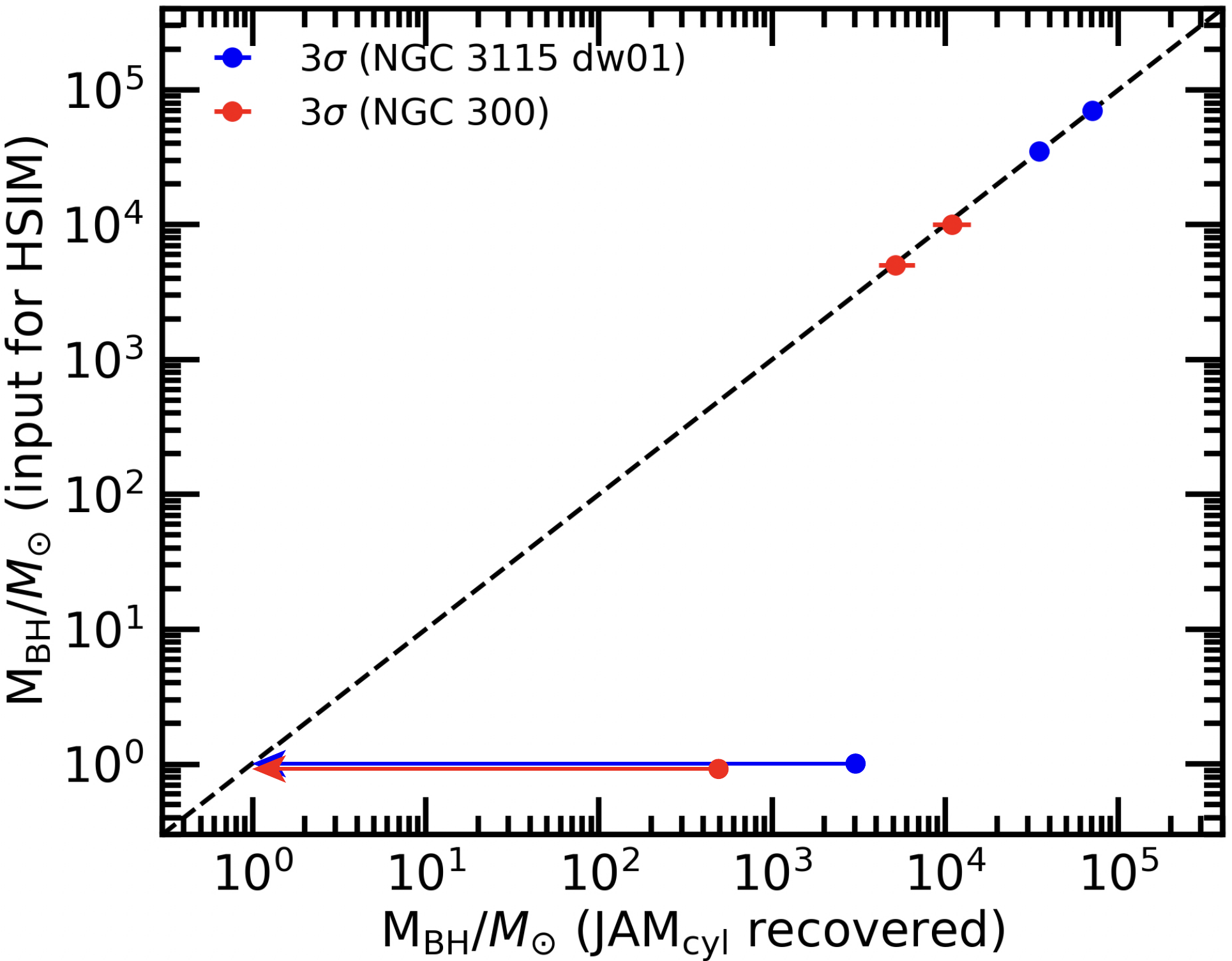}
    \caption{Comparison between the input IMBH masses used for the HARMONI simulations and the recovered masses from the JAM$_{\rm cyl}$ fits. The points show the median recovered $M_{\rm BH}$ across all six gratings for each input mass. Error bars represent the $3\sigma$ confidence interval (0.14th to 99.86th percentiles from the MCMC chains, see \autoref{tabA3} and \autoref{tabA4}). The dashed line indicates perfect recovery ($M_{\rm BH, recovered} = M_{\rm BH, input}$).   {\bf In this log--log scale plot, the (0, 0) coordinates are replaced by ($10^0$, $10^0$)}.}
    \label{summaryBH}
\end{figure}

\autoref{summaryBH} summarizes the comparison between the input $M_{\rm BH}$ values and the recovered values from the JAM$_{\rm cyl}$ modeling across all six gratings. The plot shows the median recovered mass for each input mass, with error bars indicating the $3\sigma$ confidence range. The results demonstrate excellent recovery, consistent with the input values within the uncertainties.

\subsection{Caveat: Distinguishing IMBHs from Mass Segregation}\label{mass-segregation}

A potential complication in dynamically detecting IMBHs is the effect of mass segregation. Stellar-mass black holes (sBHs), being more massive than typical stars, can sink towards the center of a dense stellar system like an NSC over time. This process increases the central mass density and, consequently, the central velocity dispersion, potentially mimicking the kinematic signature of a single, more massive IMBH \citep{Bahcall77, Bianchini17}.

Theoretical studies suggest that IMBHs down to $M_{\rm BH}\approx3\times10^3$ \Msun\ might be dynamically detectable in dense stellar systems like Galactic globular clusters, even considering mass segregation \citep[][section 3.2]{Greene20}. Our simulations explore IMBHs with masses of 0.5\% of their host NSC's mass. For NGC~300 ($M_{\rm NSC} \approx 10^6$ \Msun) and NGC~3115 dw01 ($M_{\rm NSC} \approx 7\times10^6$ \Msun), this corresponds to $M_{\rm BH} \approx 5\times10^3$ \Msun\ and $M_{\rm BH} \approx 3.5\times10^4$ \Msun, respectively. Since these masses exceed the $\approx$$3\times10^3$ \Msun\ threshold, detecting them with HARMONI appears feasible based solely on mass.

However, the central concentration of sBHs due to mass segregation can elevate the central \ml, potentially leading to a false IMBH detection \citep[e.g.,][]{denBrok14a}. To assess the importance of this effect, we can estimate the system's relaxation timescale, $t_{\rm relax}(r)$, which characterizes the time needed for stellar orbits to change significantly due to two-body interactions. Following \citet{Bahcall77} and \citet{Valluri05}, we use:
\begin{equation} 
    t_{\rm relax}({\emph r}) \approx (1.4\times10^8 \mathrm{yr})\,\sigma^3_{20}(r)\,\rho_{5}^{-1}(r)(\ln\Lambda_{10})^{-1}.
\end{equation} 
Here, $\sigma_{20}=\sigma_\star/(20\,\kms)$ is the stellar velocity dispersion normalized to 20 \kms, $\rho_{5}=\rho/(10^5\,\Msun\,\text{pc}^{-3})$ is the intrinsic stellar mass density normalized to $10^5\,\Msun\,\text{pc}^{-3}$, and $\ln\Lambda_{10}=\ln(\Lambda)/10$ is the Coulomb logarithm, typically assumed to be $\approx 10$. If $t_{\rm relax}$ at a given radius is comparable to or longer than the age of the Universe ($\sim$13.8 Gyr), significant mass segregation of sBHs towards the center is unlikely to have occurred. In such cases, an observed central rise in velocity dispersion is more likely a genuine signature of an IMBH.

\begin{figure}
    \centering\includegraphics[width=0.47\textwidth]{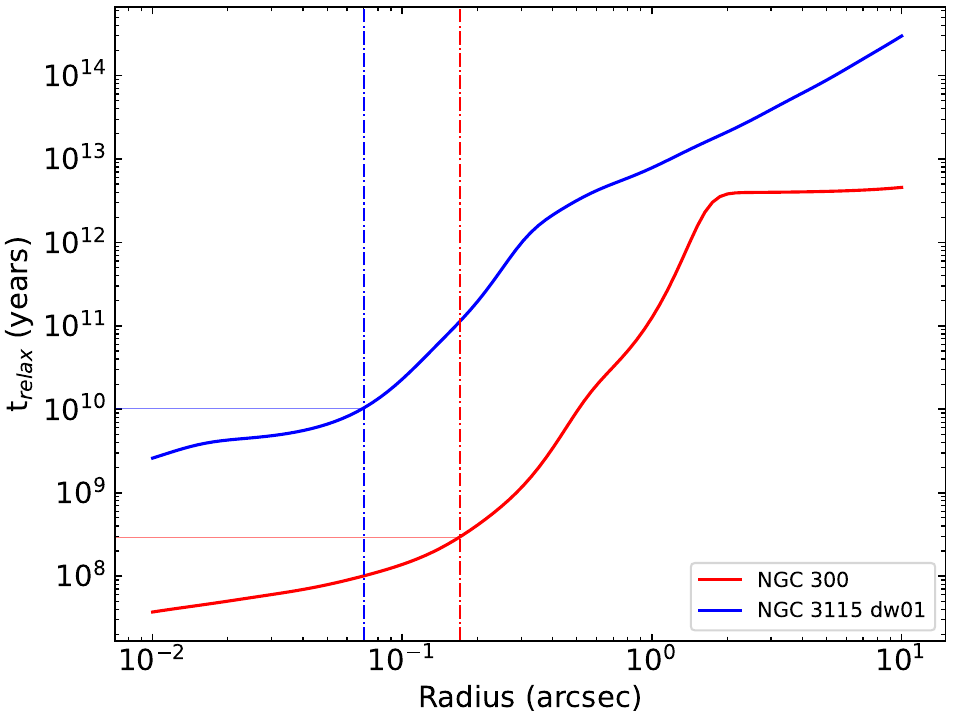}
    \caption{Relaxation time $t_{\rm relax}(r)$ as a function of radius for the NSC models of NGC~300 (red solid line) and NGC~3115 dw01 (blue solid line), assuming no central IMBH. The vertical dash-dotted lines indicate the effective radii ($R_{\rm e}$) of the respective NSCs.}
    \label{Trelax}
\end{figure}

We calculated $t_{\rm relax}(r)$ for the NSCs of NGC~300 and NGC~3115 dw01 using their observed central velocity dispersions ($\sigma_{\star}=13.3$ \kms\ for NGC~300 \citep{Kacharov18}; $\sigma_{\star}=32$ \kms\ for NGC~3115 dw01 \citep{Peterson93}) and intrinsic mass densities derived from the deprojected MGE models of the NSC components (\autoref{mgetab}, bold entries) assuming spherical symmetry and the literature $M/L$ values (\autoref{sb}). The results are shown in \autoref{Trelax}. At the effective radius of the NSC ($R_{\rm e}$), we find $t_{\rm relax}(R_e\approx1.9\,\text{pc})\approx3$ Myr for NGC~300, but $t_{\rm relax}(R_e\approx3.2\,\text{pc})\approx10$ Gyr for NGC~3115 dw01.

The very short relaxation time for NGC~300's NSC suggests that mass segregation could be significant, potentially complicating the interpretation of its central kinematics. Conversely, the long relaxation time for NGC~3115 dw01's NSC implies that mass segregation is likely unimportant, making it a cleaner target for IMBH detection. However, significant uncertainties remain regarding the efficiency of sBH segregation and retention in NSCs, requiring more detailed simulations \citep{Nguyen18}.

While alternative methods like gravitational microlensing have been proposed to detect populations of sBHs \citep{Bennett02}, they are observationally challenging for extragalactic systems and beyond the scope of this paper. It is worth noting that while IMBH formation in globular clusters faces theoretical challenges and lacks strong observational support \citep{vanderMarel10, Zocchi17, Zocchi19, Baumgardt19}, NSCs often possess higher velocity dispersions and longer relaxation times, making them more plausible environments for hosting IMBHs \citep[see review by][section 10.1]{Neumayer20}.

\subsection{Sensitivity Limits}\label{sensitivity_limits}

To determine the minimum exposure time required for reliable kinematic measurements, we performed additional simulations for both NGC~300 and NGC~3115 dw01. In these tests, we progressively reduced the simulated exposure time, which correspondingly lowered the S/N in the mock data cubes. We continued this process until the stellar kinematics could only marginally be extracted after applying Voronoi binning ({\tt Vorbin}) to achieve a target S/N of $\approx$20 per bin.

The minimum required exposure times determined through this process are listed in Column 4 of \autoref{hsimmock}. These times are relatively short (less than three hours per grating), suggesting that such observations are feasible with the ELT.

However, it is important to note that these sensitivity limits are based on simulations using photometric models derived from current \hst\ data. Future, higher-resolution imaging (e.g., from \jwst/NIRCam or ELT/MICADO) might resolve the NSCs more finely, potentially revealing lower central surface brightness than assumed in our current models. If the actual surface brightness is lower, the S/N achieved in a given exposure time will be reduced, and longer exposure times than estimated here might be necessary for real observations. Despite this caveat, our simulations demonstrate the fundamental feasibility of detecting IMBH kinematic signatures with HARMONI within reasonable exposure times.

\section{Conclusions}\label{conclusions}

We investigated the capability of the ELT/HARMONI instrument to detect and measure the masses of IMBHs residing in the NSCs of nearby dwarf galaxies ($D \lesssim 10$ Mpc). This mass range represents a critical link between sBHs and SMBHs. We defined the ``HARMONI IMBH survey," a sample of 44 galaxies selected based on criteria ensuring observability and suitability for IMBH searches, notably the presence of luminous NSCs and low central velocity dispersions ($\sigma_{\star, \rm c} < 70$ \kms).

To assess HARMONI's performance, we conducted realistic simulations for two representative galaxies (NGC~300 and NGC~3115 dw01). We combined photometric models derived from \hst\ data with JAM to predict intrinsic stellar kinematics for scenarios including IMBHs with masses ranging from 0\% to 1\% of the NSC mass. These predictions served as input for the {\tt HSIM} simulator to generate mock HARMONI IFS data cubes across six different gratings. We subsequently analyzed these mock observations using standard techniques: extracting kinematics with {\tt pPXF} and recovering the IMBH mass and other dynamical parameters using JAM within a Bayesian MCMC framework.

Our key findings are:
\begin{itemize}
    \item HARMONI's high angular resolution (simulated at 10 mas) can resolve the kinematic SOI of IMBHs with masses down to $\approx 0.5\%$ of their host NSC mass, clearly distinguishing them from models without a central BH.
    \item The recovered $M_{\rm BH}$ and stellar \ml$_{\rm F814W}$ accurately matched the input simulation values within $\lesssim 5\%$ uncertainty, demonstrating the robustness of the measurement technique with HARMONI data.
    \item Consistent kinematic results and successful parameter recovery were achieved across different HARMONI gratings, including both medium ($\lambda/\Delta\lambda \approx 7100$) and high ($\lambda/\Delta\lambda \approx 17400$) spectral resolutions.
    \item The simulations indicate that high-quality IFS data, sufficient for these demanding measurements, can be obtained within feasible exposure times (typically less than four hours per target under median conditions).
\end{itemize}
These results confirm that HARMONI will be a powerful instrument for constraining the elusive IMBH population in nearby galaxies. Such observations will provide crucial tests for BH seeding models, refine our understanding of galaxy--BH scaling relations at the lowest mass scales, and inform predictions for future gravitational wave observatories.

\section*{Acknowledgements}
The authors would like to thank the anonymous referee for their careful reading and useful comments, that helped to improve the paper greatly. 
Research conducted by HNN is funded by University of Science, VNU-HCM under grant number T2013-105.  TQTL's  and TNL's works are partially supported by a grant from the Simons Foundation to IFIRSE, ICISE (916424, N.H.). NT would like to acknowledge partial support from UKRI grant ST/X002322/1 for UK ELT Instrument Development at Oxford. MPS acknowledges support under grants RYC2021-033094-I, CNS2023-145506 and PID2023-146667NB-I00 funded by MCIN/AEI/10.13039/501100011033 and the European Union NextGenerationEU/PRTR.
 
{\it Facilities:} \hst, 2MASS, SDSS DR12, Gaia, DSS.

{\it Software:} {\tt python 3.12}, {\tt Matplotlib 3.6.0}, {\tt numpy 1.22}, {\tt scipy 1.3.1}, {\tt photutils 0.7}, {\tt MPFIT}, {\tt plotbin 3.1.3}, {\tt astropy 5.1}    \citep{AstropyCollaboration22}, {\tt adamet 2.0.9}  \citep{Cappellari13a}, {\tt jampy 7.2.0} \citep{Cappellari20}, {\tt pPXF 8.2.1} \citep{Cappellari23}, {\tt vorbin 3.1.5}   \citep{Cappellari03}, {\tt MgeFit 5.0.14} \citep{Cappellari02} and {\tt HSIM 3.11} \citep{Zieleniewski15}.



\bibliographystyle{aasjournalv7}
\bibliography{imbh} 

\begin{thebibliography}{}
\expandafter\ifx\csname natexlab\endcsname\relax\def\natexlab#1{#1}\fi
\providecommand{\url}[1]{\href{#1}{#1}}
\providecommand{\dodoi}[1]{doi:~\href{http://doi.org/#1}{\nolinkurl{#1}}}
\providecommand{\doeprint}[1]{\href{http://ascl.net/#1}{\nolinkurl{http://ascl.net/#1}}}
\providecommand{\doarXiv}[1]{\href{https://arxiv.org/abs/#1}{\nolinkurl{https://arxiv.org/abs/#1}}}

\bibitem[{J.~H. An \& N.~W. Evans(2006)An \& Evans}]{An06}
An, J.~H., \& Evans, N.~W. 2006, \bibinfo{title}{A Cusp Slope-Central
  Anisotropy Theorem,} \apj, 642, 752, \dodoi{10.1086/501040}

\bibitem[{A. {Ashok} {et~al.}(2023){Ashok}, {Seth}, {Erwin}, {Debattista}, {de
  Lorenzo-C{\'a}ceres}, {Gadotti}, {M{\'e}ndez-Abreu}, {Beckman}, {Bender},
  {Drory}, {Fisher}, {Hopp}, {Kluge}, {Kolcu}, {Maciejewski}, {Mehrgan},
  {Parikh}, {Saglia}, {Seidel}, \& {Thomas}}]{Ashok23}
{Ashok}, A., {Seth}, A., {Erwin}, P., {et~al.} 2023, \bibinfo{title}{{Composite
  Bulges. III. A Study of Nuclear Star Clusters in Nearby Spiral Galaxies},}
  \apj, 958, 100, \dodoi{10.3847/1538-4357/ace341}

\bibitem[{ {Astropy Collaboration} {et~al.}(2022){Astropy Collaboration},
  {Price-Whelan}, {Lim}, {Earl}, {Starkman}, {Bradley}, {Shupe}, {Patil},
  {Corrales}, {Brasseur}, {N{\"o}the}, {Donath}, {Tollerud}, {Morris},
  {Ginsburg}, {Vaher}, {Weaver}, {Tocknell}, {Jamieson}, {van Kerkwijk},
  {Robitaille}, {Merry}, {Bachetti}, {G{\"u}nther}, {Aldcroft},
  {Alvarado-Montes}, {Archibald}, {B{\'o}di}, {Bapat}, {Barentsen},
  {Baz{\'a}n}, {Biswas}, {Boquien}, {Burke}, {Cara}, {Cara}, {Conroy},
  {Conseil}, {Craig}, {Cross}, {Cruz}, {D'Eugenio}, {Dencheva}, {Devillepoix},
  {Dietrich}, {Eigenbrot}, {Erben}, {Ferreira}, {Foreman-Mackey}, {Fox},
  {Freij}, {Garg}, {Geda}, {Glattly}, {Gondhalekar}, {Gordon}, {Grant},
  {Greenfield}, {Groener}, {Guest}, {Gurovich}, {Handberg}, {Hart},
  {Hatfield-Dodds}, {Homeier}, {Hosseinzadeh}, {Jenness}, {Jones}, {Joseph},
  {Kalmbach}, {Karamehmetoglu}, {Ka{\l}uszy{\'n}ski}, {Kelley}, {Kern},
  {Kerzendorf}, {Koch}, {Kulumani}, {Lee}, {Ly}, {Ma}, {MacBride}, {Maljaars},
  {Muna}, {Murphy}, {Norman}, {O'Steen}, {Oman}, {Pacifici}, {Pascual},
  {Pascual-Granado}, {Patil}, {Perren}, {Pickering}, {Rastogi}, {Roulston},
  {Ryan}, {Rykoff}, {Sabater}, {Sakurikar}, {Salgado}, {Sanghi}, {Saunders},
  {Savchenko}, {Schwardt}, {Seifert-Eckert}, {Shih}, {Jain}, {Shukla}, {Sick},
  {Simpson}, {Singanamalla}, {Singer}, {Singhal}, {Sinha}, {Sip{\H{o}}cz},
  {Spitler}, {Stansby}, {Streicher}, {{\v{S}}umak}, {Swinbank}, {Taranu},
  {Tewary}, {Tremblay}, {de Val-Borro}, {Van Kooten}, {Vasovi{\'c}}, {Verma},
  {de Miranda Cardoso}, {Williams}, {Wilson}, {Winkel}, {Wood-Vasey}, {Xue},
  {Yoachim}, {Zhang}, {Zonca}, \& {Astropy Project
  Contributors}}]{AstropyCollaboration22}
{Astropy Collaboration}, {Price-Whelan}, A.~M., {Lim}, P.~L., {et~al.} 2022,
  \bibinfo{title}{{The Astropy Project: Sustaining and Growing a
  Community-oriented Open-source Project and the Latest Major Release (v5.0) of
  the Core Package},} \apj, 935, 167, \dodoi{10.3847/1538-4357/ac7c74}

\bibitem[{R.~J. {Avila} {et~al.}(2012){Avila}, {Hack}, \& {STScI AstroDrizzle
  Team}}]{Avila12}
{Avila}, R.~J., {Hack}, W.~J., \& {STScI AstroDrizzle Team}. 2012, in American
  Astronomical Society Meeting Abstracts, Vol. 220, American Astronomical
  Society Meeting Abstracts \#220, 135.13

\bibitem[{J.~N. {Bahcall} \& R.~A. {Wolf}(1977){Bahcall} \& {Wolf}}]{Bahcall77}
{Bahcall}, J.~N., \& {Wolf}, R.~A. 1977, \bibinfo{title}{{The star distribution
  around a massive black hole in a globular cluster. II. Unequal star
  masses.},} \apj, 216, 883, \dodoi{10.1086/155534}

\bibitem[{M. {Bailes} {et~al.}(2021){Bailes}, {Berger}, {Brady}, {Branchesi},
  {Danzmann}, {Evans}, {Holley-Bockelmann}, {Iyer}, {Kajita}, {Katsanevas},
  {Kramer}, {Lazzarini}, {Lehner}, {Losurdo}, {L{\"u}ck}, {McClelland},
  {McLaughlin}, {Punturo}, {Ransom}, {Raychaudhury}, {Reitze}, {Ricci},
  {Rowan}, {Saito}, {Sanders}, {Sathyaprakash}, {Schutz}, {Sesana}, {Shinkai},
  {Siemens}, {Shoemaker}, {Thorpe}, {van den Brand}, \& {Vitale}}]{Bailes21}
{Bailes}, M., {Berger}, B.~K., {Brady}, P.~R., {et~al.} 2021,
  \bibinfo{title}{{Gravitational-wave physics and astronomy in the 2020s and
  2030s},} Nature Reviews Physics, 3, 344, \dodoi{10.1038/s42254-021-00303-8}

\bibitem[{V.~F. {Baldassare} {et~al.}(2022){Baldassare}, {Stone}, {Foord},
  {Gallo}, \& {Ostriker}}]{Baldassare22}
{Baldassare}, V.~F., {Stone}, N.~C., {Foord}, A., {Gallo}, E., \& {Ostriker},
  J.~P. 2022, \bibinfo{title}{{Massive Black Hole Formation in Dense Stellar
  Environments: Enhanced X-Ray Detection Rates in High-velocity Dispersion
  Nuclear Star Clusters},} \apj, 929, 84, \dodoi{10.3847/1538-4357/ac5f51}

\bibitem[{A.~J. {Barth} {et~al.}(2002){Barth}, {Ho}, \& {Sargent}}]{Barth02}
{Barth}, A.~J., {Ho}, L.~C., \& {Sargent}, W. L.~W. 2002, \bibinfo{title}{{A
  Study of the Direct Fitting Method for Measurement of Galaxy Velocity
  Dispersions},} \aj, 124, 2607, \dodoi{10.1086/343840}

\bibitem[{A.~J. Barth {et~al.}(2001)Barth, Sarzi, Rix, Ho, Filippenko, \&
  Sargent}]{Barth2001}
Barth, A.~J., Sarzi, M., Rix, H.-W., {et~al.} 2001, \bibinfo{title}{Evidence
  for a Supermassive Black Hole in the S0 Galaxy NGC 3245,} \apj, 555, 685,
  \dodoi{10.1086/321523}

\bibitem[{A.~J. {Barth} {et~al.}(2009){Barth}, {Strigari}, {Bentz}, {Greene},
  \& {Ho}}]{Barth09}
{Barth}, A.~J., {Strigari}, L.~E., {Bentz}, M.~C., {Greene}, J.~E., \& {Ho},
  L.~C. 2009, \bibinfo{title}{{Dynamical Constraints on the Masses of the
  Nuclear Star Cluster and Black Hole in the Late-Type Spiral Galaxy NGC
  3621},} \apj, 690, 1031, \dodoi{10.1088/0004-637X/690/1/1031}

\bibitem[{H. {Baumgardt} {et~al.}(2019){Baumgardt}, {He}, {Sweet},
  {Drinkwater}, {Sollima}, {Hurley}, {Usher}, {Kamann}, {Dalgleish},
  {Dreizler}, \& {Husser}}]{Baumgardt19}
{Baumgardt}, H., {He}, C., {Sweet}, S.~M., {et~al.} 2019, \bibinfo{title}{{No
  evidence for intermediate-mass black holes in the globular clusters
  {\ensuremath{\omega}} Cen and NGC 6624},} \mnras, 488, 5340,
  \dodoi{10.1093/mnras/stz2060}

\bibitem[{D.~P. {Bennett} {et~al.}(2002){Bennett}, {Becker}, {Quinn},
  {Tomaney}, {Alcock}, {Allsman}, {Alves}, {Axelrod}, {Calitz}, {Cook},
  {Drake}, {Fragile}, {Freeman}, {Geha}, {Griest}, {Johnson}, {Keller}, {Laws},
  {Lehner}, {Marshall}, {Minniti}, {Nelson}, {Peterson}, {Popowski}, {Pratt},
  {Quinn}, {Rhie}, {Stubbs}, {Sutherland}, {Vandehei}, {Welch}, {MACHO
  Collaboration}, \& {MPS Collaboration}}]{Bennett02}
{Bennett}, D.~P., {Becker}, A.~C., {Quinn}, J.~L., {et~al.} 2002,
  \bibinfo{title}{{Gravitational Microlensing Events Due to Stellar-Mass Black
  Holes},} \apj, 579, 639, \dodoi{10.1086/342225}

\bibitem[{P. {Bianchini} {et~al.}(2017){Bianchini}, {Sills}, {van de Ven}, \&
  {Sippel}}]{Bianchini17}
{Bianchini}, P., {Sills}, A., {van de Ven}, G., \& {Sippel}, A.~C. 2017,
  \bibinfo{title}{{The relation between the mass-to-light ratio and the
  relaxation state of globular clusters},} \mnras, 469, 4359,
  \dodoi{10.1093/mnras/stx1114}

\bibitem[{J. Binney(1980)Binney}]{Binney80}
Binney, J. 1980, \bibinfo{title}{{The radius-dependence of velocity dispersion
  in elliptical galaxies},} \mnras, 190, 873, \dodoi{10.1093/mnras/190.4.873}

\bibitem[{T. {B{\"o}ker} {et~al.}(2002){B{\"o}ker}, {Laine}, {van der Marel},
  {Sarzi}, {Rix}, {Ho}, \& {Shields}}]{Boker02}
{B{\"o}ker}, T., {Laine}, S., {van der Marel}, R.~P., {et~al.} 2002,
  \bibinfo{title}{{A Hubble Space Telescope Census of Nuclear Star Clusters in
  Late-Type Spiral Galaxies. I. Observations and Image Analysis},} \aj, 123,
  1389, \dodoi{10.1086/339025}

\bibitem[{G. {Bono} {et~al.}(2010){Bono}, {Caputo}, {Marconi}, \&
  {Musella}}]{Bono10}
{Bono}, G., {Caputo}, F., {Marconi}, M., \& {Musella}, I. 2010,
  \bibinfo{title}{{Insights into the Cepheid Distance Scale},} \apj, 715, 277,
  \dodoi{10.1088/0004-637X/715/1/277}

\bibitem[{S. {Bonoli} {et~al.}(2014){Bonoli}, {Mayer}, \&
  {Callegari}}]{Bonoli14}
{Bonoli}, S., {Mayer}, L., \& {Callegari}, S. 2014, \bibinfo{title}{{Massive
  black hole seeds born via direct gas collapse in galaxy mergers: their
  properties, statistics and environment},} \mnras, 437, 1576,
  \dodoi{10.1093/mnras/stt1990}

\bibitem[{G. {Busso} {et~al.}(2022){Busso}, {Cacciari}, {Bellazzini},
  {Carrasco}, {De Angeli}, {Evans}, {Fabricius}, {Montegriffo}, {Pancino},
  {Rainer}, \& {Sanna}}]{Busso22}
{Busso}, G., {Cacciari}, C., {Bellazzini}, M., {et~al.} 2022,
  \bibinfo{title}{{Gaia DR3 documentation Chapter 5: Photometric data},}, Gaia
  DR3 documentation, European Space Agency; Gaia Data Processing and Analysis
  Consortium. id. 5
  \dodoi{https://gea.esac.esa.int/archive/documentation/GDR3/index.html}

\bibitem[{M. Cappellari(2002)Cappellari}]{Cappellari02}
Cappellari, M. 2002, \bibinfo{title}{Efficient multi-Gaussian expansion of
  galaxies,} \mnras, 333, 400, \dodoi{10.1046/j.1365-8711.2002.05412.x}

\bibitem[{M. Cappellari(2008)Cappellari}]{Cappellari08}
Cappellari, M. 2008, \bibinfo{title}{Measuring the inclination and
  mass-to-light ratio of axisymmetric galaxies via anisotropic Jeans models of
  stellar kinematics,} \mnras, 390, 71,
  \dodoi{10.1111/j.1365-2966.2008.13754.x}

\bibitem[{M. {Cappellari}(2016){Cappellari}}]{Cappellari16}
{Cappellari}, M. 2016, \bibinfo{title}{{Structure and Kinematics of Early-Type
  Galaxies from Integral Field Spectroscopy},} \araa, 54, 597,
  \dodoi{10.1146/annurev-astro-082214-122432}

\bibitem[{M. Cappellari(2020)Cappellari}]{Cappellari20}
Cappellari, M. 2020, \bibinfo{title}{{Efficient solution of the anisotropic
  spherically aligned axisymmetric Jeans equations of stellar hydrodynamics for
  galactic dynamics},} \mnras, 494, 4819, \dodoi{10.1093/mnras/staa959}

\bibitem[{M. {Cappellari}(2023){Cappellari}}]{Cappellari23}
{Cappellari}, M. 2023, \bibinfo{title}{{Full spectrum fitting with photometry
  in PPXF: stellar population versus dynamical masses, non-parametric star
  formation history and metallicity for 3200 LEGA-C galaxies at redshift z
  {\ensuremath{\approx}} 0.8},} \mnras, 526, 3273,
  \dodoi{10.1093/mnras/stad2597}

\bibitem[{M. Cappellari(2025)Cappellari}]{Cappellari2025}
Cappellari, M. 2025, \bibinfo{title}{Early-Type Galaxies: Elliptical and S0
  Galaxies, or Fast and Slow Rotators,} arXiv e-prints, arXiv:2503.02746,
  \dodoi{10.48550/arXiv.2503.02746}

\bibitem[{M. {Cappellari} \& Y. {Copin}(2003){Cappellari} \&
  {Copin}}]{Cappellari03}
{Cappellari}, M., \& {Copin}, Y. 2003, \bibinfo{title}{{Adaptive spatial
  binning of integral-field spectroscopic data using Voronoi tessellations},}
  \mnras, 342, 345, \dodoi{10.1046/j.1365-8711.2003.06541.x}

\bibitem[{M. Cappellari {et~al.}(2006)Cappellari, Bacon, Bureau, Damen, Davies,
  de~Zeeuw, Emsellem, Falc{\'o}n-Barroso, Krajnovi{\'c}, Kuntschner, McDermid,
  Peletier, Sarzi, van~den Bosch, \& van~de Ven}]{Cappellari2006}
Cappellari, M., Bacon, R., Bureau, M., {et~al.} 2006, \bibinfo{title}{The
  SAURON project - IV. The mass-to-light ratio, the virial mass estimator and
  the Fundamental Plane of elliptical and lenticular galaxies,} \mnras, 366,
  1126, \dodoi{10.1111/j.1365-2966.2005.09981.x}

\bibitem[{M. {Cappellari} {et~al.}(2011){Cappellari}, {Emsellem},
  {Krajnovi{\'c}}, {McDermid}, {Scott}, {Verdoes Kleijn}, {Young}, {Alatalo},
  {Bacon}, {Blitz}, {Bois}, {Bournaud}, {Bureau}, {Davies}, {Davis}, {de
  Zeeuw}, {Duc}, {Khochfar}, {Kuntschner}, {Lablanche}, {Morganti}, {Naab},
  {Oosterloo}, {Sarzi}, {Serra}, \& {Weijmans}}]{Cappellari11}
{Cappellari}, M., {Emsellem}, E., {Krajnovi{\'c}}, D., {et~al.} 2011,
  \bibinfo{title}{{The ATLAS$^{3D}$ project - I. A volume-limited sample of 260
  nearby early-type galaxies: science goals and selection criteria},} \mnras,
  413, 813, \dodoi{10.1111/j.1365-2966.2010.18174.x}

\bibitem[{M. {Cappellari} {et~al.}(2013{\natexlab{a}}){Cappellari}, {McDermid},
  {Alatalo}, {Blitz}, {Bois}, {Bournaud}, {Bureau}, {Crocker}, {Davies},
  {Davis}, {de Zeeuw}, {Duc}, {Emsellem}, {Khochfar}, {Krajnovi{\'c}},
  {Kuntschner}, {Morganti}, {Naab}, {Oosterloo}, {Sarzi}, {Scott}, {Serra},
  {Weijmans}, \& {Young}}]{Cappellari13b}
{Cappellari}, M., {McDermid}, R.~M., {Alatalo}, K., {et~al.}
  2013{\natexlab{a}}, \bibinfo{title}{{The ATLAS$^{3D}$ project - XX. Mass-size
  and mass-{\ensuremath{\sigma}} distributions of early-type galaxies: bulge
  fraction drives kinematics, mass-to-light ratio, molecular gas fraction and
  stellar initial mass function},} \mnras, 432, 1862,
  \dodoi{10.1093/mnras/stt644}

\bibitem[{M. {Cappellari} {et~al.}(2013{\natexlab{b}}){Cappellari}, {Scott},
  {Alatalo}, {Blitz}, {Bois}, {Bournaud}, {Bureau}, {Crocker}, {Davies},
  {Davis}, {de Zeeuw}, {Duc}, {Emsellem}, {Khochfar}, {Krajnovi{\'c}},
  {Kuntschner}, {McDermid}, {Morganti}, {Naab}, {Oosterloo}, {Sarzi}, {Serra},
  {Weijmans}, \& {Young}}]{Cappellari13a}
{Cappellari}, M., {Scott}, N., {Alatalo}, K., {et~al.} 2013{\natexlab{b}},
  \bibinfo{title}{{The ATLAS$^{3D}$ project - XV. Benchmark for early-type
  galaxies scaling relations from 260 dynamical models: mass-to-light ratio,
  dark matter, Fundamental Plane and Mass Plane},} \mnras, 432, 1709,
  \dodoi{10.1093/mnras/stt562}

\bibitem[{J.~A. {Cardelli} {et~al.}(1989){Cardelli}, {Clayton}, \&
  {Mathis}}]{Cardelli89}
{Cardelli}, J.~A., {Clayton}, G.~C., \& {Mathis}, J.~S. 1989,
  \bibinfo{title}{{The Relationship between Infrared, Optical, and Ultraviolet
  Extinction},} \apj, 345, 245, \dodoi{10.1086/167900}

\bibitem[{S.~G. {Carlsten} {et~al.}(2022){Carlsten}, {Greene}, {Beaton}, \&
  {Greco}}]{Carlsten22a}
{Carlsten}, S.~G., {Greene}, J.~E., {Beaton}, R.~L., \& {Greco}, J.~P. 2022,
  \bibinfo{title}{{ELVES II: Globular Clusters and Nuclear Star Clusters of
  Dwarf Galaxies: the Importance of Environment},} \apj, 927, 44,
  \dodoi{10.3847/1538-4357/ac457e}

\bibitem[{D.~J. {Carson} {et~al.}(2015){Carson}, {Barth}, {Seth}, {den Brok},
  {Cappellari}, {Greene}, {Ho}, \& {Neumayer}}]{Carson15}
{Carson}, D.~J., {Barth}, A.~J., {Seth}, A.~C., {et~al.} 2015,
  \bibinfo{title}{{The Structure of Nuclear Star Clusters in Nearby Late-type
  Spiral Galaxies from Hubble Space Telescope Wide Field Camera 3 Imaging},}
  \aj, 149, 170, \dodoi{10.1088/0004-6256/149/5/170}

\bibitem[{C. {Conroy} {et~al.}(2018){Conroy}, {Villaume}, {van Dokkum}, \&
  {Lind}}]{Conroy18}
{Conroy}, C., {Villaume}, A., {van Dokkum}, P.~G., \& {Lind}, K. 2018,
  \bibinfo{title}{{Metal-rich, Metal-poor: Updated Stellar Population Models
  for Old Stellar Systems},} \apj, 854, 139, \dodoi{10.3847/1538-4357/aaab49}

\bibitem[{P. {C{\^o}t{\'e}} {et~al.}(2006){C{\^o}t{\'e}}, {Piatek},
  {Ferrarese}, {Jord{\'a}n}, {Merritt}, {Peng}, {Ha{\c{s}}egan}, {Blakeslee},
  {Mei}, {West}, {Milosavljevi{\'c}}, \& {Tonry}}]{Cote06}
{C{\^o}t{\'e}}, P., {Piatek}, S., {Ferrarese}, L., {et~al.} 2006,
  \bibinfo{title}{{The ACS Virgo Cluster Survey. VIII. The Nuclei of Early-Type
  Galaxies},} \apjs, 165, 57, \dodoi{10.1086/504042}

\bibitem[{A. {Crespo G{\'o}mez} {et~al.}(2021){Crespo G{\'o}mez}, {Piqueras
  L{\'o}pez}, {Arribas}, {Pereira-Santaella}, {Colina}, \& {Rodr{\'\i}guez del
  Pino}}]{CrespoGomez21}
{Crespo G{\'o}mez}, A., {Piqueras L{\'o}pez}, J., {Arribas}, S., {et~al.} 2021,
  \bibinfo{title}{{Stellar kinematics in the nuclear regions of nearby LIRGs
  with VLT-SINFONI. Comparison with gas phases and implications for dynamical
  mass estimations},} \aap, 650, A149, \dodoi{10.1051/0004-6361/202039472}

\bibitem[{T.~A. {Davis} {et~al.}(2013){Davis}, {Bureau}, {Cappellari}, {Sarzi},
  \& {Blitz}}]{Davis13Nature}
{Davis}, T.~A., {Bureau}, M., {Cappellari}, M., {Sarzi}, M., \& {Blitz}, L.
  2013, \bibinfo{title}{{A black-hole mass measurement from molecular gas
  kinematics in NGC4526},} \nat, 494, 328, \dodoi{10.1038/nature11819}

\bibitem[{B. {de Swardt} {et~al.}(2010){de Swardt}, {Kraan-Korteweg}, \&
  {Jerjen}}]{deSwardt10}
{de Swardt}, B., {Kraan-Korteweg}, R.~C., \& {Jerjen}, H. 2010,
  \bibinfo{title}{{Photometric properties of six Local Volume dwarf galaxies
  from deep near-infrared observations},} \mnras, 407, 955,
  \dodoi{10.1111/j.1365-2966.2010.16986.x}

\bibitem[{T. {de Zeeuw}(2001){de Zeeuw}}]{deZeeuw01}
{de Zeeuw}, T. 2001, in Black Holes in Binaries and Galactic Nuclei, ed.
  L.~{Kaper}, E.~P.~J. V.~D. {Heuvel}, \& P.~A. {Woudt}, 78,
  \dodoi{10.1007/10720995_12}

\bibitem[{W. Dehnen(1993)Dehnen}]{Dehnen93}
Dehnen, W. 1993, \bibinfo{title}{A Family of Potential-Density Pairs for
  Spherical Galaxies and Bulges,} \mnras, 265, 250,
  \dodoi{10.1093/mnras/265.1.250}

\bibitem[{M. {den Brok} {et~al.}(2014){den Brok}, {van de Ven}, {van den
  Bosch}, \& {Watkins}}]{denBrok14a}
{den Brok}, M., {van de Ven}, G., {van den Bosch}, R., \& {Watkins}, L. 2014,
  \bibinfo{title}{{The central mass and mass-to-light profile of the Galactic
  globular cluster M15},} \mnras, 438, 487, \dodoi{10.1093/mnras/stt2221}

\bibitem[{T. {Di Matteo} {et~al.}(2008){Di Matteo}, {Colberg}, {Springel},
  {Hernquist}, \& {Sijacki}}]{DiMatteo08}
{Di Matteo}, T., {Colberg}, J., {Springel}, V., {Hernquist}, L., \& {Sijacki},
  D. 2008, \bibinfo{title}{{Direct Cosmological Simulations of the Growth of
  Black Holes and Galaxies},} \apj, 676, 33, \dodoi{10.1086/524921}

\bibitem[{E. {Emsellem} {et~al.}(1994){Emsellem}, {Monnet}, \&
  {Bacon}}]{Emsellem94}
{Emsellem}, E., {Monnet}, G., \& {Bacon}, R. 1994, \bibinfo{title}{{The
  multi-gaussian expansion method: a tool for building realistic photometric
  and kinematical models of stellar systems I. The formalism},} \aap, 285, 723

\bibitem[{P. {Erwin} \& D.~A. {Gadotti}(2012){Erwin} \& {Gadotti}}]{Erwin12}
{Erwin}, P., \& {Gadotti}, D.~A. 2012, \bibinfo{title}{{Do Nuclear Star
  Clusters and Supermassive Black Holes Follow the Same Host-Galaxy
  Correlations?},} Advances in Astronomy, 2012, 946368,
  \dodoi{10.1155/2012/946368}

\bibitem[{C.~J. {Evans} {et~al.}(2011){Evans}, {Davies}, {Kudritzki}, {Puech},
  {Yang}, {Cuby}, {Figer}, {Lehnert}, {Morris}, \& {Rousset}}]{Evans11}
{Evans}, C.~J., {Davies}, B., {Kudritzki}, R.~P., {et~al.} 2011,
  \bibinfo{title}{{Stellar metallicities beyond the Local Group: the potential
  of J-band spectroscopy with extremely large telescopes},} \aap, 527, A50,
  \dodoi{10.1051/0004-6361/201015986}

\bibitem[{A.~C. {Fabian}(2012){Fabian}}]{Fabian12}
{Fabian}, A.~C. 2012, \bibinfo{title}{{Observational Evidence of Active
  Galactic Nuclei Feedback},} \araa, 50, 455,
  \dodoi{10.1146/annurev-astro-081811-125521}

\bibitem[{K. {Fahrion} {et~al.}(2020){Fahrion}, {Lyubenova}, {Hilker}, {van de
  Ven}, {Falc{\'o}n-Barroso}, {Leaman}, {Mart{\'\i}n-Navarro}, {Bittner},
  {Coccato}, {Corsini}, {Gadotti}, {Iodice}, {McDermid}, {Pinna}, {Sarzi},
  {Viaene}, {de Zeeuw}, \& {Zhu}}]{Fahrion20}
{Fahrion}, K., {Lyubenova}, M., {Hilker}, M., {et~al.} 2020,
  \bibinfo{title}{{The Fornax 3D project: Non-linear colour-metallicity
  relation of globular clusters},} \aap, 637, A27,
  \dodoi{10.1051/0004-6361/202037686}

\bibitem[{K. {Fahrion} {et~al.}(2022){Fahrion}, {Bulichi}, {Hilker}, {Leaman},
  {Lyubenova}, {M{\"u}ller}, {Neumayer}, {Pinna}, {Rejkuba}, \& {van de
  Ven}}]{Fahrion22}
{Fahrion}, K., {Bulichi}, T.-E., {Hilker}, M., {et~al.} 2022,
  \bibinfo{title}{{Nuclear star cluster formation in star-forming dwarf
  galaxies},} \aap, 667, A101, \dodoi{10.1051/0004-6361/202244932}

\bibitem[{L. {Ferrarese} \& D. {Merritt}(2000){Ferrarese} \&
  {Merritt}}]{Ferrarese00}
{Ferrarese}, L., \& {Merritt}, D. 2000, \bibinfo{title}{{A Fundamental Relation
  between Supermassive Black Holes and Their Host Galaxies},} \apjl, 539, L9,
  \dodoi{10.1086/312838}

\bibitem[{E. {Gallo} {et~al.}(2008){Gallo}, {Treu}, {Jacob}, {Woo}, {Marshall},
  \& {Antonucci}}]{Gallo08}
{Gallo}, E., {Treu}, T., {Jacob}, J., {et~al.} 2008,
  \bibinfo{title}{{AMUSE-Virgo. I. Supermassive Black Holes in Low-Mass
  Spheroids},} \apj, 680, 154, \dodoi{10.1086/588012}

\bibitem[{K. {Gebhardt} {et~al.}(2005){Gebhardt}, {Rich}, \& {Ho}}]{Gebhardt05}
{Gebhardt}, K., {Rich}, R.~M., \& {Ho}, L.~C. 2005, \bibinfo{title}{{An
  Intermediate-Mass Black Hole in the Globular Cluster G1: Improved
  Significance from New Keck and Hubble Space Telescope Observations},} \apj,
  634, 1093, \dodoi{10.1086/497023}

\bibitem[{K. {Gebhardt} {et~al.}(2000){Gebhardt}, {Bender}, {Bower},
  {Dressler}, {Faber}, {Filippenko}, {Green}, {Grillmair}, {Ho}, {Kormendy},
  {Lauer}, {Magorrian}, {Pinkney}, {Richstone}, \& {Tremaine}}]{Gebhardt00}
{Gebhardt}, K., {Bender}, R., {Bower}, G., {et~al.} 2000, \bibinfo{title}{{A
  Relationship between Nuclear Black Hole Mass and Galaxy Velocity
  Dispersion},} \apjl, 539, L13, \dodoi{10.1086/312840}

\bibitem[{K. {Gebhardt} {et~al.}(2001){Gebhardt}, {Lauer}, {Kormendy},
  {Pinkney}, {Bower}, {Green}, {Gull}, {Hutchings}, {Kaiser}, {Nelson},
  {Richstone}, \& {Weistrop}}]{Gebhardt01}
{Gebhardt}, K., {Lauer}, T.~R., {Kormendy}, J., {et~al.} 2001,
  \bibinfo{title}{{M33: A Galaxy with No Supermassive Black Hole},} \aj, 122,
  2469, \dodoi{10.1086/323481}

\bibitem[{K. Gebhardt {et~al.}(2003)Gebhardt, Richstone, Tremaine, Lauer,
  Bender, Bower, Dressler, Faber, Filippenko, Green, Grillmair, Ho, Kormendy,
  Magorrian, \& Pinkney}]{Gebhardt2003}
Gebhardt, K., Richstone, D., Tremaine, S., {et~al.} 2003,
  \bibinfo{title}{{Axisymmetric Dynamical Models of the Central Regions of
  Galaxies},} \apj, 583, 92, \dodoi{10.1086/345081}

\bibitem[{I.~Y. {Georgiev} \& T. {B{\"o}ker}(2014){Georgiev} \&
  {B{\"o}ker}}]{Georgiev14}
{Georgiev}, I.~Y., \& {B{\"o}ker}, T. 2014, \bibinfo{title}{{Nuclear star
  clusters in 228 spiral galaxies in the HST/WFPC2 archive: catalogue and
  comparison to other stellar systems},} \mnras, 441, 3570,
  \dodoi{10.1093/mnras/stu797}

\bibitem[{I.~Y. {Georgiev} {et~al.}(2016){Georgiev}, {B{\"o}ker}, {Leigh},
  {L{\"u}tzgendorf}, \& {Neumayer}}]{Georgiev16}
{Georgiev}, I.~Y., {B{\"o}ker}, T., {Leigh}, N., {L{\"u}tzgendorf}, N., \&
  {Neumayer}, N. 2016, \bibinfo{title}{{Masses and scaling relations for
  nuclear star clusters, and their co-existence with central black holes},}
  \mnras, 457, 2122, \dodoi{10.1093/mnras/stw093}

\bibitem[{E. {Gonz{\'a}lez-Alfonso} {et~al.}(2023){Gonz{\'a}lez-Alfonso},
  {Garc{\'\i}a-Bernete}, {Pereira-Santaella}, {Neufeld}, {Fischer}, \&
  {Donnan}}]{Gonzalez-Alfonso}
{Gonz{\'a}lez-Alfonso}, E., {Garc{\'\i}a-Bernete}, I., {Pereira-Santaella}, M.,
  {et~al.} 2023, \bibinfo{title}{{JWST detection of extremely excited
  outflowing CO and H2O in VV 114 E SW: a possible rapidly accreting IMBH},}
  arXiv e-prints, arXiv:2312.04914, \dodoi{10.48550/arXiv.2312.04914}

\bibitem[{A.~W. Graham {et~al.}(2003)Graham, Erwin, Trujillo, \&
  Asensio~Ramos}]{Graham03}
Graham, A.~W., Erwin, P., Trujillo, I., \& Asensio~Ramos, A. 2003,
  \bibinfo{title}{{A New Empirical Model for the Structural Analysis of
  Early-Type Galaxies, and A Critical Review of the Nuker Model},} \aj, 125,
  2951, \dodoi{10.1086/375320}

\bibitem[{M.~J. {Graham} {et~al.}(2020){Graham}, {Ford}, {McKernan}, {Ross},
  {Stern}, {Burdge}, {Coughlin}, {Djorgovski}, {Drake}, {Duev}, {Kasliwal},
  {Mahabal}, {van Velzen}, {Belecki}, {Bellm}, {Burruss}, {Cenko},
  {Cunningham}, {Helou}, {Kulkarni}, {Masci}, {Prince}, {Reiley}, {Rodriguez},
  {Rusholme}, {Smith}, \& {Soumagnac}}]{Graham20}
{Graham}, M.~J., {Ford}, K.~E.~S., {McKernan}, B., {et~al.} 2020,
  \bibinfo{title}{{Candidate Electromagnetic Counterpart to the Binary Black
  Hole Merger Gravitational-Wave Event S190521g$^{*}$},} \prl, 124, 251102,
  \dodoi{10.1103/PhysRevLett.124.251102}

\bibitem[{J.~E. {Greene}(2012){Greene}}]{Greene12}
{Greene}, J.~E. 2012, \bibinfo{title}{{Low-mass black holes as the remnants of
  primordial black hole formation},} Nature Communications, 3, 1304,
  \dodoi{10.1038/ncomms2314}

\bibitem[{J.~E. {Greene} {et~al.}(2020){Greene}, {Strader}, \& {Ho}}]{Greene20}
{Greene}, J.~E., {Strader}, J., \& {Ho}, L.~C. 2020,
  \bibinfo{title}{{Intermediate-Mass Black Holes},} \araa, 58, 257,
  \dodoi{10.1146/annurev-astro-032620-021835}

\bibitem[{K. {G{\"u}ltekin} {et~al.}(2009){G{\"u}ltekin}, {Richstone},
  {Gebhardt}, {Lauer}, {Tremaine}, {Aller}, {Bender}, {Dressler}, {Faber},
  {Filippenko}, {Green}, {Ho}, {Kormendy}, {Magorrian}, {Pinkney}, \&
  {Siopis}}]{Gultekin09}
{G{\"u}ltekin}, K., {Richstone}, D.~O., {Gebhardt}, K., {et~al.} 2009,
  \bibinfo{title}{{The M-{\ensuremath{\sigma}} and M-L Relations in Galactic
  Bulges, and Determinations of Their Intrinsic Scatter},} \apj, 698, 198,
  \dodoi{10.1088/0004-637X/698/1/198}

\bibitem[{B. {Gustafsson} {et~al.}(2008){Gustafsson}, {Edvardsson}, {Eriksson},
  {J{\o}rgensen}, {Nordlund}, \& {Plez}}]{Gustafsson08}
{Gustafsson}, B., {Edvardsson}, B., {Eriksson}, K., {et~al.} 2008,
  \bibinfo{title}{{A grid of MARCS model atmospheres for late-type stars. I.
  Methods and general properties},} \aap, 486, 951,
  \dodoi{10.1051/0004-6361:200809724}

\bibitem[{H. Haario {et~al.}(2001)Haario, Saksman, \& Tamminen}]{Haario01}
Haario, H., Saksman, E., \& Tamminen, J. 2001, \bibinfo{title}{{An adaptive
  Metropolis algorithm},} Bernoulli, 7, 223

\bibitem[{G.~F. {H{\"a}gele} {et~al.}(2007){H{\"a}gele}, {D{\'\i}az},
  {Cardaci}, {Terlevich}, \& {Terlevich}}]{Hagele07}
{H{\"a}gele}, G.~F., {D{\'\i}az}, {\'A}.~I., {Cardaci}, M.~V., {Terlevich}, E.,
  \& {Terlevich}, R. 2007, \bibinfo{title}{{Kinematics of gas and stars in the
  circumnuclear star-forming ring of NGC3351},} \mnras, 378, 163,
  \dodoi{10.1111/j.1365-2966.2007.11751.x}

\bibitem[{M. Hartmann {et~al.}(2011)Hartmann, Debattista, Seth, Cappellari, \&
  Quinn}]{Hartmann2011}
Hartmann, M., Debattista, V.~P., Seth, A., Cappellari, M., \& Quinn, T.~R.
  2011, \bibinfo{title}{Constraining the role of star cluster mergers in
  nuclear cluster formation: simulations confront integral-field data,} \mnras,
  418, 2697, \dodoi{10.1111/j.1365-2966.2011.19659.x}

\bibitem[{L.~C. {Ho} {et~al.}(2009){Ho}, {Greene}, {Filippenko}, \&
  {Sargent}}]{Ho09}
{Ho}, L.~C., {Greene}, J.~E., {Filippenko}, A.~V., \& {Sargent}, W. L.~W. 2009,
  \bibinfo{title}{{A Search for ``Dwarf'' Seyfert Nuclei. VII. A Catalog of
  Central Stellar Velocity Dispersions of Nearby Galaxies},} \apjs, 183, 1,
  \dodoi{10.1088/0067-0049/183/1/1}

\bibitem[{N. {Hoyer} {et~al.}(2023){Hoyer}, {Neumayer}, {Seth}, {Georgiev}, \&
  {Greene}}]{Hoyer23}
{Hoyer}, N., {Neumayer}, N., {Seth}, A.~C., {Georgiev}, I.~Y., \& {Greene},
  J.~E. 2023, \bibinfo{title}{{Photometric and structural parameters of newly
  discovered nuclear star clusters in Local Volume galaxies},} \mnras, 520,
  4664, \dodoi{10.1093/mnras/stad220}

\bibitem[{J.~P. {Huchra} {et~al.}(2012){Huchra}, {Macri}, {Masters}, {Jarrett},
  {Berlind}, {Calkins}, {Crook}, {Cutri}, {Erdo{\v{g}}du}, {Falco}, {George},
  {Hutcheson}, {Lahav}, {Mader}, {Mink}, {Martimbeau}, {Schneider},
  {Skrutskie}, {Tokarz}, \& {Westover}}]{Huchra12}
{Huchra}, J.~P., {Macri}, L.~M., {Masters}, K.~L., {et~al.} 2012,
  \bibinfo{title}{{The 2MASS Redshift Survey{\textemdash}Description and Data
  Release},} \apjs, 199, 26, \dodoi{10.1088/0067-0049/199/2/26}

\bibitem[{K. {Inayoshi} {et~al.}(2020){Inayoshi}, {Visbal}, \&
  {Haiman}}]{Inayoshi20}
{Inayoshi}, K., {Visbal}, E., \& {Haiman}, Z. 2020, \bibinfo{title}{{The
  Assembly of the First Massive Black Holes},} \araa, 58, 27,
  \dodoi{10.1146/annurev-astro-120419-014455}

\bibitem[{T.~H. {Jarrett} {et~al.}(2003){Jarrett}, {Chester}, {Cutri},
  {Schneider}, \& {Huchra}}]{Jarrett03}
{Jarrett}, T.~H., {Chester}, T., {Cutri}, R., {Schneider}, S.~E., \& {Huchra},
  J.~P. 2003, \bibinfo{title}{{The 2MASS Large Galaxy Atlas},} \aj, 125, 525,
  \dodoi{10.1086/345794}

\bibitem[{R.~I. {Jedrzejewski}(1987){Jedrzejewski}}]{Jedrzejewski87}
{Jedrzejewski}, R.~I. 1987, \bibinfo{title}{{CCD surface photometry of
  elliptical galaxies - I. Observations, reduction and results.},} \mnras, 226,
  747, \dodoi{10.1093/mnras/226.4.747}

\bibitem[{H. {Jerjen} {et~al.}(2000{\natexlab{a}}){Jerjen}, {Binggeli}, \&
  {Freeman}}]{Jerjen00b}
{Jerjen}, H., {Binggeli}, B., \& {Freeman}, K.~C. 2000{\natexlab{a}},
  \bibinfo{title}{{Surface BR Photometry of Newly Discovered Dwarf Elliptical
  Galaxies in the Nearby Sculptor and Centaurus A Groups},} \aj, 119, 593,
  \dodoi{10.1086/301216}

\bibitem[{H. {Jerjen} {et~al.}(2000{\natexlab{b}}){Jerjen}, {Freeman}, \&
  {Binggeli}}]{Jerjen00a}
{Jerjen}, H., {Freeman}, K.~C., \& {Binggeli}, B. 2000{\natexlab{b}},
  \bibinfo{title}{{Testing the Surface Brightness Fluctuations Method for Dwarf
  Elliptical Galaxies in the Centaurus A Group},} \aj, 119, 166,
  \dodoi{10.1086/301188}

\bibitem[{N. {Kacharov} {et~al.}(2018){Kacharov}, {Neumayer}, {Seth},
  {Cappellari}, {McDermid}, {Walcher}, \& {B{\"o}ker}}]{Kacharov18}
{Kacharov}, N., {Neumayer}, N., {Seth}, A.~C., {et~al.} 2018,
  \bibinfo{title}{{Stellar populations and star formation histories of the
  nuclear star clusters in six nearby galaxies},} \mnras, 480, 1973,
  \dodoi{10.1093/mnras/sty1985}

\bibitem[{I.~D. {Karachentsev} {et~al.}(2004){Karachentsev}, {Karachentseva},
  {Huchtmeier}, \& {Makarov}}]{Karachentsev04}
{Karachentsev}, I.~D., {Karachentseva}, V.~E., {Huchtmeier}, W.~K., \&
  {Makarov}, D.~I. 2004, \bibinfo{title}{{A Catalog of Neighboring Galaxies},}
  \aj, 127, 2031, \dodoi{10.1086/382905}

\bibitem[{I.~D. {Karachentsev} {et~al.}(2013){Karachentsev}, {Makarov}, \&
  {Kaisina}}]{Karachentsev13}
{Karachentsev}, I.~D., {Makarov}, D.~I., \& {Kaisina}, E.~I. 2013,
  \bibinfo{title}{{Updated Nearby Galaxy Catalog},} \aj, 145, 101,
  \dodoi{10.1088/0004-6256/145/4/101}

\bibitem[{S.~C. {Kim} {et~al.}(2004){Kim}, {Sung}, {Park}, \& {Sung}}]{Kim04}
{Kim}, S.~C., {Sung}, H., {Park}, H.~S., \& {Sung}, E.-C. 2004,
  \bibinfo{title}{{UBVI Surface Photometry of the Spiral Galaxy NGC 300 in the
  Sculptor Group},} \cjaa, 4, 299, \dodoi{10.1088/1009-9271/4/4/299}

\bibitem[{C.~S. {Kochanek}(2016){Kochanek}}]{Kochanek16}
{Kochanek}, C.~S. 2016, \bibinfo{title}{{Tidal disruption event demographics},}
  \mnras, 461, 371, \dodoi{10.1093/mnras/stw1290}

\bibitem[{J. Kormendy \& L.~C. Ho(2013)Kormendy \& Ho}]{Kormendy13}
Kormendy, J., \& Ho, L.~C. 2013, \bibinfo{title}{{Coevolution (Or Not) of
  Supermassive Black Holes and Host Galaxies},} \araa, 51, 511,
  \dodoi{10.1146/annurev-astro-082708-101811}

\bibitem[{J. {Kormendy} \& D. {Richstone}(1995){Kormendy} \&
  {Richstone}}]{Kormendy95}
{Kormendy}, J., \& {Richstone}, D. 1995, \bibinfo{title}{{Inward Bound---The
  Search For Supermassive Black Holes In Galactic Nuclei},} \araa, 33, 581,
  \dodoi{10.1146/annurev.aa.33.090195.003053}

\bibitem[{D. {Krajnovi{\'c}} {et~al.}(2018){Krajnovi{\'c}}, {Cappellari}, \&
  {McDermid}}]{Krajnovic18a}
{Krajnovi{\'c}}, D., {Cappellari}, M., \& {McDermid}, R.~M. 2018,
  \bibinfo{title}{{Two channels of supermassive black hole growth as seen on
  the galaxies mass-size plane},} \mnras, 473, 5237,
  \dodoi{10.1093/mnras/stx2704}

\bibitem[{J. {Krist}(1995){Krist}}]{Krist95}
{Krist}, J. 1995, in Astronomical Society of the Pacific Conference Series,
  Vol.~77, Astronomical Data Analysis Software and Systems IV, ed. R.~A.
  {Shaw}, H.~E. {Payne}, \& J.~J.~E. {Hayes}, 349

\bibitem[{J.~E. {Krist} {et~al.}(2011){Krist}, {Hook}, \& {Stoehr}}]{Krist11}
{Krist}, J.~E., {Hook}, R.~N., \& {Stoehr}, F. 2011, in Society of
  Photo-Optical Instrumentation Engineers (SPIE) Conference Series, Vol. 8127,
  Optical Modeling and Performance Predictions V, ed. M.~A. {Kahan}, 81270J,
  \dodoi{10.1117/12.892762}

\bibitem[{M. {Lyubenova} {et~al.}(2012){Lyubenova}, {Kuntschner}, {Rejkuba},
  {Silva}, {Kissler-Patig}, \& {Tacconi-Garman}}]{Lyubenova12}
{Lyubenova}, M., {Kuntschner}, H., {Rejkuba}, M., {et~al.} 2012,
  \bibinfo{title}{{Integrated J- and H-band spectra of globular clusters in the
  LMC: implications for stellar population models and galaxy age dating},}
  \aap, 543, A75, \dodoi{10.1051/0004-6361/201218847}

\bibitem[{C.-P. {Ma} {et~al.}(2014){Ma}, {Greene}, {McConnell}, {Janish},
  {Blakeslee}, {Thomas}, \& {Murphy}}]{Ma14}
{Ma}, C.-P., {Greene}, J.~E., {McConnell}, N., {et~al.} 2014,
  \bibinfo{title}{{The MASSIVE Survey. I. A Volume-limited Integral-field
  Spectroscopic Study of the Most Massive Early-type Galaxies within 108 Mpc},}
  \apj, 795, 158, \dodoi{10.1088/0004-637X/795/2/158}

\bibitem[{J. {Magorrian} {et~al.}(1998){Magorrian}, {Tremaine}, {Richstone},
  {Bender}, {Bower}, {Dressler}, {Faber}, {Gebhardt}, {Green}, {Grillmair},
  {Kormendy}, \& {Lauer}}]{Magorrian98}
{Magorrian}, J., {Tremaine}, S., {Richstone}, D., {et~al.} 1998,
  \bibinfo{title}{{The Demography of Massive Dark Objects in Galaxy Centers},}
  \aj, 115, 2285, \dodoi{10.1086/300353}

\bibitem[{C. {Maraston} \& G. {Str{\"o}mb{\"a}ck}(2011){Maraston} \&
  {Str{\"o}mb{\"a}ck}}]{Maraston11}
{Maraston}, C., \& {Str{\"o}mb{\"a}ck}, G. 2011, \bibinfo{title}{{Stellar
  population models at high spectral resolution},} \mnras, 418, 2785,
  \dodoi{10.1111/j.1365-2966.2011.19738.x}

\bibitem[{C.~B. {Markwardt}(2009){Markwardt}}]{Markwardt09}
{Markwardt}, C.~B. 2009, in Astronomical Society of the Pacific Conference
  Series, Vol. 411, Astronomical Data Analysis Software and Systems XVIII, ed.
  D.~A. {Bohlender}, D.~{Durand}, \& P.~{Dowler}, 251,
  \dodoi{10.48550/arXiv.0902.2850}

\bibitem[{N.~J. McConnell \& C.-P. Ma(2013)McConnell \& Ma}]{McConnell2013}
McConnell, N.~J., \& Ma, C.-P. 2013, \bibinfo{title}{Revisiting the Scaling
  Relations of Black Hole Masses and Host Galaxy Properties,} \apj, 764, 184,
  \dodoi{10.1088/0004-637X/764/2/184}

\bibitem[{N.~J. {McConnell} {et~al.}(2011){McConnell}, {Ma}, {Gebhardt},
  {Wright}, {Murphy}, {Lauer}, {Graham}, \& {Richstone}}]{McConnell11}
{McConnell}, N.~J., {Ma}, C.-P., {Gebhardt}, K., {et~al.} 2011,
  \bibinfo{title}{{Two ten-billion-solar-mass black holes at the centres of
  giant elliptical galaxies},} \nat, 480, 215, \dodoi{10.1038/nature10636}

\bibitem[{D. Merritt {et~al.}(2001)Merritt, Ferrarese, \& Joseph}]{Merritt2001}
Merritt, D., Ferrarese, L., \& Joseph, C.~L. 2001, \bibinfo{title}{No
  Supermassive Black Hole in M33?} Science, 293, 1116,
  \dodoi{10.1126/science.1063896}

\bibitem[{M. {Mezcua}(2017){Mezcua}}]{Mezcua17}
{Mezcua}, M. 2017, \bibinfo{title}{{Observational evidence for
  intermediate-mass black holes},} International Journal of Modern Physics D,
  26, 1730021, \dodoi{10.1142/S021827181730021X}

\bibitem[{M. {Mitzkus} {et~al.}(2017){Mitzkus}, {Cappellari}, \&
  {Walcher}}]{Mitzkus17}
{Mitzkus}, M., {Cappellari}, M., \& {Walcher}, C.~J. 2017,
  \bibinfo{title}{{Dominant dark matter and a counter-rotating disc: MUSE view
  of the low-luminosity S0 galaxy NGC 5102},} \mnras, 464, 4789,
  \dodoi{10.1093/mnras/stw2677}

\bibitem[{M. {Miyoshi} {et~al.}(1995){Miyoshi}, {Moran}, {Herrnstein},
  {Greenhill}, {Nakai}, {Diamond}, \& {Inoue}}]{Miyoshi95}
{Miyoshi}, M., {Moran}, J., {Herrnstein}, J., {et~al.} 1995,
  \bibinfo{title}{{Evidence for a black hole from high rotation velocities in a
  sub-parsec region of NGC4258},} \nat, 373, 127, \dodoi{10.1038/373127a0}

\bibitem[{O. {M{\"u}ller} {et~al.}(2021){M{\"u}ller}, {Durrell}, {Marleau},
  {Duc}, {Lim}, {Posti}, {Agnello}, {S{\'a}nchez-Janssen}, {Poulain}, {Habas},
  {Emsellem}, {Paudel}, {van der Burg}, \& {Fensch}}]{Muller21}
{M{\"u}ller}, O., {Durrell}, P.~R., {Marleau}, F.~R., {et~al.} 2021,
  \bibinfo{title}{{Dwarf Galaxies in the MATLAS Survey: Hubble Space Telescope
  Observations of the Globular Cluster System in the Ultra-diffuse Galaxy
  MATLAS-2019},} \apj, 923, 9, \dodoi{10.3847/1538-4357/ac2831}

\bibitem[{J.~F. {Navarro} {et~al.}(1996){Navarro}, {Frenk}, \&
  {White}}]{Navarro96}
{Navarro}, J.~F., {Frenk}, C.~S., \& {White}, S. D.~M. 1996,
  \bibinfo{title}{{The Structure of Cold Dark Matter Halos},} \apj, 462, 563,
  \dodoi{10.1086/177173}

\bibitem[{H. {Netzer}(2015){Netzer}}]{Netzer15}
{Netzer}, H. 2015, \bibinfo{title}{{Revisiting the Unified Model of Active
  Galactic Nuclei},} \araa, 53, 365,
  \dodoi{10.1146/annurev-astro-082214-122302}

\bibitem[{N. Neumayer {et~al.}(2020)Neumayer, Seth, \& B{\"o}ker}]{Neumayer20}
Neumayer, N., Seth, A., \& B{\"o}ker, T. 2020, \bibinfo{title}{Nuclear star
  clusters,} \aapr, 28, 4, \dodoi{10.1007/s00159-020-00125-0}

\bibitem[{N. {Neumayer} \& C.~J. {Walcher}(2012){Neumayer} \&
  {Walcher}}]{Neumayer12}
{Neumayer}, N., \& {Walcher}, C.~J. 2012, \bibinfo{title}{{Are Nuclear Star
  Clusters the Precursors of Massive Black Holes?},} Advances in Astronomy,
  2012, 709038, \dodoi{10.1155/2012/709038}

\bibitem[{D.~D. {Nguyen}(2017){Nguyen}}]{Nguyen17conf}
{Nguyen}, D.~D. 2017, \bibinfo{title}{{Improved dynamical constraints on the
  mass of the central black hole in NGC 404},} arXiv e-prints,
  arXiv:1712.02470, \dodoi{10.48550/arXiv.1712.02470}

\bibitem[{D.~D. {Nguyen}(2019){Nguyen}}]{Nguyen19conf}
{Nguyen}, D.~D. 2019, in ALMA2019: Science Results and Cross-Facility
  Synergies, 106, \dodoi{10.5281/zenodo.3585410}

\bibitem[{D.~D. {Nguyen} {et~al.}(2023){Nguyen}, {Cappellari}, \&
  {Pereira-Santaella}}]{Nguyen23}
{Nguyen}, D.~D., {Cappellari}, M., \& {Pereira-Santaella}, M. 2023,
  \bibinfo{title}{{Simulating supermassive black hole mass measurements for a
  sample of ultramassive galaxies using ELT/HARMONI high-spatial-resolution
  integral-field stellar kinematics},} \mnras, 526, 3548,
  \dodoi{10.1093/mnras/stad2860}

\bibitem[{D.~D. {Nguyen} {et~al.}(2014){Nguyen}, {Seth}, {Reines}, {den Brok},
  {Sand}, \& {McLeod}}]{Nguyen14}
{Nguyen}, D.~D., {Seth}, A.~C., {Reines}, A.~E., {et~al.} 2014,
  \bibinfo{title}{{Extended Structure and Fate of the Nucleus in Henize 2-10},}
  \apj, 794, 34, \dodoi{10.1088/0004-637X/794/1/34}

\bibitem[{D.~D. {Nguyen} {et~al.}(2017){Nguyen}, {Seth}, {den Brok},
  {Neumayer}, {Cappellari}, {Barth}, {Caldwell}, {Williams}, \&
  {Binder}}]{Nguyen17}
{Nguyen}, D.~D., {Seth}, A.~C., {den Brok}, M., {et~al.} 2017,
  \bibinfo{title}{{Improved Dynamical Constraints on the Mass of the Central
  Black Hole in NGC 404},} \apj, 836, 237, \dodoi{10.3847/1538-4357/aa5cb4}

\bibitem[{D.~D. {Nguyen} {et~al.}(2018){Nguyen}, {Seth}, {Neumayer}, {Kamann},
  {Voggel}, {Cappellari}, {Picotti}, {Nguyen}, {B{\"o}ker}, {Debattista},
  {Caldwell}, {McDermid}, {Bastian}, {Ahn}, \& {Pechetti}}]{Nguyen18}
{Nguyen}, D.~D., {Seth}, A.~C., {Neumayer}, N., {et~al.} 2018,
  \bibinfo{title}{{Nearby Early-type Galactic Nuclei at High Resolution:
  Dynamical Black Hole and Nuclear Star Cluster Mass Measurements},} \apj, 858,
  118, \dodoi{10.3847/1538-4357/aabe28}

\bibitem[{D.~D. {Nguyen} {et~al.}(2019){Nguyen}, {Seth}, {Neumayer}, {Iguchi},
  {Cappellari}, {Strader}, {Chomiuk}, {Tremou}, {Pacucci}, {Nakanishi},
  {Bahramian}, {Nguyen}, {den Brok}, {Ahn}, {Voggel}, {Kacharov}, {Tsukui},
  {Ly}, {Dumont}, \& {Pechetti}}]{Nguyen19}
{Nguyen}, D.~D., {Seth}, A.~C., {Neumayer}, N., {et~al.} 2019,
  \bibinfo{title}{{Improved Dynamical Constraints on the Masses of the Central
  Black Holes in Nearby Low-mass Early-type Galactic Nuclei and the First Black
  Hole Determination for NGC 205},} \apj, 872, 104,
  \dodoi{10.3847/1538-4357/aafe7a}

\bibitem[{D.~D. {Nguyen} {et~al.}(2020){Nguyen}, {den Brok}, {Seth}, {Davis},
  {Greene}, {Cappellari}, {Jensen}, {Thater}, {Iguchi}, {Imanishi}, {Izumi},
  {Nyland}, {Neumayer}, {Nakanishi}, {Nguyen}, {Tsukui}, {Bureau}, {Onishi},
  {Nguyen}, \& {Le}}]{Nguyen20}
{Nguyen}, D.~D., {den Brok}, M., {Seth}, A.~C., {et~al.} 2020,
  \bibinfo{title}{{The MBHBM$_{{\ensuremath{\star}}}$ Project. I. Measurement
  of the Central Black Hole Mass in Spiral Galaxy NGC 3504 Using Molecular Gas
  Kinematics},} \apj, 892, 68, \dodoi{10.3847/1538-4357/ab77aa}

\bibitem[{D.~D. {Nguyen} {et~al.}(2021){Nguyen}, {Izumi}, {Thater}, {Imanishi},
  {Kawamuro}, {Baba}, {Nakano}, {Turner}, {Kohno}, {Matsushita}, {Mart{\'\i}n},
  {Meier}, {Nguyen}, \& {Nguyen}}]{Nguyen21}
{Nguyen}, D.~D., {Izumi}, T., {Thater}, S., {et~al.} 2021,
  \bibinfo{title}{{Black hole mass measurement using ALMA observations of [CI]
  and CO emissions in the Seyfert 1 galaxy NGC 7469},} \mnras, 504, 4123,
  \dodoi{10.1093/mnras/stab1002}

\bibitem[{D.~D. {Nguyen} {et~al.}(2022){Nguyen}, {Bureau}, {Thater}, {Nyland},
  {den Brok}, {Cappellari}, {Davis}, {Greene}, {Neumayer}, {Imanishi}, {Izumi},
  {Kawamuro}, {Baba}, {Nguyen}, {Iguchi}, {Tsukui}, {Lam}, \& {Ho}}]{Nguyen22}
{Nguyen}, D.~D., {Bureau}, M., {Thater}, S., {et~al.} 2022,
  \bibinfo{title}{{The MBHBM$^{{\ensuremath{\star}}}$ Project - II. Molecular
  gas kinematics in the lenticular galaxy NGC 3593 reveal a supermassive black
  hole},} \mnras, 509, 2920, \dodoi{10.1093/mnras/stab3016}

\bibitem[{D.~D. {Nguyen} {et~al.}(2025){Nguyen}, {Ngo}, {Le}, {Graham},
  {Soria}, {Chilingarian}, {Thatte}, {Phuong}, {Hoang}, {Pereira-Santaella},
  {Durre}, {Pham}, {Ngoc Tram}, {Ngoc}, \& {L{\^e}}}]{Nguyen2025a}
{Nguyen}, D.~D., {Ngo}, H.~N., {Le}, T. Q.~T., {et~al.} 2025,
  \bibinfo{title}{{Supermassive black hole mass measurement in the spiral
  galaxy NGC 4736 using JWST/NIRSpec stellar kinematics},} \aap, 698, L9,
  \dodoi{10.1051/0004-6361/202554672}

\bibitem[{J.~W. {Nightingale} {et~al.}(2023){Nightingale}, {Smith}, {He},
  {O'Riordan}, {Kegerreis}, {Amvrosiadis}, {Edge}, {Etherington}, {Hayes},
  {Kelly}, {Lucey}, \& {Massey}}]{Nightingale23}
{Nightingale}, J.~W., {Smith}, R.~J., {He}, Q., {et~al.} 2023,
  \bibinfo{title}{{Abell 1201: detection of an ultramassive black hole in a
  strong gravitational lens},} \mnras, 521, 3298, \dodoi{10.1093/mnras/stad587}

\bibitem[{M.~A. {Norris} {et~al.}(2014){Norris}, {Kannappan}, {Forbes},
  {Romanowsky}, {Brodie}, {Faifer}, {Huxor}, {Maraston}, {Moffett}, {Penny},
  {Pota}, {Smith-Castelli}, {Strader}, {Bradley}, {Eckert}, {Fohring},
  {McBride}, {Stark}, \& {Vaduvescu}}]{Norris14}
{Norris}, M.~A., {Kannappan}, S.~J., {Forbes}, D.~A., {et~al.} 2014,
  \bibinfo{title}{{The AIMSS Project - I. Bridging the star cluster-galaxy
  divide$^{★}${\textdagger}{\textdaggerdbl}{\textsection}{\textparagraph}},}
  \mnras, 443, 1151, \dodoi{10.1093/mnras/stu1186}

\bibitem[{E. {Noyola} {et~al.}(2010){Noyola}, {Gebhardt}, {Kissler-Patig},
  {L{\"u}tzgendorf}, {Jalali}, {de Zeeuw}, \& {Baumgardt}}]{Noyola10}
{Noyola}, E., {Gebhardt}, K., {Kissler-Patig}, M., {et~al.} 2010,
  \bibinfo{title}{{Very Large Telescope Kinematics for Omega Centauri: Further
  Support for a Central Black Hole},} \apjl, 719, L60,
  \dodoi{10.1088/2041-8205/719/1/L60}

\bibitem[{J.~B. {Oke}(1974){Oke}}]{Oke74}
{Oke}, J.~B. 1974, \bibinfo{title}{{Absolute Spectral Energy Distributions for
  White Dwarfs},} \apjs, 27, 21, \dodoi{10.1086/190287}

\bibitem[{Y. {Ordenes-Brice{\~n}o} {et~al.}(2018){Ordenes-Brice{\~n}o},
  {Puzia}, {Eigenthaler}, {Taylor}, {Mu{\~n}oz}, {Zhang},
  {Alamo-Mart{\'\i}nez}, {Ribbeck}, {Grebel}, {{\'A}ngel}, {C{\^o}t{\'e}},
  {Ferrarese}, {Hilker}, {Lan{\c{c}}on}, {Mieske}, {Miller}, {Rong}, \&
  {S{\'a}nchez-Janssen}}]{Ordenes-Briceno18b}
{Ordenes-Brice{\~n}o}, Y., {Puzia}, T.~H., {Eigenthaler}, P., {et~al.} 2018,
  \bibinfo{title}{{The Next Generation Fornax Survey (NGFS). IV. Mass and Age
  Bimodality of Nuclear Clusters in the Fornax Core Region},} \apj, 860, 4,
  \dodoi{10.3847/1538-4357/aac1b8}

\bibitem[{B.~R. {Parodi} {et~al.}(2002){Parodi}, {Barazza}, \&
  {Binggeli}}]{Parodi02}
{Parodi}, B.~R., {Barazza}, F.~D., \& {Binggeli}, B. 2002,
  \bibinfo{title}{{Structure and stellar content of dwarf galaxies. VII. B and
  R photometry of 25 southern field dwarfs and a disk parameter analysis of the
  complete sample of nearby irregulars},} \aap, 388, 29,
  \dodoi{10.1051/0004-6361:20020432}

\bibitem[{R. {Pechetti} {et~al.}(2020){Pechetti}, {Seth}, {Neumayer},
  {Georgiev}, {Kacharov}, \& {den Brok}}]{Pechetti20}
{Pechetti}, R., {Seth}, A., {Neumayer}, N., {et~al.} 2020,
  \bibinfo{title}{{Luminosity Models and Density Profiles for Nuclear Star
  Clusters for a Nearby Volume-limited Sample of 29 Galaxies},} \apj, 900, 32,
  \dodoi{10.3847/1538-4357/abaaa7}

\bibitem[{R. {Pechetti} {et~al.}(2022){Pechetti}, {Seth}, {Kamann}, {Caldwell},
  {Strader}, {den Brok}, {Luetzgendorf}, {Neumayer}, \& {Voggel}}]{Pechetti22}
{Pechetti}, R., {Seth}, A., {Kamann}, S., {et~al.} 2022,
  \bibinfo{title}{{Detection of a 100,000 M $_{{\ensuremath{\odot}}}$ black
  hole in M31's Most Massive Globular Cluster: A Tidally Stripped Nucleus},}
  \apj, 924, 48, \dodoi{10.3847/1538-4357/ac339f}

\bibitem[{R.~C. {Peterson} \& N. {Caldwell}(1993){Peterson} \&
  {Caldwell}}]{Peterson93}
{Peterson}, R.~C., \& {Caldwell}, N. 1993, \bibinfo{title}{{Stellar Velocity
  Dispersions of Dwarf Elliptical Galaxies},} \aj, 105, 1411,
  \dodoi{10.1086/116520}

\bibitem[{F. {Pinna} {et~al.}(2021){Pinna}, {Neumayer}, {Seth}, {Emsellem},
  {Nguyen}, {B{\"o}ker}, {Cappellari}, {McDermid}, {Voggel}, \&
  {Walcher}}]{Pinna21}
{Pinna}, F., {Neumayer}, N., {Seth}, A., {et~al.} 2021,
  \bibinfo{title}{{Resolved Nuclear Kinematics Link the Formation and Growth of
  Nuclear Star Clusters with the Evolution of Their Early- and Late-type
  Hosts},} \apj, 921, 8, \dodoi{10.3847/1538-4357/ac158f}

\bibitem[{J.~T. {Rayner} {et~al.}(2009){Rayner}, {Cushing}, \&
  {Vacca}}]{Rayner09}
{Rayner}, J.~T., {Cushing}, M.~C., \& {Vacca}, W.~D. 2009, \bibinfo{title}{{The
  Infrared Telescope Facility (IRTF) Spectral Library: Cool Stars},} \apjs,
  185, 289, \dodoi{10.1088/0067-0049/185/2/289}

\bibitem[{J. {Rossa} {et~al.}(2006){Rossa}, {van der Marel}, {B{\"o}ker},
  {Gerssen}, {Ho}, {Rix}, {Shields}, \& {Walcher}}]{Rossa06}
{Rossa}, J., {van der Marel}, R.~P., {B{\"o}ker}, T., {et~al.} 2006,
  \bibinfo{title}{{Hubble Space Telescope STIS Spectra of Nuclear Star Clusters
  in Spiral Galaxies: Dependence of Age and Mass on Hubble Type},} \aj, 132,
  1074, \dodoi{10.1086/505968}

\bibitem[{R.~P. {Saglia} {et~al.}(2016){Saglia}, {Opitsch}, {Erwin}, {Thomas},
  {Beifiori}, {Fabricius}, {Mazzalay}, {Nowak}, {Rusli}, \&
  {Bender}}]{Saglia16}
{Saglia}, R.~P., {Opitsch}, M., {Erwin}, P., {et~al.} 2016,
  \bibinfo{title}{{The SINFONI Black Hole Survey: The Black Hole Fundamental
  Plane Revisited and the Paths of (Co)evolution of Supermassive Black Holes
  and Bulges},} \apj, 818, 47, \dodoi{10.3847/0004-637X/818/1/47}

\bibitem[{N. {Sahu} {et~al.}(2019){Sahu}, {Graham}, \& {Davis}}]{Sahu19a}
{Sahu}, N., {Graham}, A.~W., \& {Davis}, B.~L. 2019, \bibinfo{title}{{Black
  Hole Mass Scaling Relations for Early-type Galaxies. I. M $_{BH}$-M $_{*,}$
  $_{sph}$ and M $_{BH}$-M $_{*,gal}$},} \apj, 876, 155,
  \dodoi{10.3847/1538-4357/ab0f32}

\bibitem[{R. {S{\'a}nchez-Janssen} {et~al.}(2019){S{\'a}nchez-Janssen},
  {C{\^o}t{\'e}}, {Ferrarese}, {Peng}, {Roediger}, {Blakeslee}, {Emsellem},
  {Puzia}, {Spengler}, {Taylor}, {{\'A}lamo-Mart{\'\i}nez}, {Boselli},
  {Cantiello}, {Cuillandre}, {Duc}, {Durrell}, {Gwyn}, {MacArthur},
  {Lan{\c{c}}on}, {Lim}, {Liu}, {Mei}, {Miller}, {Mu{\~n}oz}, {Mihos},
  {Paudel}, {Powalka}, \& {Toloba}}]{Sanchez-Janssen19}
{S{\'a}nchez-Janssen}, R., {C{\^o}t{\'e}}, P., {Ferrarese}, L., {et~al.} 2019,
  \bibinfo{title}{{The Next Generation Virgo Cluster Survey. XXIII.
  Fundamentals of Nuclear Star Clusters over Seven Decades in Galaxy Mass},}
  \apj, 878, 18, \dodoi{10.3847/1538-4357/aaf4fd}

\bibitem[{R.~P. {Schiavon}(2007){Schiavon}}]{Schiavon07}
{Schiavon}, R.~P. 2007, \bibinfo{title}{{Population Synthesis in the Blue. IV.
  Accurate Model Predictions for Lick Indices and UBV Colors in Single Stellar
  Populations},} \apjs, 171, 146, \dodoi{10.1086/511753}

\bibitem[{E.~F. {Schlafly} \& D.~P. {Finkbeiner}(2011){Schlafly} \&
  {Finkbeiner}}]{Schlafly11}
{Schlafly}, E.~F., \& {Finkbeiner}, D.~P. 2011, \bibinfo{title}{{Measuring
  Reddening with Sloan Digital Sky Survey Stellar Spectra and Recalibrating
  SFD},} \apj, 737, 103, \dodoi{10.1088/0004-637X/737/2/103}

\bibitem[{J. {Schombert} \& A.~K. {Smith}(2012){Schombert} \&
  {Smith}}]{Schombert12}
{Schombert}, J., \& {Smith}, A.~K. 2012, \bibinfo{title}{{The Structure of
  Galaxies I: Surface Photometry Techniques},} \pasa, 29, 174,
  \dodoi{10.1071/AS11059}

\bibitem[{J.~L. Sersic(1968)Sersic}]{Sersic68}
Sersic, J.~L. 1968, Atlas de galaxias australes (C\'ordoba: Obs. Astron. Univ.
  Nacional de C\'ordoba)

\bibitem[{A. {Seth} {et~al.}(2008){Seth}, {Ag{\"u}eros}, {Lee}, \&
  {Basu-Zych}}]{Seth08}
{Seth}, A., {Ag{\"u}eros}, M., {Lee}, D., \& {Basu-Zych}, A. 2008,
  \bibinfo{title}{{The Coincidence of Nuclear Star Clusters and Active Galactic
  Nuclei},} \apj, 678, 116, \dodoi{10.1086/528955}

\bibitem[{A.~C. {Seth} {et~al.}(2010){Seth}, {Cappellari}, {Neumayer},
  {Caldwell}, {Bastian}, {Olsen}, {Blum}, {Debattista}, {McDermid}, {Puzia}, \&
  {Stephens}}]{Seth10}
{Seth}, A.~C., {Cappellari}, M., {Neumayer}, N., {et~al.} 2010,
  \bibinfo{title}{{The NGC 404 Nucleus: Star Cluster and Possible
  Intermediate-mass Black Hole},} \apj, 714, 713,
  \dodoi{10.1088/0004-637X/714/1/713}

\bibitem[{A.~C. {Seth} {et~al.}(2014){Seth}, {van den Bosch}, {Mieske},
  {Baumgardt}, {Brok}, {Strader}, {Neumayer}, {Chilingarian}, {Hilker},
  {McDermid}, {Spitler}, {Brodie}, {Frank}, \& {Walsh}}]{Seth14}
{Seth}, A.~C., {van den Bosch}, R., {Mieske}, S., {et~al.} 2014,
  \bibinfo{title}{{A supermassive black hole in an ultra-compact dwarf
  galaxy},} \nat, 513, 398, \dodoi{10.1038/nature13762}

\bibitem[{J.~C. {Shields} {et~al.}(2008){Shields}, {Walcher}, {B{\"o}ker},
  {Ho}, {Rix}, \& {van der Marel}}]{Shields08}
{Shields}, J.~C., {Walcher}, C.~J., {B{\"o}ker}, T., {et~al.} 2008,
  \bibinfo{title}{{An Accreting Black Hole in the Nuclear Star Cluster of the
  Bulgeless Galaxy NGC 1042},} \apj, 682, 104, \dodoi{10.1086/589680}

\bibitem[{J. {Silk} \& M.~J. {Rees}(1998){Silk} \& {Rees}}]{Silk98}
{Silk}, J., \& {Rees}, M.~J. 1998, \bibinfo{title}{{Quasars and galaxy
  formation},} \aap, 331, L1, \dodoi{10.48550/arXiv.astro-ph/9801013}

\bibitem[{M.~F. {Skrutskie} {et~al.}(2006){Skrutskie}, {Cutri}, {Stiening},
  {Weinberg}, {Schneider}, {Carpenter}, {Beichman}, {Capps}, {Chester},
  {Elias}, {Huchra}, {Liebert}, {Lonsdale}, {Monet}, {Price}, {Seitzer},
  {Jarrett}, {Kirkpatrick}, {Gizis}, {Howard}, {Evans}, {Fowler}, {Fullmer},
  {Hurt}, {Light}, {Kopan}, {Marsh}, {McCallon}, {Tam}, {Van Dyk}, \&
  {Wheelock}}]{Skrutskie06}
{Skrutskie}, M.~F., {Cutri}, R.~M., {Stiening}, R., {et~al.} 2006,
  \bibinfo{title}{{The Two Micron All Sky Survey (2MASS)},} \aj, 131, 1163,
  \dodoi{10.1086/498708}

\bibitem[{C. {Spengler} {et~al.}(2017){Spengler}, {C{\^o}t{\'e}}, {Roediger},
  {Ferrarese}, {S{\'a}nchez-Janssen}, {Toloba}, {Liu}, {Guhathakurta},
  {Cuillandre}, {Gwyn}, {Zirm}, {Mu{\~n}oz}, {Puzia}, {Lan{\c{c}}on}, {Peng},
  {Mei}, \& {Powalka}}]{Spengler17}
{Spengler}, C., {C{\^o}t{\'e}}, P., {Roediger}, J., {et~al.} 2017,
  \bibinfo{title}{{Virgo Redux: The Masses and Stellar Content of Nuclei in
  Early-type Galaxies from Multiband Photometry and Spectroscopy},} \apj, 849,
  55, \dodoi{10.3847/1538-4357/aa8a78}

\bibitem[{O. {Straub} {et~al.}(2014){Straub}, {Godet}, {Webb}, {Servillat}, \&
  {Barret}}]{Straub14}
{Straub}, O., {Godet}, O., {Webb}, N., {Servillat}, M., \& {Barret}, D. 2014,
  \bibinfo{title}{{Investigating the mass of the intermediate mass black hole
  candidate HLX-1 with the slimbh model},} \aap, 569, A116,
  \dodoi{10.1051/0004-6361/201423874}

\bibitem[{ {STScI}(2020){STScI}}]{STScI2020}
{STScI}. 2020, \bibinfo{title}{Digitized Sky Survey,} IPAC,
  \dodoi{10.26131/IRSA441}

\bibitem[{S. {Thater} {et~al.}(2023){Thater}, {Lyubenova}, {Fahrion},
  {Mart{\'\i}n-Navarro}, {Jethwa}, {Nguyen}, \& {van de Ven}}]{Thater23}
{Thater}, S., {Lyubenova}, M., {Fahrion}, K., {et~al.} 2023,
  \bibinfo{title}{{Effect of the initial mass function on the dynamical SMBH
  mass estimate in the nucleated early-type galaxy FCC 47},} \aap, 675, A18,
  \dodoi{10.1051/0004-6361/202245362}

\bibitem[{N.~A. {Thatte} {et~al.}(2016){Thatte}, {Clarke}, {Bryson},
  {Shnetler}, {Tecza}, {Fusco}, {Bacon}, {Richard}, {Mediavilla}, {Neichel},
  {Arribas}, {Garcia-Lorenzo}, {Evans}, {Remillieux}, {El Madi}, {Herreros},
  {Melotte}, {O'Brien}, {Tosh}, {Vernet}, {Hammersley}, {Ives}, {Finger},
  {Houghton}, {Rigopoulou}, {Lynn}, {Allen}, {Zieleniewski}, {Kendrew},
  {Ferraro-Wood}, {P{\'e}contal-Rousset}, {Kosmalski}, {Laurent}, {Loupias},
  {Piqueras}, {Renault}, {Blaizot}, {Daguis{\'e}}, {Migniau}, {Jarno}, {Born},
  {Gallie}, {Montgomery}, {Henry}, {Schwartz}, {Taylor}, {Zins},
  {Rodr{\'\i}guez-Ramos}, {Cagigas}, {Battaglia}, {Rebolo L{\'o}pez},
  {Hern{\'a}ndez Su{\'a}rez}, {Gigante-Ripoll}, {Piqueras L{\'o}pez}, {Villar
  Martin}, {Correia}, {Pascal}, {Blanco}, {Vola}, {Epinat}, {Peroux}, {Vigan},
  {Dohlen}, {Sauvage}, {Lee}, {Carlotti}, {Verinaud}, {Morris}, {Myers},
  {Reeves}, {Swinbank}, {Calcines}, \& {Larrieu}}]{Thatte16}
{Thatte}, N.~A., {Clarke}, F., {Bryson}, I., {et~al.} 2016, in Society of
  Photo-Optical Instrumentation Engineers (SPIE) Conference Series, Vol. 9908,
  Ground-based and Airborne Instrumentation for Astronomy VI, ed. C.~J.
  {Evans}, L.~{Simard}, \& H.~{Takami}, 99081X, \dodoi{10.1117/12.2230629}

\bibitem[{N.~A. {Thatte} {et~al.}(2020){Thatte}, {Bryson}, {Clarke},
  {Ferraro-Wood}, {Fusco}, {Le Mignant}, {Melotte}, {Neichel}, {Schnetler},
  {Tecza}, {Arribas}, {Crespo}, {Estrada Piqueras}, {Garc{\'\i}a Garc{\'\i}a},
  {Pereira Santaella}, {Piqueras L{\'o}pez}, {Blaizot}, {Bouch{\'e}}, {Boudon},
  {Chapuis}, {Daguise}, {Disseau}, {Guibert}, {Jarno}, {Jeanneau}, {Laurent},
  {Loupias}, {Migniau}, {Piqueras}, {Remillieux}, {Richard},
  {P{\'e}contal-Rousset}, {Bardou}, {Close}, {Deshmukh}, {Dimoudi},
  {Dubbledam}, {King}, {Morris}, {Morris}, {O'Brien}, {Staykov}, {Swinbank},
  {Townson}, {Younger}, {Accardo}, {Avarez Mendez}, {Conzelmann}, {Egner},
  {George}, {Gont{\'e}}, {Hopgood}, {Ives}, {Mehrgan}, {Mueller}, {Peroux},
  {Vernet}, {Alonso Sanchez}, {Giuseppina}, {Cagigas}, {Delgado}, {Fernandez
  Izquierdo}, {Fragoso L{\'o}pez}, {Garcia-Lorenzo}, {Hernandez Suarez},
  {Herreros Linares}, {Joven}, {L{\'o}pez}, {Mart{\'\i}n Hernando},
  {Mediavilla}, {Monreal}, {Pe{\~n}ate Castro}, {Rasilla}, {Rebolo},
  {Rodr{\'\i}guez-Ramos}, {Vega Moreno}, {Viera}, {Carlotti}, {Correia},
  {Delboulbe}, {Guieu}, {Hours}, {Hubert}, {Jocou}, {Magnard}, {Moulin},
  {Pancher}, {Rabou}, {Stadler}, {Contini}, {Larrieu}, {Fantei-Caujolle},
  {Lecron}, {Rousseau}, {Beltramo-Martin}, {Bon}, {Bonnefoi}, {Ceria},
  {Choquet}, {Correia}, {Costille}, {Dohlen}, {Ducret}, {El-Hadi}, {Epinat},
  {Fetick}, {Gach}, {Groussin}, {Jaafar}, {Le Merrer}, {Llored}, {Pedreros},
  {Renault}, {Sanchez}, {Vigan}, {Vola}, {Lim}, {Vedrenne}, {Petit}, {Sauvage},
  {Bagci}, {Cann}, {Chao Ortiz}, {Elliott}, {Seitis}, {Tosh}, {Anderson},
  {Black}, {Bond}, {Born}, {Campbell}, {Campbell}, {Carruthers}, {Cochrane},
  {Dobson}, {Evans}, {Gallie}, {Gonzalez}, {Harman}, {Henry}, {Humphreys},
  {Louth}, {Miller}, {Montgomery}, {Murray}, {O'Malley}, {Ritchie},
  {Sanchez-Janssen}, {Schwartz}, {Smith}, {Watt}, {Wells}, {Wilson},
  {Gultekin}, {Mateo}, {Meyer}, {Valluri}, {Ahmad}, {Booth}, {Capone},
  {Cappellari}, {Gooding}, {Grisdale}, {Hidalgo}, {Kariuki}, {Lewis}, {Lowe},
  {Lynn}, {Menduina}, {Ozer}, {Preece}, {Rigopoulou}, {Rodrigues}, \&
  {Routledge}}]{Thatte20}
{Thatte}, N.~A., {Bryson}, I., {Clarke}, F., {et~al.} 2020, in Society of
  Photo-Optical Instrumentation Engineers (SPIE) Conference Series, Vol. 11447,
  Ground-based and Airborne Instrumentation for Astronomy VIII, ed. C.~J.
  {Evans}, J.~J. {Bryant}, \& K.~{Motohara}, 114471W,
  \dodoi{10.1117/12.2562144}

\bibitem[{N.~A. Thatte {et~al.}(2024)Thatte, Melotte, Neichel, Le~Mignant,
  Rees, Clarke, Ferraro-Wood, Gonzalez, Jones, {\'A}lvarez~Urue{\~n}a,
  Argelaguet~Vilaseca, Arribas~Mocoroa, Caballero, Carracedo~Carballal,
  Estrada~Piqueras, Ferro, Garc{\'\i}a~Garc{\'\i}a, Lamperti,
  Pereira~Santaella, Perna, Piqueras~Lopez, Bouch{\'e}, Boudon, Daguise,
  Domenis, Fensch, Olivier~Flasseur, Giroud, Guibert, Jarno, Jeanneau,
  Krogager, Langlois, Laurent, Loupias, Migniau, Nguyen, Piqueras, Remillieux,
  Richard, Pecontal, Bardou, Barr, Cetre, Dimoudi, Dubbeldam, Dunn, Gadotti,
  Guy, King, McLeod, Morris, Morris, O'Brien, Ronson, Smith, Staykov, Swinbank,
  Accardo, Alvarez~Mendez, Fuerte~Rodriguez, George, Ives, Mehrgan, Mueller,
  Reyes, Conzelmann, Gutierrez~Cheetham, Alonso~Sanchez, Battaglia, Cagigas,
  Castro-Almaz{\'a}n, Chulani, Delgado-Garc{\'\i}a, Fernandez~Izquierdo,
  Esparza-Arredondo, Garc{\'\i}a-Lorenzo, Hern{\'a}ndez~Gonz{\'a}lez,
  Hern{\'a}ndez~Su{\'a}rez, Licandro, Joven, L{\'o}pez~L{\'o}pez,
  Lujan~Gonzalez, Mart{\'\i}n~Hernando, Mart{\'\i}n-Navarro, Mediavilla,
  Men{\'e}ndez~Mendoza, Montoya~Mart{\'\i}nez, Pe{\~n}ate~Castro, Murgas,
  Pall{\'e}, P{\'e}rez, Rasilla, Rebolo, Rodr{\'\i}guez, Rodr{\'\i}guez~Ramos,
  S{\'a}nchez~B{\'e}jar, Shahbaz, Vega~Moreno, Viera, Bonnefoy, Bret, Carlotti,
  Correia, Curaba, Delboulb{\'e}, Guieu, Hours, Hubert, Jocou, Magnard,
  Michaud, Moulin, Pancher, Rabou, Rochat, Stadler, Contini, Larrieu,
  Mamessier, Boebion, Fantei-Caujolle, Lecron, Amram, Blanchard, Bon, Bonnefoi,
  Bozier, Ceria, Challita, Charles, Choquet, Costille, Delsanti, Dohlen,
  Ducret, El~Hadi, Foulon, Gimenez, Groussin, Jaquet, Renault, Rouquette,
  Sanchez, Vigan, Zavagno, F{\'e}tick, Fusco, H{\'e}ritier, Sauvage, Vedrenne,
  Aksoy, Caldwell, Fitzpatrick, Geddert, Hiscock, Johnson, Nalagatla, Saraff,
  Shreeves, Tildesley, Wells, Aretos, Barrett, Black, Bond, Brierley, Bryson,
  Calderhead, Campbell, Carruthers, Chapman, Cochrane, Gillespie, Harman,
  Harvey, Harvey, Johnson, Louth, MacIntosh, MacIver, Miller, Montgomery,
  Murali, Murray, O'Malley, Sanchez-Janssen, Schwartz, Smith, Strachan, Todd,
  Wasley, Wilson, Zhou, Bell, Gnedin, Gultekin, Mateo, Meyer, \&
  Birkby}]{Thatte2024}
Thatte, N.~A., Melotte, D., Neichel, B., {et~al.} 2024, in Society of
  Photo-Optical Instrumentation Engineers (SPIE) Conference Series, Vol. 13096,
  Ground-based and Airborne Instrumentation for Astronomy X, ed. J.~J.
  {Bryant}, K.~{Motohara}, \& J.~R.~D. {Vernet}, 1309614,
  \dodoi{10.1117/12.3018520}

\bibitem[{S. Tremaine {et~al.}(1994)Tremaine, Richstone, Byun, Dressler, Faber,
  Grillmair, Kormendy, \& Lauer}]{Tremaine94}
Tremaine, S., Richstone, D.~O., Byun, Y.-I., {et~al.} 1994, \bibinfo{title}{A
  family of models for spherical stellar systems,} \aj, 107, 634,
  \dodoi{10.1086/116883}

\bibitem[{I. Trujillo {et~al.}(2004)Trujillo, Erwin, Asensio~Ramos, \&
  Graham}]{Trujillo04}
Trujillo, I., Erwin, P., Asensio~Ramos, A., \& Graham, A.~W. 2004,
  \bibinfo{title}{{Evidence for a New Elliptical-Galaxy Paradigm: S{\'e}rsic
  and Core Galaxies},} \aj, 127, 1917, \dodoi{10.1086/382712}

\bibitem[{M. {Valluri} {et~al.}(2005){Valluri}, {Ferrarese}, {Merritt}, \&
  {Joseph}}]{Valluri05}
{Valluri}, M., {Ferrarese}, L., {Merritt}, D., \& {Joseph}, C.~L. 2005,
  \bibinfo{title}{{The Low End of the Supermassive Black Hole Mass Function:
  Constraining the Mass of a Nuclear Black Hole in NGC 205 via Stellar
  Kinematics},} \apj, 628, 137, \dodoi{10.1086/430752}

\bibitem[{R.~P. van~der Marel(1999)van~der Marel}]{vanderMarel99}
van~der Marel, R.~P. 1999, \bibinfo{title}{The Black Hole Mass Distribution in
  Early-Type Galaxies: Cusps in Hubble Space Telescope Photometry Interpreted
  through Adiabatic Black Hole Growth,} \aj, 117, 744, \dodoi{10.1086/300730}

\bibitem[{R.~P. {van der Marel} \& J. {Anderson}(2010){van der Marel} \&
  {Anderson}}]{vanderMarel10}
{van der Marel}, R.~P., \& {Anderson}, J. 2010, \bibinfo{title}{{New Limits on
  an Intermediate-Mass Black Hole in Omega Centauri. II. Dynamical Models},}
  \apj, 710, 1063, \dodoi{10.1088/0004-637X/710/2/1063}

\bibitem[{S. {van Wassenhove} {et~al.}(2010){van Wassenhove}, {Volonteri},
  {Walker}, \& {Gair}}]{vanWassenhove10}
{van Wassenhove}, S., {Volonteri}, M., {Walker}, M.~G., \& {Gair}, J.~R. 2010,
  \bibinfo{title}{{Massive black holes lurking in Milky Way satellites},}
  \mnras, 408, 1139, \dodoi{10.1111/j.1365-2966.2010.17189.x}

\bibitem[{E.~K. {Verolme} {et~al.}(2002){Verolme}, {Cappellari}, {Copin}, {van
  der Marel}, {Bacon}, {Bureau}, {Davies}, {Miller}, \& {de Zeeuw}}]{Verolme02}
{Verolme}, E.~K., {Cappellari}, M., {Copin}, Y., {et~al.} 2002,
  \bibinfo{title}{{A SAURON study of M32: measuring the intrinsic flattening
  and the central black hole mass},} \mnras, 335, 517,
  \dodoi{10.1046/j.1365-8711.2002.05664.x}

\bibitem[{M. {Volonteri}(2012){Volonteri}}]{Volonteri12}
{Volonteri}, M. 2012, \bibinfo{title}{{The Formation and Evolution of Massive
  Black Holes},} Science, 337, 544, \dodoi{10.1126/science.1220843}

\bibitem[{M. Volonteri {et~al.}(2021)Volonteri, Habouzit, \&
  Colpi}]{Volonteri2021}
Volonteri, M., Habouzit, M., \& Colpi, M. 2021, \bibinfo{title}{The origins of
  massive black holes,} Nature Reviews Physics, 3, 732,
  \dodoi{10.1038/s42254-021-00364-9}

\bibitem[{C.~J. {Walcher} {et~al.}(2006){Walcher}, {B{\"o}ker}, {Charlot},
  {Ho}, {Rix}, {Rossa}, {Shields}, \& {van der Marel}}]{Walcher06}
{Walcher}, C.~J., {B{\"o}ker}, T., {Charlot}, S., {et~al.} 2006,
  \bibinfo{title}{{Stellar Populations in the Nuclei of Late-Type Spiral
  Galaxies},} \apj, 649, 692, \dodoi{10.1086/505166}

\bibitem[{C.~J. {Walcher} {et~al.}(2005){Walcher}, {van der Marel},
  {McLaughlin}, {Rix}, {B{\"o}ker}, {H{\"a}ring}, {Ho}, {Sarzi}, \&
  {Shields}}]{Walcher05}
{Walcher}, C.~J., {van der Marel}, R.~P., {McLaughlin}, D., {et~al.} 2005,
  \bibinfo{title}{{Masses of Star Clusters in the Nuclei of Bulgeless Spiral
  Galaxies},} \apj, 618, 237, \dodoi{10.1086/425977}

\bibitem[{L. {Wallace} \& K. {Hinkle}(1996){Wallace} \& {Hinkle}}]{Wallace96}
{Wallace}, L., \& {Hinkle}, K. 1996, \bibinfo{title}{{High-Resolution Spectra
  of Ordinary Cool Stars in the K Band},} \apjs, 107, 312,
  \dodoi{10.1086/192367}

\bibitem[{L. {Wallace} \& K. {Hinkle}(1997){Wallace} \& {Hinkle}}]{Wallace97}
{Wallace}, L., \& {Hinkle}, K. 1997, \bibinfo{title}{{Medium-Resolution Spectra
  of Normal Stars in the K Band},} \apjs, 111, 445, \dodoi{10.1086/313020}

\bibitem[{J.~L. {Walsh} {et~al.}(2013){Walsh}, {Barth}, {Ho}, \&
  {Sarzi}}]{Walsh13}
{Walsh}, J.~L., {Barth}, A.~J., {Ho}, L.~C., \& {Sarzi}, M. 2013,
  \bibinfo{title}{{The M87 Black Hole Mass from Gas-dynamical Models of Space
  Telescope Imaging Spectrograph Observations},} \apj, 770, 86,
  \dodoi{10.1088/0004-637X/770/2/86}

\bibitem[{B.~F. {Williams} {et~al.}(2013){Williams}, {Dalcanton}, {Stilp},
  {Dolphin}, {Skillman}, \& {Radburn-Smith}}]{Williams13}
{Williams}, B.~F., {Dalcanton}, J.~J., {Stilp}, A., {et~al.} 2013,
  \bibinfo{title}{{The ACS Nearby Galaxy Survey Treasury. XI. The Remarkably
  Undisturbed NGC 2403 Disk},} \apj, 765, 120,
  \dodoi{10.1088/0004-637X/765/2/120}

\bibitem[{K. Zhu {et~al.}(2025)Zhu, Cappellari, Mao, Lu, Li, Shi, Simon, Fu, \&
  Wang}]{Zhu2025}
Zhu, K., Cappellari, M., Mao, S., {et~al.} 2025, \bibinfo{title}{MaNGA DynPop
  -- VII. A Unified Bulge-Disk-Halo Model for Explaining Diversity in Circular
  Velocity Curves of 6000 Spiral and Early-Type Galaxies,} arXiv e-prints,
  arXiv:2503.06968, \dodoi{10.48550/arXiv.2503.06968}

\bibitem[{S. {Zieleniewski} {et~al.}(2015){Zieleniewski}, {Thatte}, {Kendrew},
  {Houghton}, {Swinbank}, {Tecza}, {Clarke}, \& {Fusco}}]{Zieleniewski15}
{Zieleniewski}, S., {Thatte}, N., {Kendrew}, S., {et~al.} 2015,
  \bibinfo{title}{{HSIM: a simulation pipeline for the HARMONI integral field
  spectrograph on the European ELT},} \mnras, 453, 3754,
  \dodoi{10.1093/mnras/stv1860}

\bibitem[{A. {Zocchi} {et~al.}(2017){Zocchi}, {Gieles}, \&
  {H{\'e}nault-Brunet}}]{Zocchi17}
{Zocchi}, A., {Gieles}, M., \& {H{\'e}nault-Brunet}, V. 2017,
  \bibinfo{title}{{Radial anisotropy in {\ensuremath{\omega}} Cen limiting the
  room for an intermediate-mass black hole},} \mnras, 468, 4429,
  \dodoi{10.1093/mnras/stx316}

\bibitem[{A. {Zocchi} {et~al.}(2019){Zocchi}, {Gieles}, \&
  {H{\'e}nault-Brunet}}]{Zocchi19}
{Zocchi}, A., {Gieles}, M., \& {H{\'e}nault-Brunet}, V. 2019,
  \bibinfo{title}{{The effect of stellar-mass black holes on the central
  kinematics of {\ensuremath{\omega}} Cen: a cautionary tale for IMBH
  interpretations},} \mnras, 482, 4713, \dodoi{10.1093/mnras/sty1508}

\end{thebibliography}

\appendix

\section{SUPPLEMENTARY FIGURES and Tables}\label{appendixa}

\begin{table*}
\caption{Full list of our HARMONI IMBH survey: host galaxies' and their nuclear star clusters' properties}
\footnotesize
\begin{tabular}{ccccccccccccc}
\hline\hline   
No. & Galaxy & R.A. (J2000) & Decl. (J2000) & D  & Hubble  & $L_K$ & $M_{\star, \rm gal.}$ & $\sigma_{\star, \rm NSC}$ & $R_{e,\rm gal}$ & $M_{\rm NSC}$ & $R_{e,\rm NSC}$ & ref. \\  
      &  Name & (h:m:s) & (d:m:s)  & (Mpc)  & Type & (\Lsun) & (\Msun) & (\kms) & (kpc) & (\Msun) & (pc) & \\
(1)& (2) & (3) & (4)  & (5)  & (6) & (7) & (8) & (9) & (10) & (11) & (12) & (13) \\
\hline

1 &	PGC046680   &13:22:02.048 &$-$42:32:07.34   &3.77& $-3.0$&$1.32\times10^8$&$4.0\times10^7$& 13.2 & 0.73 & $4.30\times10^6$& 3.0 &(1)\\
2 &	NGC~5206    &13:33:43.961 &$-$48:09:03.95   &3.50& $-3.0$&$1.05\times10^9$&$2.5\times10^9$& 46	.0 & 0.99 & $1.54\times10^7$& 3.4&(2)\\
3 &	ESO269-066  &13:13:09.156 &$-$44:53:23.63   &3.66& $-$2.1& $2.75\times10^8$&$1.0\times10^8$& --     & 0.70 & $1.55\times10^6$& --& --\\
4 &	ESO269-068  &13:13:11.917 &$-$43:15:54.69  &3.77&$-2.1$&$8.91\times10^7$&$1.7\times10^7$& --	    & 0.41 & $6.10\times10^5$ & 15.2 & (11)\\
5 &	NGC~4600	&12:40:22.957 &   03:07:03.91	    &8.15&  0.0	&$1.32\times10^9$&$1.2\times10^9$& 73.6 & 0.94 & $1.03\times10^7$& 8.3&(2)\\
6 &	NGC~5102	&13:21:57.610 &$-$36:37:48.37   &3.20&	0.0	&$4.27\times10^9$&	$6.9\times10^9$& 61.1 &1.20 & $7.30\times10^7$& 26.3&(3)\\
7 &	NGC~300	    &00:54:53.444 &$-$37:41:03.22   &2.01&	6.9	&$5.00\times10^8$     &$1.0\times10^9$& 47.0 &2.94 & $1.05\times10^6$& 8.0&(4, 5)\\
8 &	NGC~7793	&23:57:49.753 &$-$32:35:27.74.  &3.40&	7.4	&$7.94\times10^7$     &$2.8\times10^9$& 24.6 &2.19 & $7.76\times10^6$& 7.9&(4, 5)\\
9 &	NGC~3621	&11:18:16.511 &$-$32:48:50.55   &6.60&	7.0    &$2.24\times10^{10}$&$7.9\times10^9$& 43.0 &2.60 & $1.01\times10^7$& 4.2&(6)\\
10&	NGC~247	    &00:47:08.470 &$-$20:45:36.71   &3.38&	6.9	&$3.16\times10^9$     &$3.0\times10^9$& 15.1 &2.46 & $2.51\times10^6$& 3.4&(7)\\
11&	IC3077	    &12:24:22.180 &   21:09:35.95	    &9.12&	5.8	&$1.45\times10^8$&	$1.3\times10^9$& 25.0 &1.40 & --& --& --\\
12&	NGC~5237	&13:37:39.050 &$-$42:50:49.099&3.33&$-$3.5	&$2.82\times10^8$&$1.6\times10^8$& 48.5 &0.34 & -- & -- & --\\
13&	PGC029300	&10:05:41.599 &$-$07:58:53.40  &9.70&$-$1.9	&$1.86\times10^9$&$2.8\times10^9$& 32.0	    &1.20 & $7.01\times10^6$ & 6.6&(3)\\
14&	ESO384-016	&13:57:01.370 &$-$35 19 58.699&4.49&$-$4.9	&$6.76\times10^7$&$6.0\times10^7$&	    --&0.69 & --& --& --\\
15&	ESO174-001	&13:33:19.680 &$-$53:21:16.88 &3.60&$-$1.7	&$7.94\times10^7$&$7.9\times10^7$& -- &0.42 & --& --& --\\
16&	MESSIER083	&13:37:00.950 &$-$29:51:55.50 &4.85&	5.0	&     $7.24\times10^{10}$     &$4.5\times10^{10}$&40.0&5.29 & $1.67\times10^6$ & 4.9& --\\
17&	NGC~3593	&11:14:37.001 &12:49:03.61	  &9.20  &$-$4.8	&  $2.69\times10^{10}$   &$1.5\times10^{10}$&60.0&1.90 & $1.58\times10^8$ & 6&(3, 8)\\
18&	ESO274-G001	&15:14:14.554 &$-$46:48:19.32&3.15&	6.7	&   $8.32\times10^8$ &$5.0\times10^8$&17.2  &1.45 & $2.54\times10^6$ & 2.0&(3)\\
19&	NGC~4713	&12:49:57.874 &05:18:41.04	  &10.7&	6.8	&	--  &$3.0\times10^{10}$&23.2&2.12 & -- & -- &--\\
20&	NGC~3351	&10:43:57.701 &11:42:13.72	  &9.46&	3.1	&$2.00\times10^{10}$&$2.0\times10^{10}$&67.0&3.95 & $1.45\times10^{6}$ & 4.4&(9) \\
21&	IC 5332	    &23:34:27.490 &$-$36:06:03.89&4.63&	6.8	&   $5.50\times10^9$ &$4.7\times10^9$&58.3  &	1.18 & $6.95\times10^6$ & 23.2&(3)\\
22&	NGC~1493	&03:57:27.430 &$-$46:12:38.52&11.3&	6.0	&	--  &   $4.0\times10^9$   &25.0     & --	& $1.51\times10^6$ & 4.2& --\\
23&	NGC~1042	&02:40:23.966 &$-$08:26:00.74&4.21&	6.0	&   -- &$2.1\times10^9$&32.0     & 1.09 & $2.12\times10^6$ & 1.3&(4)\\
24&	NGC~628	    &01:36:41.747 & 15:47:01.18   &4.21&	5.2	&  $3.98\times10^{10}$  &$1.4\times10^{10}$& 72.2  & 6.60 & $1.13\times10^7$ & 5.4&(10)\\
25&	ESO059-01	& 07:31:18.00 &	$-$68:11:14  &	4.57& 9.8  &	$1.48\times10^8$  &$5.1\times10^7$& -- &	--	&$1.45\times10^6$ & --& --\\
26&	UGC 3755	& 07:13:51.57 &	10:31:16.53  &	4.99& 9.9  &	$2.69\times10^8$  &$9.3\times10^7$& -- &	--	&$6.03\times10^4$ & 0.5 & (10)\\
27&	IC 1959	    & 03:33:12.53 &	$-$50:24:52.19  &	6.05& 8.4  &	$3.09\times10^8$  &$1.4\times10^8$& -- &	--	&$1.35\times10^6$ & --& --\\
28&	UGC 5889    & 10:47:22.4 &	14:04:16.58  &	6.89 & 8.9  &	$3.55\times10^8$  &$1.6\times10^8$& -- &	--	&$1.23\times10^6$ & 1.0& (10)\\
39&	NGC~4592    & 12:39:18.76 &	$-$00:31:55.61  &	10.6 & 7.9  &	$1.45\times10^9$  &$1.0\times10^9$& 38.9 &	--	&$6.31\times10^5$ & 1.1& (10)\\
30&	NGC~1796    & 05:02:41.27 &	$-$61 08 18.79  &	10.6 & 5.3  &	--  &$1.2\times10^9$& -- &	--	&$6.61\times10^6$ & 2.6& --\\
31&	ESO359-G029 & 04:12:50.52 &	$-$33:00:10.34  &	10.1 & 9.9  &	--  &$2.3\times10^8$& 29.5 &	--	&$1.48\times10^5$ & 0.5& (10)\\
32&	IC 4710     & 18:28:40.9 &	$-$66:59:10.0  &	8.91 & 8.9  &	$1.74\times10^9$  &$1.3\times10^9$& 67.2 &	--	&$2.49\times10^6$ & 1.0& (10)\\
33&	NGC~1566    &  04:20:0.4 &	$-$54:56:16.5  &	 9.95  & 4.0  &	--  &$1.5\times10^{10}$& 116 &	--	&$1.49\times10^8$ & 2.3& (12)\\
34&	NGC~2835    & 09:17:52.85 &	$-$22:21:16.79  &	10.8 & 5.0  &$2.19\times10^{10}$  &$5.1\times10^9$& 70.7 &	--	&$4.07\times10^6$ & 3.4& (10)\\
35&	NGC~4204    & 12:15:14.44 &	20:39:30.14  &	7.8 & 7.8  &$1.32\times10^9$  &$4.9\times10^8$& 20.1 &	--	&$1.44\times10^5$ & 0.9& (10)\\
36&	NGC~4517    & 12:32:45.51 &	00:06:54.9  &	10.6 & 6.0  &$1.86\times10^{10}$  &$1.2\times10^{10}$& 28.3 &	--	&$7.01\times10^5$ & 2.3& (10)\\
37&	NGC~5068    & 13:18:54.77  &	$-$21:02:19.66  & 6.03 & 6.0  &$5.37\times10^9$  &$4.83\times10^9$& 15.6 &	--	&$5.31\times10^5$ & 5.1& (10)\\
38&	NGC~5264    & --  & --  & 4.51 & 9.7  &$7.59\times10^8$  &$3.40\times10^8$& 51.0 &	--	&$5.83\times10^5$ & 0.5& (10)\\
39&	NGC~7090    & 21:36:28.19  &	$-$54 33 19.86  & 8.71 & 5.0  &$1.10\times10^{10}$  &$4.36\times10^9$& 56.3 &	--	&$3.03\times10^6$ & 2.4& (10)\\
\hline
40 & [KK2000] 03 & 02 24 44.4 & -73 30 51 & 2.12 & -4.9 & $7.70\times10^7$ & $2.30\times10^7$ & - & - & $1.35\times10^5$ & 3.8 & (11) \\ 
41 & LV J0956-0929 & 09 56 37.57  & -09 29 10.8  & 9.37 & -5 & - & $1.95\times10^8$ & - & - & $9.12\times10^5$ & 3.0 & (11) \\ 
42 & UGC 01104  & 01 32 42.53  & +18 19 01.6  & 7.55 & 9.5 & $3.28\times10^8$ & $1.00\times10^8$ & - & 8.17 & - & 3.7 & (11) \\ 
43 & PGC 154449 & 09 57 08.88  & -09 15 48.7  & 9.68 & -1 & $1.60\times10^8$ & $5.01\times10^7$ & - & 5.25 & - & - & (11) \\ 
44 & ESO 553-046 & 05 27 05.72  & -20 40 41.1  & 6.7 & 4 & $1.56\times10^8$ & $4.79\times10^7$ & - & 1.47 & $1.10\times10^6$ & 4.3 & (11) \\ 
\hline
 \end{tabular}										
  \label{tab_imbhsample}
 \parbox[t]{1.04\textwidth}{\textit{Notes:} Columns 2--5: Galaxy name, R.A., and Decl., and distance to the galaxy. Column 6: Galaxy Hubble type. Columns 7 and 8: Total galaxy’s $K$-band luminosity and total galaxy stellar mass. Columns 9--12: the nucleus stellar velocity dispersion, the galaxy effective radius that enclosed half of the galaxy mass or light, the nuclear star cluster mass, and the effective radius, respectively.  Column 13: references - (1):  \citet{Fahrion20}, (2):  \citet{Nguyen18},  (3): \citet{Pechetti20}, (4): \citet{Walcher05}, (5): \citet{Neumayer12}, (6): \citet{Barth09}, (7): \citet{Carlsten22a}, (8):  \citet{Nguyen21},  (9): \citet{Ashok23}, (10):  \citet{Georgiev14}, (11): \citet{Hoyer23}. Target numbers from 41--45 are galaxies those are not present in \autoref{imbh_sample} (main text).} 
\end{table*}

\begin{figure*}
    \centering\includegraphics[width=0.99\textwidth]{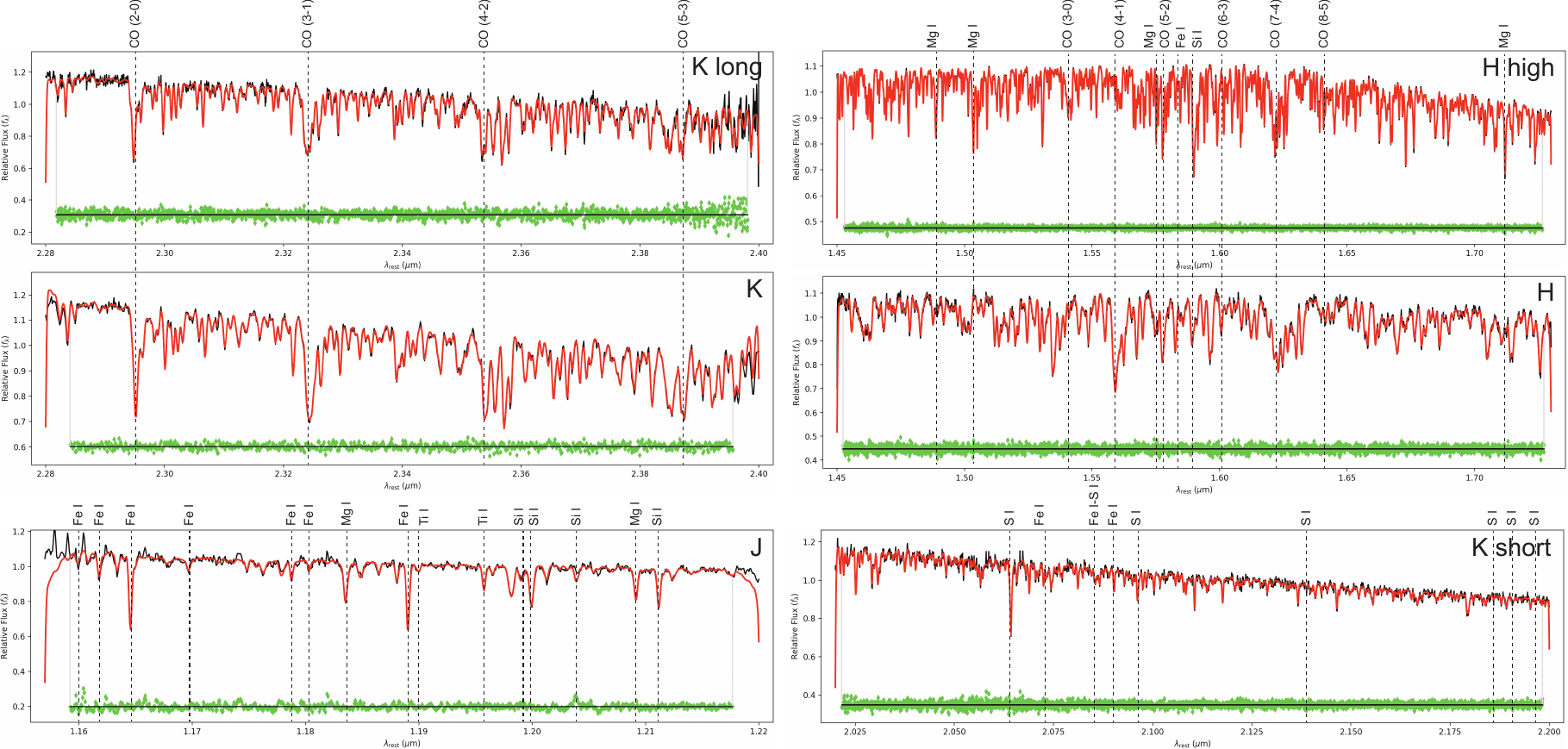}
    \caption{Parts of the simulated spectra over six {\tt HSIM} data cubes, which are fixed as the best spectral wavelength ranges for measuring the stellar kinematics for NGC~300 and NGC~3115 dw01, including $\lambda$2.28--2.40 $\mu$m for both the $K$- and $K$-long, $\lambda$1.45--1.74 $\mu$m for both the $H$- and $H$-high, and $\lambda$1.16--1.218  $\mu$m for the $J$- and $\lambda$2.05--2.20 $\mu$m for $K$-long, specifically. For the fixed spectra parts of the $K$- and $K$-long bands, we used the CO absorption bandheads \citep[e.g. $^{12}$CO(2--0) $\lambda$2.293 $\mu$m, $^{12}$CO(3--1) $\lambda$2.312 $\mu$m, $^{12}$CO(4-2) $\lambda$2.351 $\mu$m, and $^{12}$CO(5-3) $\lambda$2.386 $\mu$m; ][]{Wallace97}. For the fixed spectra parts of the $H$- and $H$-bands, we used the atomic-absorption species of \ion{Mg}{1} $\lambda$1.487, 1.503, 1.575, 1.711 $\mu$m; \ion{Fe}{1} $\lambda$1.583 $\mu$m; \ion{Si}{1} $\lambda$1.589 $\mu$m and the CO-absorptions lines of $^{12}$CO(3--0) $\lambda$1.540 $\mu$m; $^{12}$CO(4--1) $\lambda$1.561 $\mu$m; $^{12}$CO(5--2) $\lambda$1.577 $\mu$m; $^{12}$CO(6--3) $\lambda$1.602 $\mu$m; $^{12}$CO(7--4) $\lambda$1.622 $\mu$m; $^{12}$CO(8--5) $\lambda$1.641 $\mu$m. For the fixed spectrum part of the $J$-band, we used the atomic-absorption species of \ion{Ti}{1} $\lambda$1.1896, 1.1953 $\mu$m; \ion{Si}{1} $\lambda$1.1988, 1.1995, 1.2035, 1.2107 $\mu$m; \ion{Mg}{1} $\lambda$1.1831, 1.2087 $\mu$m; \ion{Fe}{1} $\lambda$1.1597, 1.1611, 1.1641, 1.1693, 1.1783, 1.1801, 1.1887, 1.1976 $\mu$m. For the fixed spectrum part of the $K$-short band, we used the atomic-absorption lines of \ion{Fe}{1} $\lambda$2.088, 2.070 $\mu$m;  \ion{S}{1} $\lambda$2.188, 2.183, 2.179, 2.137, 2.092, 2.062 $\mu$m, and the blended-absorption features of \ion{Fe}{1} -- \ion{S}{1} at $\lambda$2.070, $\lambda$2.083 $\mu$m. In each panel plot, the black vertically thin-dashed lines indicate the positions of atomic/molecule absorptions features of the stellar component extracted from one bin (black line) and its best-fit model produced by {\tt pPXF} (red line). The two gray-vertical lines limit the wavelength range where the spectrum was fit, and blue dots show the residual between the galaxy spectrum and the best-fitting model ({\tt data-model}). The kinematic results of these {\tt pPXF} fits are shown in \autoref{ngc300_mock_kin} and \autoref{ngc300_mock_kin_full} for NGC~300 and \autoref{ngc3115dw01_mock_kin} and \autoref{ngc3115dw01_mock_kin_full} for NGC~3115 dw01 in terms of the 2D maps.}
    \label{spectrum_IMBH}
\end{figure*}

\begin{table*}
\caption{The stellar-light MGE models of NGC~300 and NGC~3115 dw01}    
\centering\begin{tabular}{ccccccc}
 \hline\hline
&$\log_{10} \Sigma_{\star,j} \, $(\Lsun${\rm pc^{-2}})$ &$\sigma_j$ ($\arcsec$) &$q'_j=b_j/a_i$&$\log_{10} \Sigma_{\star,j}/$(\Lsun${\rm pc^{-2}})$ &$\sigma_j$ ($\arcsec$) &$q'_j=b_j/a_i$\\
(1) & (2) & (3) & (4) & (5) & (6) & (7) \\	                
\hline
$j$&NGC~300&NGC~300&NGC~300&NGC~3115 dw01 &NGC~3115 dw01&NGC~3115 dw01\\
\hline
 
1 & {\bf 3.954} & {\bf 0.0007} & {\bf 0.90} & {\bf 4.298} & {\bf 0.0069} & {\bf 0.90} \\ 
2 & {\bf 3.745} & {\bf 0.0012} & {\bf 0.90} & {\bf 4.791} & {\bf 0.0512} & {\bf 0.90} \\ 
3 & {\bf 3.714} & {\bf 0.0020} & {\bf 0.90} & {\bf 4.339} & {\bf 0.1044} & {\bf 0.90} \\ 
4 & {\bf 3.377} & {\bf 0.0028} & {\bf 0.90} & 3.201 & 0.2293 & 0.90 \\ 
5 & {\bf 3.195} & {\bf 0.0039} & {\bf 0.90} & 2.886 & 0.6191 & 0.90 \\ 
6 & {\bf 3.774} & {\bf 0.0052} & {\bf 0.90} & 2.739 & 1.3815 & 0.90 \\ 
7 & {\bf 3.971} & {\bf 0.0119} & {\bf 0.90} & 2.570 & 2.8272 & 0.90 \\ 
8 & {\bf 3.159} & {\bf 0.0174} & {\bf 0.90} & 2.376 & 5.4128 & 0.90 \\ 
9 & {\bf 3.456} & {\bf 0.0267} & {\bf 0.90} & 2.154 & 9.8186 & 0.90 \\ 
10& {\bf 4.334} & {\bf 0.0358} & {\bf 0.90} & 1.905 & 17.040 & 0.90 \\ 
11& {\bf 4.778} & {\bf 0.0877} & {\bf 0.90} & 1.632 & 28.594 & 0.90 \\ 
12& {\bf 3.620} & {\bf 0.1103} & {\bf 0.90} & 1.332 & 47.157 & 0.90 \\ 
13& {\bf 4.814} & {\bf 0.1774} & {\bf 0.90} & 0.994 & 78.302 & 0.90 \\ 
14&      4.518  &      0.5033  &      0.75 & 0.587 & 136.755 & 0.90\\ 
15&      3.312  &      57.735  &      0.75 &$-$0.019& 288.675& 0.90\\ 
\hline
\end{tabular}\\
\label{mgetab}
\vspace{2mm}
\parbox[t]{\textwidth}{\textit{Notes:} The MGE models used in JAM$_{\rm cyl}$ model to create the input cubes for {\tt HSIM} (\autoref{mocks}) and recover the IMBH mass (\autoref{bhrecovering}). Each MGE model has 15 Gaussian components as shown in Column 1. Columns 2, 3 and 4: The MGE models that represented the mass models of the galaxies when scaled with a constant \ml$_{\rm F814W}$, the dispersion of Gaussians, and the ratio between the semimajor and semiminor axes for NGC~300, respectively. Columns 5, 6 and 7: Similarities of Columns 2, 3 and 4 but for NGC~3115 dw01. Bold-face numbers are representative parameters for the Gaussians decomposed from the Core-Sérsic profile, which describes the NSC.} 
\end{table*}

\begin{figure*}
    \centering\includegraphics[width=0.92\textwidth]{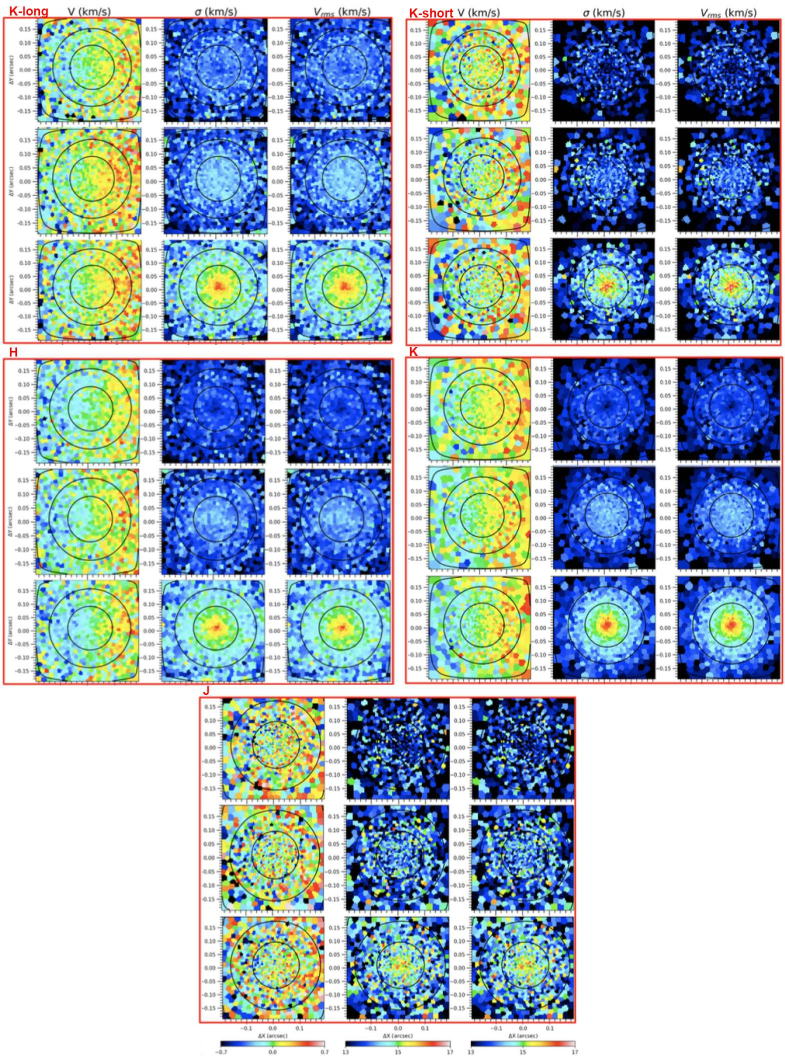}
    \caption{The stellar kinematic maps of NGC~300 extracted from the spectral part of $\lambda$2.28--2.40 $\mu$m ($K$ and $K$-long band), $\lambda$2.05--2.20 $\mu$m ($K$-short band), $\lambda$1.15--1.22 $\mu$m ($H$ band), $\lambda$1.15--1.22 $\mu$m ($J$ band) {\tt HSIM} IFS produced from JAM$_{\rm cyl}$ \citep{Cappellari20} using {\tt pPXF} \citep{Cappellari23}. In each red rectangular, these maps are present with three $M_{\rm BH}$: $ = 0$ \Msun\ (first row), $ = 5\times10^3$ \Msun\ (second row), and $ = 10^4$ \Msun\ (third row). On each row, the kinematic maps are listed orderly with $V$, $\sigma_{\star}$, and $V_{\rm rms}$. The contours indicate the isophotes from the collapsed {\tt HSIM} IFS cubes spaced by 1 mag arcsec$^{-2}$. The color bars are fixed at the same scale for all three $M_{\rm BH}$ to illustrate the kinematic effects of the central BHs and also indicate the robustness of our proposed kinematic measurements at the centers of these dwarf galaxies hosting bright NSCs as the kinematic signatures for IMBHs.} 
    \label{ngc300_mock_kin_full}
\end{figure*}

\begin{figure*}
    \centering\includegraphics[width=0.92\textwidth]{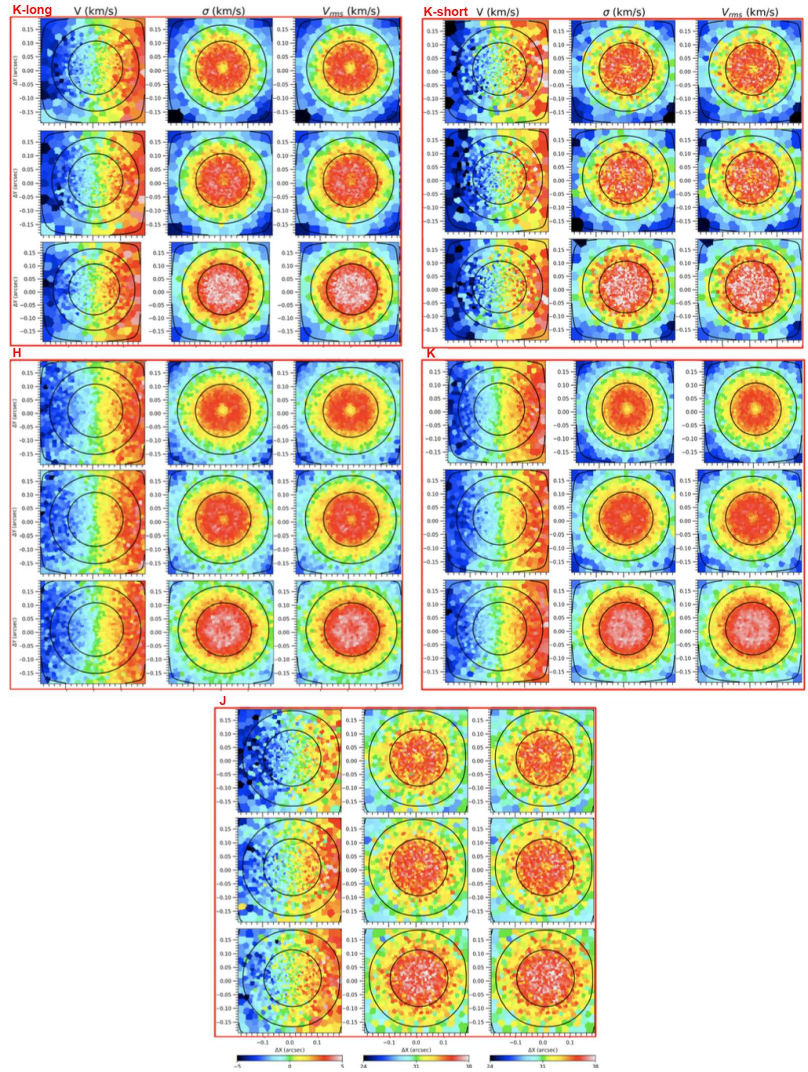}
    \caption{The stellar kinematic maps of NGC~3115 dw01 extracted from the spectral part of $\lambda$2.28--2.40 $\mu$m ($K$ and $K$-long band), $\lambda$2.05--2.20 $\mu$m ($K$-short band), $\lambda$1.15--1.22 $\mu$m ($H$ band), $\lambda$1.15--1.22 $\mu$m ($J$ band) {\tt HSIM} IFS produced from JAM$_{\rm cyl}$ \citep{Cappellari20} using {\tt pPXF} \citep{Cappellari23}. In each red rectangular, these maps are present with three $M_{\rm BH}$: $ = 0$ \Msun\ (first row), $ = 3.5\times10^4$ \Msun\ (second row), and $ = 7\times10^4$ \Msun\ (third row). On each row, the kinematic maps are listed orderly with $V$, $\sigma_{\star}$, and $V_{\rm rms}$. The contours indicate the isophotes from the collapsed {\tt HSIM} IFS cubes spaced by 1 mag arcsec$^{-2}$. The color bars are fixed at the same scale for all three $M_{\rm BH}$ to illustrate the kinematic effects of the central BHs and also indicate the robustness of our proposed kinematic measurements at the centers of these dwarf galaxies hosting bright NSCs as the kinematic signatures for IMBHs.} 
    \label{ngc3115dw01_mock_kin_full}
 \end{figure*}

\begin{figure*}
    \centering\includegraphics[width=0.99\textwidth]{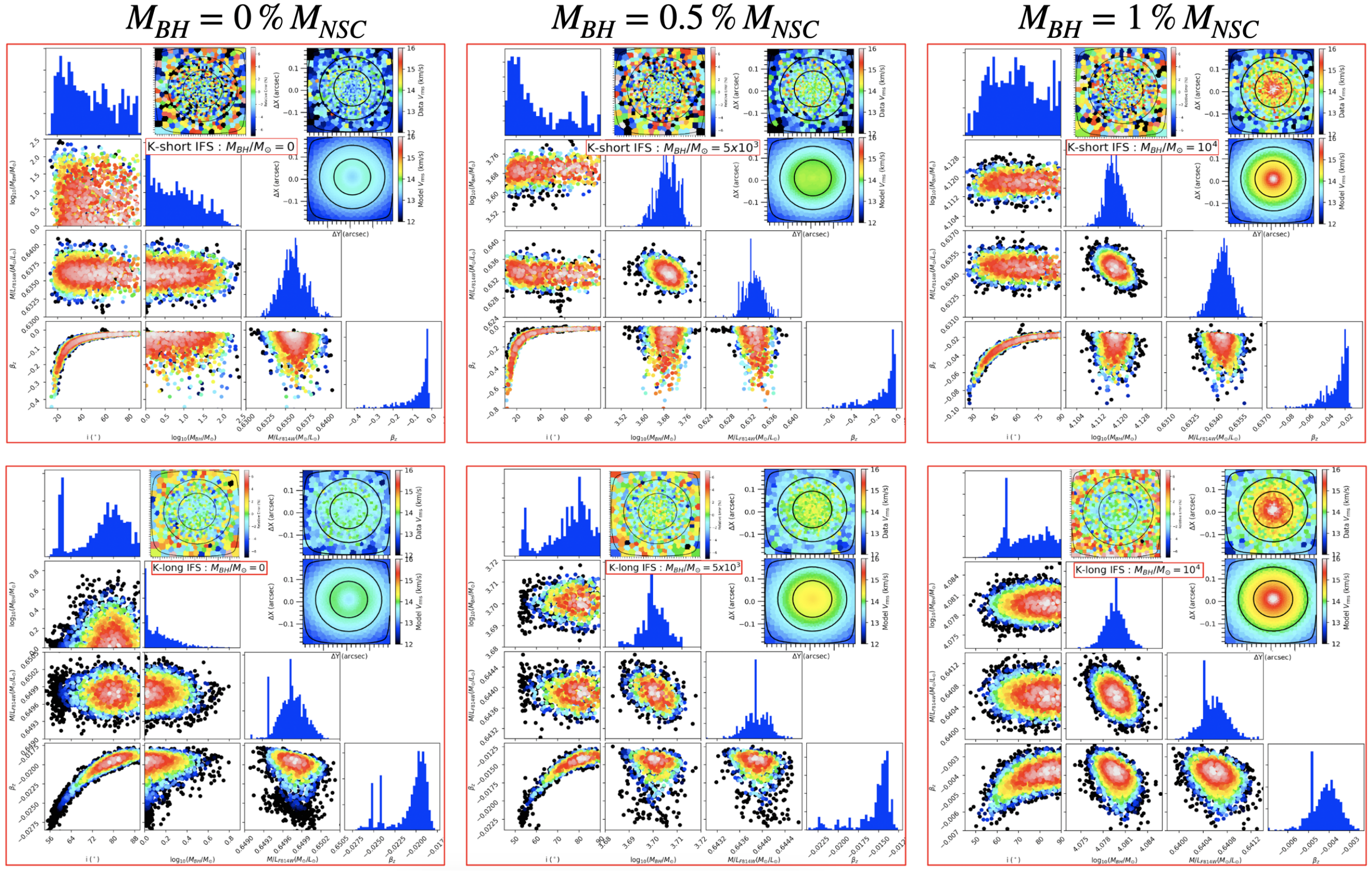}
    \caption{The posterior distributions obtained after the burn-in phase of the {\tt adamet} MCMC optimization process for the JAM$_{\rm cyl}$ models applied to the $K$-short and $K$-long {\tt HSIM} kinematics of NGC~300. These simulations were generated using the JAM$_{\rm cyl}$ models and feature three $M_{\rm BH}$: = 0 \Msun\ (left), $=5\times10^3$ \Msun\ (middle), and $=10^4$ \Msun\ (right). Each red-square panel presents a set of four parameters ($i$, $M_{\rm BH}$, $M/L_{\rm F814W}$, and $\beta_z$), depicted as scatter plots illustrating their projected 2D distributions and histograms displaying their projected 1D distributions. In the top right corner, there are inset maps that depict the $V_{\rm rms}$ values. The top maps represent the simulated kinematic maps extracted from the simulated datacubes, while the bottom maps represent the kinematic maps recovered from the best-fit JAM$_{\rm cyl}$ model. These maps visually illustrate the level of agreement or disagreement at each spaxel between the simulated data and our best-fit model. The determination of the best-fit JAM$_{\rm cyl}$ model is based on the PDF with the highest likelihood.}
    \label{ngc300_mock_kin_BHrecover11}
\end{figure*}

\begin{figure*}
    \centering\includegraphics[width=0.99\textwidth]{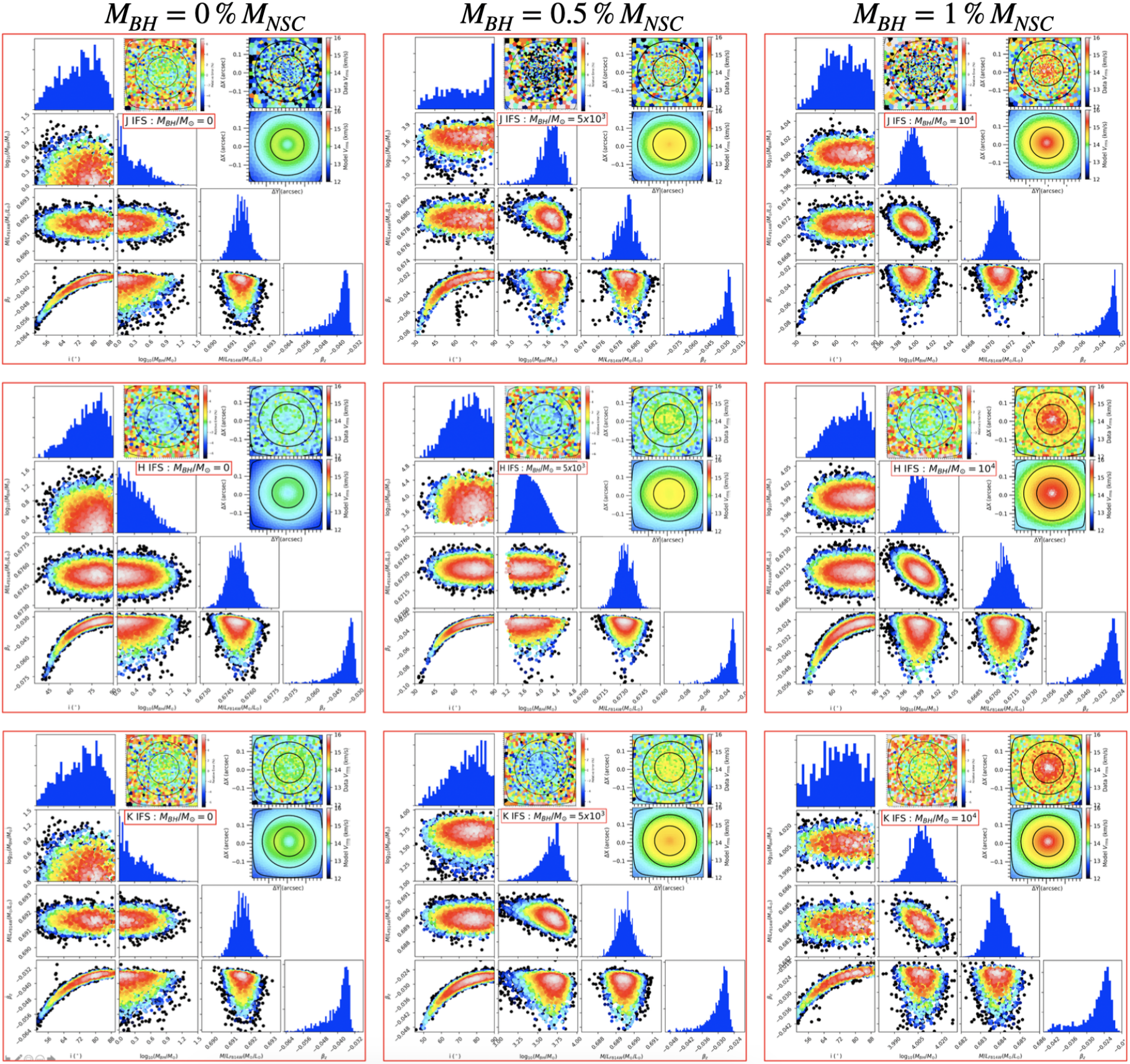}
    \caption{The posterior distributions obtained after the burn-in phase of the {\tt adamet} MCMC optimization process for the JAM$_{\rm cyl}$ models applied to the $J$, $H$, and $K$ {\tt HSIM} kinematics of NGC~300. These simulations were generated using the JAM$_{\rm cyl}$ models and feature three $M_{\rm BH}$: = 0 \Msun\ (left), $=5\times10^3$ \Msun\ (middle), and $= 10^4$ \Msun\ (right). Each red-square panel presents a set of four parameters ($i$, $M_{\rm BH}$, $M/L_{\rm F814W}$, and $\beta_z$), depicted as scatter plots illustrating their projected 2D distributions and histograms displaying their projected 1D distributions. In the top right corner, there are inset maps that depict the $V_{\rm rms}$ values. The top maps represent the simulated kinematic maps extracted from the simulated datacubes, while the bottom maps represent the kinematic maps recovered from the best-fit JAM$_{\rm cyl}$ model. These maps visually illustrate the level of agreement or disagreement at each spaxel between the simulated data and our best-fit model. The determination of the best-fit JAM$_{\rm cyl}$ model is based on the PDF with the highest likelihood.}  
    \label{ngc300_mock_kin_BHrecover22}
\end{figure*}

\begin{figure*}
    \centering\includegraphics[width=0.99\textwidth]{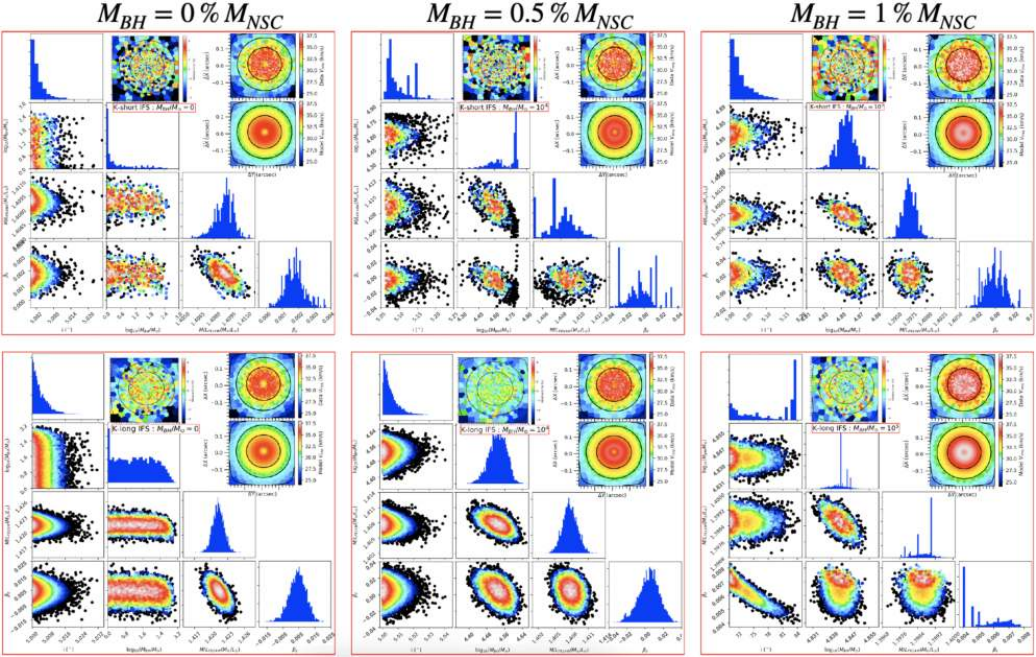}
    \caption{The posterior distributions obtained after the burn-in phase of the {\tt adamet} MCMC optimization process for the JAM$_{\rm cyl}$ models applied to the $K$-short and $K$-long {\tt HSIM} kinematics of NGC~3115 dw01. These simulations were generated using the JAM$_{\rm cyl}$ models and feature three $M_{\rm BH}$: = 0 \Msun\ (left), $=3.5\times10^4$ \Msun\ (middle), and $=7\times10^4$ \Msun\ (right). Each red-square panel presents a set of four parameters ($i$, $M_{\rm BH}$, $M/L_{\rm F814W}$, and $\beta_z$), depicted as scatter plots illustrating their projected 2D distributions and histograms displaying their projected 1D distributions. In the top right corner, there are inset maps that depict the $V_{\rm rms}$ values. The top maps represent the simulated kinematic maps extracted from the simulated datacubes, while the bottom maps represent the kinematic maps recovered from the best-fit JAM$_{\rm cyl}$ model. These maps visually illustrate the level of agreement or disagreement at each spaxel between the simulated data and our best-fit model. The determination of the best-fit JAM$_{\rm cyl}$ model is based on the PDF with the highest likelihood.}
    \label{ngc3115dw01_mock_kin_BHrecover11}
\end{figure*}

\begin{figure*}
    \centering\includegraphics[width=0.99\textwidth]{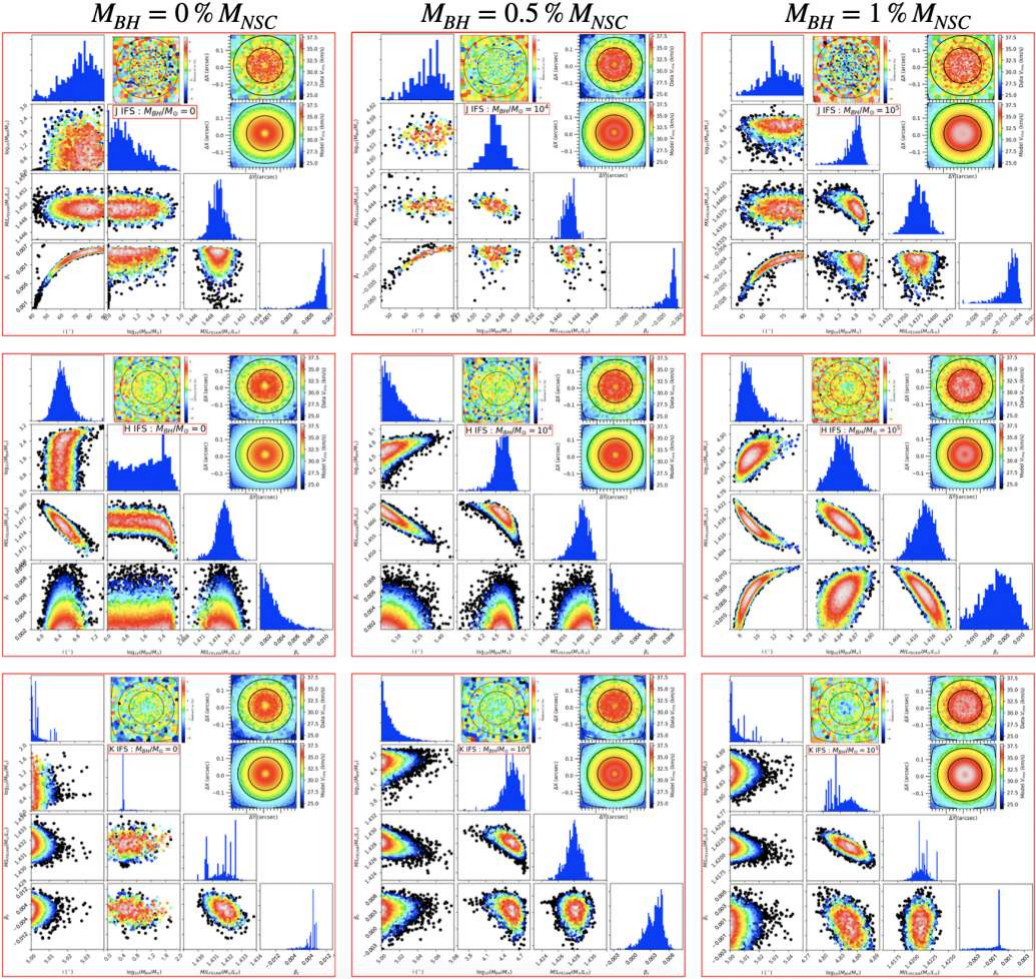}
    \caption{The posterior distributions obtained after the burn-in phase of the {\tt adamet} MCMC optimization process for the JAM$_{\rm cyl}$ models applied to the $J$, $H$, and $K$ {\tt HSIM} kinematics of NGC~3115 dw01. These simulations were generated using the JAM$_{\rm cyl}$ models and feature three $M_{\rm BH}$: = 0 \Msun\ (left), $=3.5\times10^4$ \Msun\ (middle), and $=7\times10^4$ \Msun\ (right). Each red-square panel presents a set of four parameters ($i$, $M_{\rm BH}$, $M/L_{\rm F814W}$, and $\beta_z$), depicted as scatter plots illustrating their projected 2D distributions and histograms displaying their projected 1D distributions. In the top right corner, there are inset maps that depict the $V_{\rm rms}$ values. The top maps represent the simulated kinematic maps extracted from the simulated datacubes, while the bottom maps represent the kinematic maps recovered from the best-fit JAM$_{\rm cyl}$ model. These maps visually illustrate the level of agreement or disagreement at each spaxel between the simulated data and our best-fit model. The determination of the best-fit JAM$_{\rm cyl}$ model is based on the PDF with the highest likelihood.}
    \label{ngc3115dw01_mock_kin_BHrecover22}
\end{figure*}

\begin{table*}
    \centering
     \caption{Best-fit JAM$_{ \rm cyl}$ parameters and their statistical uncertainties for six mock IFS of NGC~300}
    \begin{tabular}{|c|c|ccc|ccc|ccc|}
    \hline\hline
        Grating & Parameters & \multicolumn{3}{c|}{input $M_{\rm BH} = 0$~M$_\odot$} & \multicolumn{3}{c|}{input $\log_{10}(M_{\rm BH}/$M$_\odot) = 3.7$} & \multicolumn{3}{c|}{input $\log_{10}(M_{\rm  BH}/$M$_\odot) = 4$} \\ 
        (1) & (2) & (3) & (4) & (5) & (6) & (7) & (8) & (9) & (10) & (11) \\ 
        ~ & ~ & best-fit & 1$\sigma$ & 3$\sigma$ & best-fit & 1$\sigma$ & 3$\sigma$ & best-fit & 1$\sigma$ & 3$\sigma$ \\ 
        \hline
        ~ & ~ & ~ & (16-84\%) & (0.14-99.86\%) &  & (0.14-99.86\%) & (16-84\%) &  & (16-84\%) & (0.14-99.86\%) \\
         \hline
        $H$-high & $\log_{10}(M_{\rm BH}/$M$_\odot)$ & 0.1 & $<$0.5 & $<$0.8 & 3.710 & $\pm0.012$ & $\pm0.025$ & 4.04 & $\pm0.01$ & $\pm0.03$ \\ 
        ~ & \ml$_{\rm F814W}$ (\Msun/\Lsun)& 0.676 & $\pm0.001$ & $\pm0.003$ & 0.67 & $\pm0.001$ & $\pm0.003$ & 0.665 & $\pm0.001$ & $\pm0.002$ \\ 
        ~ & $i (^{\circ})$ & 77.2 & $\pm7.7$ & $\pm19.0$ & 80.0 & $\pm9.1$ & $\pm17.6$ & 80.0 & $\pm8.2$ & $\pm17.2$ \\ 
        ~ & $\beta_z$ & $-$0.027 & $\pm0.002$ & $\pm0.006$ & $-$0.019 & $\pm0.002$ & $\pm0.005$ & $-$0.017 & $\pm0.002$ & $\pm0.006$ \\ 
        \hline
        $K$-short & $\log_{10}(M_{\rm BH}/$M$_\odot)$  & 0.5 & $<$1.9 & $<$2.4 & 3.70 & $\pm0.04$ & $\pm0.17$ & 4.117 & $\pm0.004$ & $\pm0.013$ \\ 
        ~ & \ml$_{\rm F814W}$ (\Msun/\Lsun)& 0.636 & $\pm0.002$ & $\pm0.004$ & 0.635 & $\pm0.002$ & $\pm0.006$ & 0.634 & $\pm0.001$ & $\pm0.003$ \\ 
        ~ & $i (^{\circ})$ & 84 & $\pm26$ & $\pm37$ & 56 & $\pm30$ & $\pm40$ & 89 & $\pm20$ & $\pm32$ \\ 
        ~ & $\beta_z$ & $-$0.024 & $\pm0.069$ & $\pm0.190$ & $-$0.028 & $\pm0.120$ & $\pm0.400$ & $-$0.017 & $\pm0.01$ & $\pm0.035$ \\ 
        \hline
        $K$-long & $\log_{10}(M_{\rm BH}/$M$_\odot)$ & 0.1 & $<$0.4 & $<$0.7 & 3.703 & $\pm0.005$ & $\pm0.015$ & 4.080 & $\pm0.001$ & $\pm0.004$ \\ 
        ~ & \ml$_{\rm F814W}$ (\Msun/\Lsun)& 0.650 & $\pm0.001$ & $\pm0.002$ & 0.644 & $\pm0.001$ & $\pm0.002$ & 0.640 & $\pm0.001$ & $\pm0.002$ \\ 
        ~ & $i (^{\circ})$ & 82 & $\pm6$ & $\pm16$ & 81 & $\pm11$ & $\pm18$ & 83.0 & $\pm11$ & $\pm19$ \\ 
        ~ & $\beta_z$ & $-$0.019 & $\pm0.001$ & $\pm0.003$ & $-$0.014 & $\pm0.002$ & $\pm0.009$ & $-$0.004 & $\pm0.001$ & $\pm0.003$ \\ 
        \hline
       $J$ & $\log_{10}(M_{\rm BH}/$M$_\odot)$ & 0.3 & $<$0.6 & $<$1.2 & 3.69 & $\pm0.19$ & $\pm0.45$ & 4.00 & $\pm0.01$ & $\pm0.03$ \\ 
        ~ & \ml$_{\rm F814W}$ (\Msun/\Lsun)& 0.692 & $\pm0.003$ & $\pm0.008$ & 0.679 & $\pm0.001$ & $\pm0.003$ & 0.671 & $\pm0.001$ & $\pm0.003$ \\ 
        ~ & $i (^{\circ})$ & 80 & $\pm7$ & $\pm16$ & 82 & $\pm14$ & $\pm30$ & 75 & $\pm16$ & $\pm29$ \\ 
        ~ & $\beta_z$ & $-$0.038 & $\pm0.140$ & $\pm0.460$ & $-$0.023 & $\pm0.011$ & $\pm0.032$ & $-$0.025 & $\pm0.009$ & $\pm0.033$ \\ 
        \hline
        $H$ & $\log_{10}(M_{\rm BH}/$M$_\odot)$ & 0.1 & $<$0.8 & $<$1.6 & 3.66 & $\pm0.32$ & $\pm0.91$ & 4.00 & $\pm0.03$ & $\pm0.06$ \\ 
        ~ &\ml$_{\rm F814W}$ (\Msun/\Lsun)& 0.675 & $\pm0.001$ & $\pm0.002$ & 0.673 & $\pm0.001$ & $\pm0.002$ & 0.671 & $\pm0.001$ & $\pm0.003$ \\ 
        ~ & $i (^{\circ})$ & 81 & $\pm11$ & $\pm23$ & 82 & $\pm14$ & $\pm24$ & 82 & $\pm14$ & $\pm24$ \\ 
        ~ & $\beta_z$ & $-$0.033 & $\pm0.007$ & $\pm0.031$ & $-$0.025 & $\pm0.006$ & $\pm0.017$ & $-$0.028 & $\pm0.003$ & $\pm0.009$ \\ 
        \hline
        $K$ & $\log_{10}(M_{\rm BH}/$M$_\odot)$ & 0.1 & $<$0.7 & $<$1.2 & 3.710 & $\pm0.11$ & $\pm0.30$ & 4.005 & $\pm0.007$ & $\pm0.015$ \\ 
        ~ & \ml$_{\rm F814W}$ (\Msun/\Lsun)& 0.691 & $\pm0.001$ & $\pm0.002$ & 0.689 & $\pm0.001$ & $\pm0.002$ & 0.684 & $\pm0.001$ & $\pm0.002$ \\ 
        ~ & $i (^{\circ})$ & 80 & $\pm11$ & $\pm22$ & 79 & $\pm15$ & $\pm31$ & 80 & $\pm15$ & $\pm25$ \\ 
        ~ & $\beta_z$ & $-$0.038 & $\pm0.006$ & $\pm0.016$ & $-$0.028 & $\pm0.006$ & $\pm0.012$ & $-$0.025 & $\pm0.003$ & $\pm0.012$ \\ 
        \hline 
    \end{tabular}
\label{tabA3}
\parbox[t]{\textwidth}{\textit{Notes:} We fixed the search ranges for all four model parameters ($M_{\rm BH}$, $\ml_{\rm F814W}$, $\beta_z$, and $i$) to ensure consistency in the fitting process. These ranges were set as follows: 

• $M_{\rm BH}$: from 0 to $10^6$ \Msun\ (or $\log_{10}(M_{\rm BH}/$\Msun): from 0 to 6) 

• \ml$_{\rm F814W}$: from 0.1 to 3 (\Msun/\Lsun)

• $\beta_z$: from $-$1.0 to 0.99

• $i$: from 5$^{\circ}$ to 90$^{\circ}$

Column 1: The ELT/HARMONI grating. Column 2: The JAM$_{\rm cyl}$ model’s parameters. Columns 3, 4, and 5: The best-fit parameters, 1$\sigma$ (or 16-84\%), and 3$\sigma$ (or 0.14-99.86\%) uncertainties provided by the JAM$_{ \rm cyl}$ models, respectively, when constrained from the HARMONI mock kinematics. These model constraints are associated with the case of no input black hole ($M_{\rm BH} = 0$~M$_\odot$). Columns 6, 7, and 8: Similarities of Columns 3, 4, and 5 but for the case of input black hole mass of $M_{\rm BH}=5\times10^3$ M$_\odot$. Columns 9, 10, and 11: Similarities of Columns 3, 4, and 5 but for the case of input black hole mass of $M_{\rm BH}=10^4$ M$_\odot$.}
\end{table*}

\begin{table*}
    \centering
     \caption{Best-fit JAM$_{ \rm cyl}$ parameters and their statistical uncertainties for six mock IFS of NGC~3115 dw01}
    \begin{tabular}{|c|c|ccc|ccc|ccc|}
    \hline\hline
    	Grating & Parameters & \multicolumn{3}{c|}{input $M_{\rm BH} = 0$~M$_\odot$} & \multicolumn{3}{c|}{input $\log_{10}(M_{\rm BH}/$M$_\odot) = 4.544$} & \multicolumn{3}{c|}{input $\log_{10}(M_{\rm  BH}/$M$_\odot) = 4.845$} \\ 
        (1) & (2) & (3) & (4) & (5) & (6) & (7) & (8) & (9) & (10) & (11) \\ 
         ~ & ~ & best-fit & 1$\sigma$ & 3$\sigma$ & best-fit & 1$\sigma$ & 3$\sigma$ & best-fit & 1$\sigma$ & 3$\sigma$ \\  
         \hline
        ~ & ~ & ~ & (16-84\%) & (0.14-99.86\%) &  & (0.14-99.86\%) & (16-84\%) &  & (16-84\%) & (0.14-99.86\%) \\
        \hline
        $H$-high & $\log_{10}(M_{\rm BH}/$M$_\odot)$ & 2.4 & $<$2.5 & $<$2.6 & 4.543 & $\pm{0.031}$ & $\pm{0.091}$ & 4.850 & $\pm{0.020}$ & $\pm{0.050}$ \\
        ~ & \ml$_{\rm F814W}$ (\Msun/\Lsun)& 1.487 & $\pm{0.001}$ & $\pm{0.004}$ & 1.446 & $\pm{0.002}$ & $\pm{0.005}$ & 1.441 & $\pm{0.002}$ & $\pm{0.005}$ \\
        ~ & $i (^{\circ})$ &  5.00  & $\pm{0.010}$ & $\pm{0.034}$ & 5.001 & $\pm{0.006}$ & $\pm{0.022}$ & 5.000 & $\pm{0.01}$ & $\pm{0.03}$ \\
        ~ & $\beta_z$ & 0.003 & $\pm{0.002}$ & $\pm{0.006}$ & 0.005 & $\pm{0.005}$ & $\pm{0.018}$ & 0.003 & $\pm{0.001}$ & $\pm{0.003}$ \\ 
        \hline
        $K$-short & $\log_{10}(M_{\rm BH}/$M$_\odot)$ & 0.5 & $<$1.8 & $<$2.4 & 4.550 & $\pm{0.13}$ & $\pm{0.27}$ & 4.850 & $\pm{0.01}$ & $\pm{0.08}$ \\
        ~ & \ml$_{\rm F814W}$ (\Msun/\Lsun)  & 1.410 & $\pm{0.002}$ & $\pm{0.010}$ & 1.408 & $\pm{0.003}$ & $\pm{0.007}$ & 1.398 & $\pm{0.001}$ & $\pm{0.003}$ \\
        ~ & $i (^{\circ})$ & 5.000 & $\pm{0.020}$ & $\pm{0.099}$ & 5.07 & $\pm{0.04}$ & $\pm{0.14}$ & 5.00 & $\pm{0.03}$ & $\pm{0.10}$ \\
        ~ & $\beta_z$ & 0.002 & $\pm{0.001}$ & $\pm{0.002}$ & $-$0.002 & $\pm{0.005}$ & $\pm{0.018}$ & -0.002 & $\pm{0.007}$ & $\pm{0.022}$ \\
        \hline
        $K$-long & $\log_{10}(M_{\rm BH}/$M$_\odot)$ & 1.2 & $<$2.8 & $<$3.2 & 4.540 & $\pm{0.04}$ & $\pm{0.12}$ & 4.843 & $\pm{0.003}$ & $\pm{0.010}$ \\
        ~ & \ml$_{\rm F814W}$ (\Msun/\Lsun) & 1.422 & $\pm{0.001}$ & $\pm{0.003}$ & 1.408 & $\pm{0.001}$ & $\pm{0.003}$ & 1.398 & $\pm{0.001}$ & $\pm{0.002}$ \\
        ~ & $i (^{\circ})$ & 5.000 & $\pm{0.004}$ & $\pm{0.012}$ & 5.000 & $\pm{0.005}$ & $\pm{0.014}$ & 70.1 & $\pm{2.1}$ & $\pm{6.6}$ \\
        ~ & $\beta_z$ & 0.004 & $\pm{0.006}$ & $\pm{0.015}$ & 0.000 & $\pm{0.012}$ & $\pm{0.031}$ & 0.007 & $\pm{0.001}$ & $\pm{0.002}$ \\
        \hline
        $J$ & $\log_{10}(M_{\rm BH}/$M$_\odot)$ & 1.0 & $<$2.0 & $<$2.5 & 4.540 & $\pm{0.015}$ & $\pm{0.043}$ & 4.830 & $\pm{0.240}$ & $\pm{0.750}$ \\
        ~ & \ml$_{\rm F814W}$ (\Msun/\Lsun) & 1.449 & $\pm{0.002}$ & $\pm{0.006}$ & 1.443 & $\pm{0.001}$ & $\pm{0.009}$ & 1.438 & $\pm{0.002}$ & $\pm{0.004}$ \\
        ~ & $i (^{\circ})$ & 80 & $\pm{12}$ & $\pm{24}$ & 79 & $\pm{11}$ & $\pm{29}$ & 80 & $\pm{12}$ & $\pm{24}$ \\
        ~ & $\beta_z$ &0.006 & $\pm{0.001}$ & $\pm{0.003}$ & $-$0.007 & $\pm{0.006}$ & $\pm{0.020}$ & $-$0.005 & $\pm{0.005}$ & $\pm{0.016}$ \\
        \hline
        $H$ & $\log_{10}(M_{\rm BH}/$M$_\odot)$ & 2.4 & $<$2.0 & $<$2.5 & 4.580 & $\pm{0.15}$ & $\pm{0.59}$ & 4.846 & $\pm{0.022}$ & $\pm{0.053}$ \\
        ~ & \ml$_{\rm F814W}$ (\Msun/\Lsun) & 1.475 & $\pm{0.002}$ & $\pm{0.007}$ & 1.463 & $\pm{0.003}$ & $\pm{0.007}$ & 1.414 & $\pm{0.004}$ & $\pm{0.009}$ \\
        ~ & $i (^{\circ})$ & 6.49 & $\pm{0.18}$ & $\pm{0.69}$ & 5.006 & $\pm{0.075}$ & $\pm{0.210}$ & 9.0 & $\pm{1.2}$ & $\pm{3.5}$ \\
        ~ & $\beta_z$ & 0.002 & $\pm{0.002}$ & $\pm{0.006}$ & 0.001 & $\pm{0.003}$ & $\pm{0.008}$ & -0.001& $\pm{0.008}$ & $\pm{0.019}$ \\
         \hline
        $K$ & $\log_{10}(M_{\rm BH}/$M$_\odot)$ & 0.5 & $<$1.2 & $<$1.6 & 4.550 & $\pm{0.16}$ & $\pm{0.54}$ & 4.845 & $\pm{0.023}$ & $\pm{0.045}$ \\
        ~ & \ml$_{\rm F814W}$ (\Msun/\Lsun) & 1.432 & $\pm{0.001}$ & $\pm{0.002}$ & 1.428 & $\pm{0.001}$ & $\pm{0.004}$ & 1.421 & $\pm{0.001}$ & $\pm{0.003}$ \\
        ~ & $i (^{\circ})$ & 5.000 & $\pm{0.004}$ & $\pm{0.010}$ & 5.000 & $\pm{0.01}$ & $\pm{0.04}$ & 5.000 & $\pm{0.007}$ & $\pm{0.018}$ \\
        ~ & $\beta_z$ & 0.001 & $\pm{0.004}$ & $\pm{0.012}$ & 0.003 & $\pm{0.002}$ & $\pm{0.005}$ & $-$0.002 & $\pm{0.001}$ & $\pm{0.004}$ \\
        \hline
        
    \end{tabular}
   \label{tabA4}
\parbox[t]{\textwidth}{\textit{Notes:} We fixed the search ranges for all four model parameters ($M_{\rm BH}$, $\ml_{\rm F814W}$, $\beta_z$, and $i$) to ensure consistency in the fitting process. These ranges were set as follows: 

• $M_{\rm BH}$: from 0 to $10^6$ \Msun\ (or $\log_{10}(M_{\rm BH}/$\Msun): from 0 to 6) 

• \ml$_{\rm F814W}$: from 0.1 to 3 (\Msun/\Lsun)

• $\beta_z$: from $-$1.0 to 0.99

• $i$: from 5$^{\circ}$ to 90$^{\circ}$

Column 1: The ELT/HARMONI grating. Column 2: The JAM$_{ \rm cyl}$ model’s parameters. Columns 3, 4, and 5: The best-fit parameters, 1$\sigma$ (or 16-84\%), and 3$\sigma$ (or 0.14-99.86\%) uncertainties provided by the JAM$_{ \rm cyl}$ models, respectively, when constrained from the HARMONI mock kinematics. These model constraints are associated with the case of no input black hole ($M_{\rm BH} = 0$~M$_\odot$). Columns 6, 7, and 8: Similarities of Columns 3, 4, and 5 but for the case of input black hole mass of $M_{\rm BH}=3.5\times10^4$ M$_\odot$. Columns 9, 10, and 11: Similarities of Columns 3, 4, and 5 but for the case of input black hole mass of $M_{\rm BH}=7\times10^4$ M$_\odot$.} 
\end{table*}

\label{lastpage}
\end{document}